\documentclass[10pt,reqno,a4paper]{amsart}
\usepackage[british]{babel}
\usepackage{amssymb,amstext,amsthm,eucal,bbm,mathrsfs,amscd,bbold}
\usepackage[margin=2cm]{geometry}
\usepackage{array}
\usepackage[svgnames,table]{xcolor}
\usepackage{graphicx}
\usepackage[utf8]{inputenc}
\usepackage{enumitem}
\usepackage[small]{eulervm}
\usepackage{tgpagella}
\usepackage{tikz,pgfplots}
\usetikzlibrary{calc}
\pgfplotsset{compat=1.14}
\usepackage{booktabs}
\usepackage[unicode]{hyperref}
\hypersetup{%
  pdftitle   = {Conformal Lie algebras via deformation theory},
  pdfkeywords = {Lie algebra, deformations, conformal, graded},
  pdfauthor  = {José Figueroa-O'Farrill},
  pdfcreator = {\LaTeX\ with package \flqq hyperref\frqq},
  linkcolor=NavyBlue,
  citecolor=ForestGreen,
  urlcolor=OrangeRed,
  anchorcolor=OrangeRed,
  colorlinks=true
}
\theoremstyle{plain}

\theoremstyle{definition}
\newtheorem{definition}{Definition}
\theoremstyle{definition}

\renewcommand{\a}{\mathfrak{a}}
\renewcommand{\b}{\mathfrak{b}}
\renewcommand{\c}{\mathfrak{c}}
\renewcommand{\d}{\partial}
\renewcommand{\r}{\mathfrak{r}}
\newcommand{\g}{\mathfrak{g}}

\renewcommand{\k}{\mathfrak{k}}

\newcommand{\p}{\mathfrak{p}}
\newcommand{\co}{\mathfrak{co}}
\newcommand{\e}{\mathfrak{e}}
\newcommand{\n}{\mathfrak{n}}

\renewcommand{\sl}{\mathfrak{sl}}
\newcommand{\so}{\mathfrak{so}}

\newcommand{\h}{\mathfrak{h}}

\newcommand{\s}{\mathfrak{s}}

\newcommand{\B}{\boldsymbol{B}}

\renewcommand{\P}{\boldsymbol{P}}
\newcommand{\Bt}{\widetilde{\boldsymbol{B}}}
\newcommand{\Pt}{\widetilde{\boldsymbol{P}}}
\newcommand{\Ht}{\widetilde{H}}
\newcommand{\J}{\boldsymbol{J}}

\newcommand{\RR}{\mathbb{R}}
\newcommand{\ZZ}{\mathbb{Z}}

\newcommand{\CC}{\mathbb{C}}
\renewcommand{\S}{\mathbb{S}}

\newcommand{\bb}{\mathbb{b}}
\newcommand{\cc}{\mathbb{c}}

\newcommand{\bbbar}{\overline{\mathbb{b}}}
\newcommand{\ccbar}{\overline{\mathbb{c}}}
\newcommand{\bnu}{\mathbb{\nu}}
\newcommand{\bnubar}{\overline{\mathbb{\nu}}}
\newcommand{\tbar}{\overline{t}}
\newcommand{\ubar}{\overline{u}}
\newcommand{\xbar}{\overline{x}}
\newcommand{\ybar}{\overline{y}}
\newcommand{\alphabar}{\overline{\alpha}}

\newcommand{\lambdabar}{\overline{\lambda}}
\newcommand{\psibar}{\overline{\psi}}
\newcommand{\V}{\mathbb{V}}
\newcommand{\Vbar}{\overline{\mathbb{V}}}
\newcommand{\W}{\mathbb{W}}
\newcommand{\spn}[1]{\operatorname{span}_\RR(#1)}
\newcommand{\cspn}[1]{\operatorname{span}_\CC(#1)}

\newcommand{\GL}{\operatorname{GL}}

\newcommand{\Hom}{\operatorname{Hom}}

\newcommand{\GCA}[1]{\mathsf{GCA}_{#1}}

\newcommand{\GLA}[1]{\mathsf{GLA}_{#1}}
\newcommand{\Vp}[1]{V_{+\,#1}}
\newcommand{\Vm}[1]{V_{-\,#1}}
\newcommand{\Vz}[1]{V_{0\,#1}}
\newcommand{\Vpm}[1]{V_{\pm\,#1}}
\newcommand{\Vmp}[1]{V_{\mp\,#1}}
\newcommand{\NR}[1]{[\![#1]\!]}
\usepackage{mathtools}
\newcommand{\choice}[2]{\substack{\mathrlap{#1}\\ \mathrlap{#2}}}
\definecolor{gris}{rgb}{0.8,0.8,0.8}
\newcommand{\zero}{{\color{gris}0}}
\allowdisplaybreaks[0]
\begin{document}

\title{Conformal Lie algebras via deformation theory}
\author{José M. Figueroa-O'Farrill}
\address{Maxwell Institute and School of Mathematics, The University
  of Edinburgh, James Clerk Maxwell Building, Peter Guthrie Tait Road,
  Edinburgh EH9 3FD, United Kingdom}
\begin{abstract}
  We discuss possible notions of conformal Lie algebras, paying
  particular attention to graded conformal Lie algebras with
  $d$-dimensional space isotropy: namely, those with a
  $\mathfrak{co}(d)$ subalgebra acting in a prescribed way on the
  additional generators.  We classify those Lie algebras up to
  isomorphism for all $d\geq 2$ following the same methodology used
  recently to classify kinematical Lie algebras, as deformations of
  the ``most abelian'' graded conformal algebra.  We find 17
  isomorphism classes of Lie algebras for $d\neq 3$ and 23 classes for
  $d=3$.  Lie algebra contractions define a partial order in the set
  of isomorphism classes and this is illustrated via the corresponding
  Hesse diagram.  The only metric graded conformal Lie algebras are
  the simple Lie algebras, isomorphic to either $\so(d{+}1,2)$ or
  $\so(d{+}2,1)$.  We also work out the central extensions of the graded
  conformal algebras and study their invariant inner products. We find
  that central extensions of a Lie algebra in $d=3$ and two Lie
  algebras in $d=2$ are metric.  We then discuss several other notions
  of conformal Lie algebras (generalised conformal, Schrödinger and
  Lifshitz Lie algebras) and we present some partial results on their
  classification.  We end with some open problems suggested by our
  results.
\end{abstract}
\thanks{EMPG-18-13}
\maketitle
\tableofcontents

\section{Introduction}
\label{sec:introduction}

In a recent series of papers \cite{Figueroa-OFarrill:2017ycu,
  Figueroa-OFarrill:2017tcy, Andrzejewski:2018gmz} we classified
kinematical Lie algebras in arbitrary dimensions via deformation
theory.  This extends to arbitrary dimension the classic results for
$3+1$ dimensions by Bacry and Nuyts \cite{MR857383}, following up form
the earlier work of Bacry and Lévy-Leblond \cite{MR0238545}.  A
natural extension of the classification problem of kinematical Lie
algebras is the classification of conformal Lie algebras, but one
first has to decide what a conformal Lie algebra is.

Just like not all kinematical Lie algebras act by isometries on a
pseudo-riemannian spacetime, we do not expect all conformal Lie
algebras (regardless of the definition) to act by conformal
transformations on a conformal manifold.  But by the same token, just
like kinematical Lie algebras generalise the Lie algebras of
isometries of the maximally symmetric riemannian and lorentzian
spaces, whichever definition of conformal Lie algebra one adopts, one
would expect that the Lie algebra of conformal Killing vector fields
on one of these spaces qualifies as a conformal Lie
algebra; although as we shall see there are notions closely related to
conformal Lie algebras (e.g., Lifshitz and Schrödinger algebras) which
deviate from this requirement.

The Lie algebra of conformal Killing vector fields of
($d{+}1$)-dimensional euclidean space and Minkowski spacetime (or indeed
of any other simply-connected conformally flat riemannian or
lorentzian manifold of the same dimension) is isomorphic to
$\so(d{+}2,1)$ in the riemannian case or $\so(d{+}1,2)$ in the lorentzian
case.  We let $J_{\mu\nu}$ denote the generators, where the index
$\mu$ decomposes into $(a,+,-,0)$, where $a=1,\dots,d$.  The inner
product $\eta_{\mu\nu}$ has components
\begin{equation}
  \eta_{ab} = \delta_{ab},\qquad \eta_{+-} = 1 \qquad\text{and}\qquad
  \eta_{00} = \varepsilon,
\end{equation}
with $\varepsilon=\pm 1$ in the riemannian and lorentzian cases,
respectively.  Then the generators $J_{\mu\nu}$, which satisfy the Lie
brackets
\begin{equation}
  [J_{\mu\nu},J_{\rho\sigma}] = \eta_{\nu\rho} J_{\mu\sigma} -
  \eta_{\mu\rho} J_{\nu\sigma} - \eta_{\nu\sigma} J_{\mu\rho} +
  \eta_{\mu\sigma} J_{\nu\rho},
\end{equation}
break up into $J_{ab}$, $D := J_{+-}$, $\Vz{a} := J_{a0}$, $\Vpm{a} :=
J_{a\pm}$ and $S_\pm := J_{0\pm}$, with Lie brackets
\begin{equation}
  \label{eq:conformal}
  \begin{aligned}[m]
    [J_{ab}, J_{cd}] &= \delta_{bc} J_{ad} - \delta_{ac} J_{bd} - \delta_{bd} J_{ac} + \delta_{ad} J_{bc}\\
    [J_{ab}, \Vpm{c}] &= \delta_{bc} \Vpm{a} - \delta_{ac} \Vpm{b}\\
    [J_{ab}, \Vz{c}] &= \delta_{bc} \Vz{a} - \delta_{ac} \Vz{b}\\
    [D, \Vpm{a}] &= \pm \Vpm{a}
  \end{aligned}
  \qquad\qquad
  \begin{aligned}[m]
    [D, S_\pm] &= \pm S_\pm\\
    [J_{ab}, S_\pm] &= 0\\
    [J_{ab}, D] &= 0\\
    [D, \Vz{a}] &= 0.
  \end{aligned}
\end{equation}
and, in addition,
\begin{equation}\label{eq:simple-conformal}
  \begin{aligned}[m]
    [S_\pm, \Vz{a}] &= \Vpm{a}\\
    [S_\pm, \Vmp{a}] &= \Vz{a}\\
    [S_+,S_-] &= D
  \end{aligned}
  \qquad\text{and}\qquad
  \begin{aligned}[m]
    [\Vz{a}, \Vz{b}] &= - \varepsilon J_{ab}\\
    [\Vz{a}, \Vpm{b}] &= \varepsilon \delta_{ab} S_\pm\\
    [\Vp{a}, \Vm{b}] &= \varepsilon (J_{ab} + \delta_{ab} D).
  \end{aligned}
\end{equation}

What properties of the above Lie algebras shall we take as
characterising the notion of a ``conformal Lie algebra''?

The obvious geometrical answer is that the above Lie algebras are
isomorphic to Lie algebras of conformal vector fields on
pseudo-riemannian manifolds.  The Lie algebra of conformal Killing
vectors in a pseudo-riemannian manifold has a precise algebraic
structure shared with the Lie algebra of isometries and, more
generally, the Killing superalgebra of a supergravity background
\cite{Figueroa-OFarrill:2016khp}.  Conformal Killing vectors share,
with Killing vectors and Killing spinors in some supergravity
theories, the property that they are parallel sections of a vector
bundle with connection.  The connection in the conformal case is the
one that gives rise to the notion of conformal Killing transport
\cite{MR0250643}, an extension to conformal Killing vectors of earlier
results of Kostant's \cite{MR0084825} for Killing vectors.  All these
(super)algebras have in common that they are filtered deformations of
graded subalgebras of a graded Lie (super)algebra: the (super)algebra
associated to the ``flat'' model for the geometry.  For conformal
Killing transport, the flat model is that of any (simply-connected)
conformally flat manifold which, in the present context, would be one
of the simple conformal Lie algebras above.  Of course, if we insist
on the dimension being $\tfrac12 (d+3)(d+2)$, then the algebra
\emph{is} that of the flat model, but we could drop this requirement
and insist only that it should contain an $\so(d)$ subalgebra and
perhaps that it acted transitively.  This is an interesting problem
which we will not address in this paper.

Although this property of acting like conformal transformations on a
pseudo-riemannian manifold is the property dictated by geometric
orthodoxy, it is a sense too narrow.  Just like not all kinematical
Lie algebras are isomorphic to a Lie algebra of isometries in a
pseudo-riemannian manifold, we do not wish to restrict attention to
Lie algebras of conformal Killing vectors.  Kinematical Lie 
algebras with $d$-dimensional space isotropy, for $d\geq 2$, are
symmetries of non-metrical structures such as galilean (or
Newton--Cartan), carrollian and aristotelian
\cite{Figueroa-OFarrill:2018ilb}.   By analogy with the case of 
lorentzian and riemannian spacetimes, one could define a
\emph{conformal vector field} to be one which generates conformal
rescalings of the relevant structures.  Since Newton--Cartan and
carrollian structures are degenerate, these conformal vector fields
typically span an infinite-dimensional Lie algebra, which in at least
one carrollian case has been shown to coincide with the BMS algebra
\cite{Duval:2014uva}.

Another property of the above simple conformal Lie algebras, and one
that forms the kinematical basis for the AdS/CFT correspondence, is
that they are isomorphic to the Lie algebra of isometries of a
lorentzian manifold in one higher dimension.  This suggests
re-interpreting kinematical Lie algebras with $(d{+}1)$-dimensional
space isotropy as \emph{holographic conformal algebras} with
$d$-dimensional space isotropy.

Moving away from the geometric conformal algebras, we may instead
concentrate on algebraic properties.  At the surface lies the
observation that the simple conformal algebras above have a Lie
subalgebra isomorphic to $\so(d)$, relative to which the additional
generators transform as 3 copies of the vector $d$-dimensional
representation and 3 copies of the scalar one-dimensional
representation.  We shall call such Lie algebras \emph{generalised
  conformal algebras} (see Definition~\ref{def:genca}); although it is
not clear that they have the right to be called conformal.  We
consider them because they form a large class of Lie algebras which
encompass many of the conformal Lie algebras we consider in this
paper.

More closely related to conformality is the fact that the simple
conformal algebras above have a Lie subalgebra isomorphic to $\co(d) =
\so(d) \oplus \RR D$, where the adjoint action of the
\emph{dilatation} $D$ defines a $\ZZ$-grading, where $J_{ab}$, $V_0$ and $D$
have degree $0$, and $V_\pm$ and $S_\pm$ in degree $\pm 1$.  In this
paper we will take this gradation by ``conformal weight'' to be the
defining property of a conformal algebra.  We will not just demand
that the Lie algebra be graded by $D$, but we are also specifying the
conformal weights.

We may generalise this class of Lie algebras along at least two
directions.  Firstly, we can still demand on the $\co(d)$ subalgebra
with $D$ a grading element, but allowing for arbitrary conformal
weights.  This leads to the notion a \emph{generalised Lifshitz
  algebra} (see Definition~\ref{def:gla}), discussed briefly in
Section~\ref{sec:gener-grad-conf}, where as a first step in their
classification, we classify possible $\ZZ$-gradings on kinematical Lie
algebras.

The second generalisation is to demand that the dilatation $D$ should
be part of an $\sl(2,\RR)$ subalgebra.  In other words, the
$\co(d)$-subalgebra extends to an $\so(d) \oplus \sl(2,\RR)$
subalgebra in such a way that the additional generators transform
according to a representation $\V \otimes E$ of
$\so(d) \oplus \sl(2,\RR)$, where $\V$ is the $d$-dimensional vector
representation of $\so(d)$ and $E$ is a representation of $\sl(2,\RR)$
of dimension $2$ or $3$.  In the case where $E$ is the fundamental
representation (so $\dim E = 2$) we will show that for $d\geq 3$ the
Lie algebra is unique up to isomorphism and admits a unique central
extension which is isomorphic to the Schrödinger Lie algebra.  In
general we shall call the central extensions of such Lie algebras
\emph{generalised Schrödinger algebras} (see Definition~\ref{def:gsa})
and we will discuss their classification in
Section~\ref{sec:gener-schr-algebr}.

We shall now define the main characters in this paper: the graded
conformal Lie algebras.

Let $d \geq 2$.  Recall that $\so(d)$ is the Lie algebra of
skew-symmetric linear transformations of $d$-dimensional euclidean
space and $\co(d) = \so(d) \oplus \RR D$ is the extension of $\so(d)$
by dilatations.

\begin{definition}\label{def:gca}
  By a \textbf{graded conformal algebra} (with $d$-dimensional space
  isotropy) we mean a real Lie algebra $\g$ satisfying the following
  properties
  \begin{enumerate}
  \item $\g$ has a subalgebra $\h \cong \co(d)$, and
  \item under the adjoint action of $\h$, $\g = \h \oplus \V_+ \oplus
    \V_- \oplus \V_0 \oplus \S_+ \oplus \S_-$, where $\V_{\pm}, \V_0$
    are vectors under $\so(d)$ and have weights $\pm 1$ and $0$,
    respectively, under $D$, whereas $\S_\pm$ are scalars under
    $\so(d)$ and have weights $\pm 1$ under $D$.
  \end{enumerate}
\end{definition}

In other words, a graded conformal algebra is the real span of
$J_{ab} = - J_{ba}$, $\Vpm{a}$, $\Vz{a}$, $S_\pm$, with
$a,b \in \{1,\dots,d\}$, satisfying equation~\eqref{eq:conformal} and
any other Lie brackets not mentioned here, subject only to the axioms
of a Lie algebra.

Notice that the simple conformal Lie algebras whose brackets are
given by \eqref{eq:conformal} and \eqref{eq:simple-conformal} is not
just $\ZZ$-graded, but actually ($\ZZ_2 \times \ZZ$)-graded, with
$J_{ab},D,S_\pm$ even under $\ZZ_2$ and $\Vpm{a}, \Vz{a}$ odd under
$\ZZ_2$.  This parity is maintained in all graded conformal algebras in
$d\neq 3$, but not necessarily in $d=3$ due to the existence of the
parity-violating vector product in $\RR^3$.

All graded conformal Lie algebras share the brackets
\eqref{eq:conformal} and are distinguished by the additional brackets
between the generators $\Vpm{a}$, $\Vz{a}$, $S_\pm$.  If these
additional brackets are zero, we have the \textbf{static graded
  conformal algebra}, which we will denote $\g$ from now on.  Any
other graded conformal algebra is, by definition, a deformation of
$\g$ and this is the approach we will take towards the classification.
As in the case of kinematical Lie algebras, the problem depends on the
value of $d$, the case $d\geq 4$ being generic, whereas $d \leq 3$
needs special care.  In this paper we will deal with $d\geq 2$;
although the discussion breaks up into several cases: $d \geq 5$,
$d=4$, $d=3$ and $d=2$.  We refer to \cite{Figueroa-OFarrill:2017ycu}
for the methodology and to \cite{Andrzejewski:2018gmz} for the
rationale of working with complex Lie algebras in $d=2$.

This paper is organised as follows.  In
Section~\ref{sec:deformations-dgeq-4} we discuss the deformations of
the static graded conformal algebra for $d\geq 4$, which is the
generic range. The bulk of the discussion is about $d\geq 5$, but in
Section~\ref{sec:deformations-d=4} we simply observe that the generic
case also includes $d=4$ since there is no substantial change in the
calculations. All that happens in $d=4$ is that $\so(4)$ is not
simple, but this does not change the results. After introducing the
deformation complex in Section~\ref{sec:deformation-complex-dgeq5}, we
calculate the second cohomology and thus determine the infinitesimal
deformations in Section~\ref{sec:infin-deform}. In
Section~\ref{sec:obstructions} we work out the obstructions to
integrating the infinitesimal deformations and after solving the
resulting system of quadratic equations, in
Section~\ref{sec:isom-class-conf-dgeq5} we exploit the action of
automorphisms of $\g$ in order to classify the isomorphism classes of
deformations. The results are summarised in Section~\ref{sec:summary}
and particularly in Table~\ref{tab:summary-d-geq-4}. In
Section~\ref{sec:invar-inner-prod} we show that the only metric Lie
algebras in this classification are the simple conformal algebras:
$\so(d{+}1,2)$ and $\so(d{+}2,1)$. In Section~\ref{sec:contractions} we
show how all graded conformal algebras for $d\geq 4$ can be obtained
as contractions of the simple conformal algebras and we exhibit
explicit contractions for each case.

In Section~\ref{sec:deformations-d=3} we discuss the case of $d=3$,
which is substantially different due to the existence of an
$\so(3)$-invariant vector product (or, equivalently, the Levi-Civita
symbol $\epsilon_{abc}$) which can now appear in brackets. The
panorama of graded conformal algebras in $d=3$ is somewhat richer than
in $d\geq 4$, resulting in new graded conformal algebras which have no
analogue in $d\geq 4$.  In Section~\ref{sec:deformation-complex-deq3} we
introduce the differential complex, which is larger than the one for
$d\geq4$.  Its second cohomology is computed in
Section~\ref{sec:infin-deform-1} and in this way determine the space
of infinitesimal deformations.  The obstructions to integrability are
determined in Section~\ref{sec:obstructions-1}.  There we see that the
solution of the integrability conditions leads to several branches of
solutions: one of them being the $d=3$ analogue of the $d\geq 4$
deformations.  The other branches are unique to $d=3$ and their study
concludes in Section~\ref{sec:isom-class-lie} with the determination
of the isomorphism classes of Lie algebras for $d=3$.  The graded
conformal algebras unique to $d=3$ are listed in
Table~\ref{tab:isom-class-conf-3}.  In
Section~\ref{sec:contractions-1} we discuss the contractions and we
summarise the results in Figure~\ref{fig:contractions}, which
illustrates the Hasse diagram of the partial order defined by
contraction on the set of isomorphism classes of graded conformal
algebras.

In Section~\ref{sec:deformations-d=2} we discuss the case of $d=2$,
which although superficially different from $d>2$ actually results in
formally the same classification as $d\geq4$.  The complex for $d=2$
is again different from the generic $d$ due to the existence of the
$\so(2)$-invariant symplectic structure on the vector representation.
This manifests itself in a larger endomorphism ring (namely, $\CC$ as
opposed to $\RR$) and it is therefore convenient to work with a
complexified Lie algebra, while ensuring that at every moment the
deformations are real.  In Section~\ref{sec:complex-d=2} we introduce
the complex Lie algebra whose deformation complex is described in
Section~\ref{sec:deform-compl-d=2}.  Infinitesimal deformations are
determined in Section~\ref{sec:infin-deform-d=2} and the obstructions
in Section~\ref{sec:obstructions-2}.  Finally, the isomorphism classes
of Lie algebras are determined in Section~\ref{sec:isom-class-deform},
with results which are no different from those of $d\geq4$.  This is
in sharp contrast with the case of kinematical Lie algebras
\cite{Andrzejewski:2018gmz}, where the panorama in $d=2$ was
substantially richer than for generic $d$.

In Section~\ref{sec:central-extensions} we work out the central
extensions of the graded conformal algebras.  Again, the results depend on
$d$.  In Section~\ref{sec:centr-extens-dgeq} we classify the universal
central extensions for the Lie algebras in
Tables~\ref{tab:summary-d-geq-4} and \ref{tab:isom-class-conf-3}.  The
results are contained in Table~\ref{tab:summary-cent-ext-d-geq-3}.  In
Section~\ref{sec:metric-d-geq-3-centr-ext} we investigate whether any
central extension admits an invariant inner product.  We find that one
Lie algebra ($\hyperlink{ca16}{\GCA{16}}$), which exists only for
$d=3$, has a metric central extension.  In
Section~\ref{sec:centr-extens-d=2} we classify the universal central
extensions for the $d=2$ avatars of the Lie algebras in
Table~\ref{tab:summary-d-geq-4} and summarise the results in
Table~\ref{tab:summary-cent-ext-d-eq-2}.  The metricity of these central
extensions is investigated in Section~\ref{sec:invar-inner-prod-1},
where it is found that two $d=2$ Lie algebras admit metric central
extensions: $\hyperlink{ca1}{\GCA{1}}$ and $\hyperlink{ca8}{\GCA{8}}$.

In Section~\ref{sec:gener-conf-algebr} we discuss a generalisation of
the notion of graded conformal algebras, where we dispense with the
grading.  It is questionable to consider them as conformal algebras at
all, but they are introduce since they are a large class of algebras
with the right spectrum of generators, which encompass many of the
other kinds of algebras discussed in this paper.  We will not solve
the classification problem here, but for $d\geq 4$ we will write down
the most general deformation and the conditions for integrability, but
will not solve them in general.

In Section \ref{sec:gener-grad-conf} we discuss generalisations of the
Lifshitz algebra (extended by boosts).  To a first approximation,
these Lie algebras are essentially a graded kinematical Lie algebra
extended by the grading element.  We classify the consistent gradings
of kinematical Lie algebras where $\so(d)$ is in degree zero, a result
summarised in Table~\ref{tab:glas}.

In Section~\ref{sec:gener-schr-algebr} we discuss generalisations of
the Schrödinger algebra, which can be understood as a central
extension of a conformal algebra (which might be missing a vectorial
generator).  We define generalised Schrödinger algebras as those
containing an $\so(d) \oplus \sl(2,\RR)$ subalgebra, relative to which
the vectorial generators transform according to $V \otimes E$, where
$V$ is the vector representation of $\so(d)$ and $E$ is a
representation of $\sl(2,\RR)$ of dimension $2$ or $3$.  We classify
them for $d\geq 3$ when $\dim E = 2$ and for $d\geq 4$ when
$\dim E = 3$.

Finally, Section~\ref{sec:conclusions} we summarise the results and
make some general comments.

\section{Deformations for $d\geq 4$}
\label{sec:deformations-dgeq-4}

Let $d\geq 5$ to begin with. We will let $\W$ denote the real vector
space spanned by $\Vpm{a},\Vz{a},S_\pm$ and let $\W^*$ denote its dual,
with canonical dual basis $\upsilon^{\pm}_a,\upsilon^0_a,\sigma^\pm$. (Because
of the existence of the $\so(d)$-invariant $\delta_{ab}$ we are free
to identify vector subscripts and superscripts, explaining why we
write $\upsilon^0_a$ and not, say, $\upsilon^{0\,a}$.) Recall that
$\h \cong \co(d)$ is the Lie subalgebra spanned by $J_{ab}$ and $D$.

\subsection{The deformation complex}
\label{sec:deformation-complex-dgeq5}

The differential complex for deformations of the static graded
conformal algebra is denoted $(C^\bullet,\d)$.  The cochains are those
of the static conformal algebra $\g$ relative to $\h$ and with values
in the adjoint module $\g$:
\begin{equation}
  C^p := C^p(\g,\h;\g) \cong \Hom(\Lambda^p \W,\g)^\h \cong \left(
    \Lambda^p \W^* \otimes \g\right)^\h,
\end{equation}
which is the space of $\h$-equivariant skewsymmetric $p$-multilinear
maps $\W \times \cdots \times \W \to \g$, and the differential $\d : C^p
\to C^{p+1}$ is defined on generators by
\begin{equation}
  \begin{split}
    \d J_{ab} &= \upsilon^+_a \Vp{b} - \upsilon^+_b \Vp{a} + \upsilon^-_a \Vm{b} -
    \upsilon^-_b \Vm{a} +  \upsilon^0_a \Vz{b} - \upsilon^0_b \Vz{a}\\
    \d D &= - \upsilon^+_a \Vp{a} + \upsilon^-_a \Vm{b} - \sigma^+ S_+ +
    \sigma^- S_-,
  \end{split}
\end{equation}
and zero elsewhere, where we have omitted $\otimes$ in expressions
such as $\sigma^+ \otimes S_+$, et cetera.  The cochains do not have
any legs on $\h$ because we do not wish to deform any of the brackets
involving $J_{ab}$ or $D$.  The $\h$-equivariance ensures that the
Jacobi identity involving one generator in $\h$ is satisfied.

Deformation theory only requires only $C^p$ for $p=1,2,3$, of which
the first two are
\begin{equation}\label{eq:cochains}
  \begin{split}
    C^1 &= \spn{\upsilon^+ V_+, \upsilon^- V_-, \upsilon^0 V_0,\sigma^+S_+,\sigma^-  S_-}\\
    C^2 &= \spn{\upsilon^+\upsilon^- J, \upsilon^+\upsilon^- D, \tfrac12 \upsilon^0\upsilon^0 J,
      \sigma^+\sigma^-D, \upsilon^0\upsilon^+S_+, \upsilon^0 \upsilon^- S_-, \sigma^+
      \upsilon^- V_0, \sigma^-\upsilon^+ V_0,\sigma^+\upsilon^0 V_+,\sigma^-\upsilon^0 V_-},
  \end{split}
\end{equation}
where we have introduced shorthand notations such as
\begin{equation}
  \upsilon^+ V_+ := \upsilon^+_a \otimes \Vp{a} \qquad\text{and}\qquad \upsilon^+\upsilon^- J:=
  \upsilon^+_a \wedge \upsilon^-_b \otimes J_{ab}.
\end{equation}

\subsection{Infinitesimal deformations}
\label{sec:infin-deform}

Infinitesimal deformations are classified by $H^2$, which, since
$\d : C^1 \to C^2$ is the zero map, is given by $H^2 = \ker \left(\d :
  C^2 \to C^3\right)$.

The differential $\d : C^2 \to C^3$ has nonzero components
\begin{equation}\label{eq:b3}
  \begin{split}
    \d(\sigma^+\sigma^- D) &= \sigma^+\sigma^-\upsilon^- V_- - \sigma^+\sigma^-\upsilon^+ V_+\\
    \d(\upsilon^+\upsilon^- D) &= -\upsilon^+\upsilon^-\upsilon^+V_+ + \upsilon^+\upsilon^-\upsilon^-V_- -  \sigma^+\upsilon^+\upsilon^- S_+ + \sigma^-\upsilon^+\upsilon^- S_-\\
    \d(\upsilon^+\upsilon^- J) &= \upsilon^+\upsilon^-\upsilon^+V_+ - \upsilon^+\upsilon^-\upsilon^-V_- +  \upsilon^0 \upsilon^+ \upsilon^- V_0 + \upsilon^0 \upsilon^- \upsilon^+ V_0\\
    \d(\tfrac12 \upsilon^0 \upsilon^0 J) &= - \upsilon^0 \upsilon^+ \upsilon^0 V_+ - \upsilon^0 \upsilon^- \upsilon^0 V_-,
  \end{split}
\end{equation}
where the abbreviated notation is such that $\upsilon^+\upsilon^-\upsilon^+V_+ :=
\upsilon^+_a \wedge \upsilon^-_a \wedge \upsilon^+_b \otimes \Vp{b}$, et cetera.
This implies that the infinitesimal (i.e., first order) deformations
are classified by
\begin{equation}
  H^2 = \spn{\upsilon^0\upsilon^+S_+, \upsilon^0 \upsilon^- S_-, \sigma^+ \upsilon^- V_0,
    \sigma^-\upsilon^+ V_0,\sigma^+\upsilon^0 V_+,\sigma^-\upsilon^0 V_-}.
\end{equation}
The most general infinitesimal deformation is therefore given by
\begin{equation}
  \mu_1 = t_1^+ \upsilon^0\upsilon^+S_+ + t_1^- \upsilon^0 \upsilon^- S_- + t_2^+ \sigma^+
  \upsilon^- V_0 + t_2^- \sigma^-\upsilon^+ V_0 + t_3^+ \sigma^+\upsilon^0 V_+ +
  t_3^- \sigma^-\upsilon^0 V_-,
\end{equation}
where we have introduced parameters $t_1^\pm,t_2^\pm,t_3^\pm \in \RR$.

\subsection{Obstructions}
\label{sec:obstructions}

The obstructions to integrating the infinitesimal deformation $\mu_1$
are classes in $H^3$, the first of which is the class of
$\tfrac12 \NR{\mu_1,\mu_1}$, where
$\NR{-,-}:C^p \times C^q \to C^{p+q-1}$ is the Nijenhuis--Richardson
bracket \cite{MR0214636}, which defines a graded Lie superalgebra
structure on $A^\bullet = \bigoplus_{p\geq -1} A^p$, where $A^p =
C^{p+1}$.  If $\alpha \otimes X \in C^{p+1}$ and $\beta \otimes Y \in
C^{q+1}$, then their Nijenhuis--Richardson bracket is given by a
graded commutator
\begin{equation}
  \NR{\alpha\otimes X, \beta\otimes Y} = (\alpha \otimes X) \bullet
  (\beta \otimes Y)- (-1)^{pq}  (\beta \otimes Y) \bullet (\alpha
  \otimes X),
\end{equation}
where the operation $\bullet$, which is \emph{not} associative, is
given by
\begin{equation}
  (\alpha \otimes X) \bullet  (\beta \otimes Y) = \alpha \wedge \imath_X
  \beta \otimes Y.
\end{equation}

Table~\ref{tab:NR-dot-1} lists the $\bullet$ product $\bullet: C^2 \times
C^2 \to C^3$, from where we can read off the Nijenhuis--Richardson
brackets.  The shorthand notation in that table is such that, for example,
\begin{equation}
  \upsilon^0\upsilon^+\upsilon^-V_0 := \upsilon^0_a \wedge \upsilon^+_a \wedge \upsilon^-_b \otimes \Vz{b},
\end{equation}
et cetera.

\begin{table}[h!]
  \centering 
  \caption{Some components of the Nijenhuis--Richardson product $\bullet: C^2 \times C^2 \to C^3$}
  \label{tab:NR-dot-1}
  \setlength{\extrarowheight}{2pt}  
  \begin{tabular}{>{$}l<{$}*{6}{|>{$}c<{$}}}
    \multicolumn{1}{c|}{$\bullet$} & \upsilon^0\upsilon^+S_+ & \upsilon^0 \upsilon^- S_- & \sigma^+ \upsilon^- V_0 & \sigma^-\upsilon^+ V_0 & \sigma^+\upsilon^0 V_+ & \sigma^-\upsilon^0 V_-\\\midrule
    \upsilon^0\upsilon^+S_+ & & & \upsilon^0\upsilon^+\upsilon^-V_0 & & \upsilon^0 \upsilon^+ \upsilon^0 V_+ & \\\hline
    \upsilon^0 \upsilon^- S_- & & & & \upsilon^0\upsilon^-\upsilon^+V_0 & & \upsilon^0 \upsilon^- \upsilon^0 V_-  \\\hline
    \sigma^+ \upsilon^- V_0& - \sigma^+ \upsilon^+ \upsilon^- S_+ &&&&& \sigma^+ \sigma^- \upsilon^- V_- \\\hline
    \sigma^-\upsilon^+ V_0 & & \sigma^- \upsilon^+ \upsilon^- S_- &&&- \sigma^+ \sigma^- \upsilon^+ V_+ & \\\hline
    \sigma^+\upsilon^0 V_+ & & & & \sigma^+ \sigma^- \upsilon^0 V_0 && \\\hline
    \sigma^-\upsilon^0 V_- & & & - \sigma^+ \sigma^- \upsilon^0 V_0 &&&\\
  \end{tabular}
\end{table}

Calculating $\tfrac12 \NR{\mu_1,\mu_1} = \mu_1 \bullet \mu_1$, we find
\begin{multline}
  \tfrac12 \NR{\mu_1,\mu_1}= t_1^+ t_2^+ \upsilon^0\upsilon^+\upsilon^-V_0+t_1^+
  t_3^+ \upsilon^0\upsilon^+\upsilon^0V_+ + t_1^- t_2^- \upsilon^0\upsilon^-\upsilon^+V_0 + t_1^-
  t_3^- \upsilon^0\upsilon^-\upsilon^0V_- - t_1^+ t_2^+ \sigma^+ \upsilon^+ \upsilon^- S_+\\ +
  t_1^- t_2^-  \sigma^- \upsilon^+ \upsilon^- S_- - t_2^- t_3^+ \sigma^+
  \sigma^- \upsilon^+ V_+ + t_2^+ t_3^- \sigma^+ \sigma^- \upsilon^- V_- +
  (t_2^- t_3^+ - t_2^+ t_3^-) \sigma^+  \sigma^- \upsilon^0 V_0.
\end{multline}
This has to be a coboundary, so equal to $\d\mu_2$, where
\begin{equation}
  \mu_2 =u_1 \sigma^+\sigma^- D + u_2 \upsilon^+\upsilon^- D + u_3 \upsilon^+\upsilon^- J
  + \tfrac12 u_4 \upsilon^0 \upsilon^0 J,
\end{equation}
for some $u_1,u_2,u_3,u_4 \in \RR$.  From equation~\eqref{eq:b3}, we
see that
\begin{multline}
  \d\mu_2 = u_3 \upsilon^0\upsilon^+\upsilon^-V_0  - u_4 \upsilon^0\upsilon^+\upsilon^0V_+ + u_3
  \upsilon^0\upsilon^-\upsilon^+V_0 - u_4 \upsilon^0\upsilon^-\upsilon^0V_- - u_2 \sigma^+ \upsilon^+
  \upsilon^- S_+ + u_2 \sigma^- \upsilon^+ \upsilon^- S_-\\ - u_1 \sigma^+ \sigma^-
  \upsilon^+ V_+ + u_1 \sigma^+ \sigma^- \upsilon^- V_- + (u_2-u_3)
  (-\upsilon^+\upsilon^-\upsilon^+V_+ + \upsilon^+\upsilon^-\upsilon^-V_- )
\end{multline}

The first obstruction equation $\d\mu_2 = \frac12 \NR{\mu_1,\mu_1}$
has a solution provided that the following equations are satisfied:
\begin{equation}\label{eq:obs1}
  \begin{aligned}[m]
    u_1 &= t_2^+ t_3^-\\
    u_2 = u_3 &= t_1^+ t_2^+\\
    u_4 &= - t_1^+ t_3^+\\
  \end{aligned}
  \qquad\text{and}\qquad
  \begin{aligned}[m]
    t_1^+ t_2^+ &= t_1^- t_2^-\\
    t_1^+ t_3^+ &= t_1^- t_3^-\\
    t_2^+ t_3^- &= t_2^- t_3^+,
  \end{aligned}
\end{equation}
in which case the infinitesimal deformation $\mu_1$ integrates to
second order with quadratic terms
\begin{equation}
  \mu_2 =t_2^+ t_3^- \sigma^+\sigma^- D + t_1^+ t_2^+ (\upsilon^+\upsilon^- D
  + \upsilon^+\upsilon^- J) - \tfrac12 t_1^+ t_3^+ \upsilon^0 \upsilon^0 J.
\end{equation}

\begin{table}[h!]
  \centering 
  \caption{More components of the Nijenhuis--Richardson product $\bullet: C^2 \times C^2 \to C^3$}
  \label{tab:NR-dot-too}
  \setlength{\extrarowheight}{2pt}  
  \begin{tabular}{>{$}l<{$}*{4}{|>{$}c<{$}}}
    \multicolumn{1}{c|}{$\bullet$} & \sigma^+\sigma^- D & \upsilon^+\upsilon^- D & \upsilon^+\upsilon^- J & \tfrac12 \upsilon^0 \upsilon^0 J\\\midrule
    \upsilon^0\upsilon^+S_+ & \sigma^- \upsilon^0 \upsilon^+ D & & & \\\hline
    \upsilon^0 \upsilon^- S_- & - \sigma^+ \upsilon^0 \upsilon^- D & & &  \\\hline
    \sigma^+ \upsilon^- V_0 & & & & \sigma^+ \upsilon^0 \upsilon^- J\\\hline
    \sigma^-\upsilon^+ V_0 & & & & \sigma^- \upsilon^0 \upsilon^+ J\\\hline
    \sigma^+\upsilon^0 V_+ & & \sigma^+ \upsilon^0 \upsilon^- D &  \sigma^+ \upsilon^0 \upsilon^- J & \\\hline
    \sigma^-\upsilon^0 V_- & & - \sigma^- \upsilon^0 \upsilon^+ D & \sigma^- \upsilon^0 \upsilon^+ J &
  \end{tabular}
\end{table}

To probe integrability to third order, we must solve $\d\mu_3 =
\NR{\mu_1,\mu_2}$ for some $\mu_3$.  Since there are
no cochains with legs on $D$ or $J_{ab}$ in our complex, $\mu_2
\bullet \mu_1 = 0$ and hence $\NR{\mu_1,\mu_2} = \mu_1 \bullet \mu_2$,
which can be calculated making use of Table~\ref{tab:NR-dot-too}.
Using equation~\eqref{eq:obs1}, one finds that $\mu_1 \bullet \mu_2
=0$, so we can take $\mu_3 =0$.  We then notice that $\NR{\mu_2,\mu_2}
= 0$ identically, so that $\mu_4 =0$ and hence all higher deformations
vanish.

In summary, the most general integrable deformation is given by
\begin{multline}
   \mu = t_1^+ \upsilon^0\upsilon^+S_+ + t_1^- \upsilon^0 \upsilon^- S_- + t_2^+ \sigma^+
   \upsilon^- V_0 + t_2^- \sigma^-\upsilon^+ V_0 + t_3^+ \sigma^+\upsilon^0 V_+ +
   t_3^- \sigma^-\upsilon^0 V_-\\ + t_2^+ t_3^- \sigma^+\sigma^- D + t_1^+ t_2^+ (\upsilon^+\upsilon^- D
  + \upsilon^+\upsilon^- J) - \tfrac12 t_1^+ t_3^+ \upsilon^0 \upsilon^0 J,
\end{multline}
subject to the following quadratic equations:
\begin{equation}
    t_1^+ t_2^+ = t_1^- t_2^-, \qquad t_1^+ t_3^+ = t_1^- t_3^- \qquad\text{and}\qquad
    t_2^+ t_3^- = t_2^- t_3^+.
\end{equation}
These equations are easy to interpret: they simply state that 
$(t_1^-,t_1^+)$, $(t_2^+,t_2^-)$ and $(t_3^+,t_3^-)$ are
collinear vectors in $\RR^2$.  In other words, there exists a
\emph{nonzero} vector $(x,y) \in \RR^2$ and reals $\alpha_1,\alpha_2,
\alpha_3$ so that
\begin{equation}
  (t_1^-,t_1^+) = \alpha_1 (x,y), \qquad
  (t_2^+,t_2^-) = \alpha_2 (x,y) \qquad\text{and}\qquad
  (t_3^+,t_3^-) = \alpha_3 (x,y).
\end{equation}
Then the deformation becomes
\begin{multline}
   \mu = \alpha_1 y \upsilon^0\upsilon^+S_+ + \alpha_1 x \upsilon^0 \upsilon^- S_- +
   \alpha_2 x \sigma^+ \upsilon^- V_0 + \alpha_2 y \sigma^-\upsilon^+ V_0 +
   \alpha_3 x \sigma^+\upsilon^0 V_+ + \alpha_3 y \sigma^-\upsilon^0 V_-\\ +
   \alpha_2 \alpha_3 x y \sigma^+\sigma^- D + \alpha_1 \alpha_2 x y (\upsilon^+\upsilon^- D
  + \upsilon^+\upsilon^- J) - \tfrac12 \alpha_1 \alpha_3 x y \upsilon^0 \upsilon^0 J.
\end{multline}

This results in the following Lie brackets in addition to those in
\eqref{eq:conformal}:
\begin{equation}\label{eq:gen-def-dgeq5}
  \begin{aligned}[m]
    [S_+, \Vz{a}] &= \alpha_3 x \Vp{a}\\
    [S_-, \Vz{a}] &= \alpha_3 y \Vm{a}\\    
    [S_+ , \Vm{a}] &= \alpha_2 x \Vz{a}\\
    [S_- , \Vp{a}] &= \alpha_2 y \Vz{a}\\    
    [S_+,S_-] &= \alpha_2 \alpha_3 x y D
  \end{aligned}
  \qquad\text{and}\qquad
  \begin{aligned}[m]
    [\Vz{a}, \Vz{b}] &=  -\alpha_1 \alpha_3 x y     J_{ab}\\
    [\Vz{a}, \Vp{b}] &= \alpha_1 y \delta_{ab} S_+\\
    [\Vz{a}, \Vm{b}] &= \alpha_1 x \delta_{ab} S_-\\
    [\Vp{a}, \Vm{b}] &= \alpha_1 \alpha_2 x y (J_{ab} + \delta_{ab} D).\\
  \end{aligned}
\end{equation}

\subsection{Isomorphism classes of deformations}
\label{sec:isom-class-conf-dgeq5}

Clearly the parameters in equation~\eqref{eq:gen-def-dgeq5} are not
effective.  To bring these brackets to normal form, we need to
identify Lie algebras which are related by automorphisms of the static
graded conformal algebra, which is the ``gauge'' symmetry of the
deformation complex.  We will not need to determine the full
automorphism group, but it will suffice to consider automorphisms of
two types.  One is a $\ZZ_2$ subgroup which acts by
\begin{equation}
  \label{eq:z2-auto}
  J \mapsto J, \qquad D \mapsto -D, \qquad S_\pm \mapsto
  S_\mp,  \qquad V_\pm \mapsto V_\mp \qquad\text{and}\qquad V_0
  \mapsto V_0.
\end{equation}
Its effect on the Lie brackets in \eqref{eq:gen-def-dgeq5} is to
exchange $x$ and $y$.  The second type of automorphism acts by
rescaling some of the generators:
\begin{equation}
  \label{eq:rescale-auto}
  J \mapsto J,  \qquad D \mapsto D, \qquad S_\pm \mapsto
  \alpha_\pm S_\pm, \qquad V_\pm \mapsto \beta_\pm V_\pm \qquad
  \text{and}\qquad V_0 \mapsto \gamma V_0,
\end{equation}
for some $\alpha_\pm, \beta_\pm, \gamma \in \RR^\times$.

There are three cases to consider: $x=0$, $y=0$ and $xy \neq 0$. The
first two are actually related by the automorphism \eqref{eq:z2-auto},
so this leaves two cases to consider: $y=0$ (and hence $x \neq 0$) and
$xy \neq 0$.

\subsubsection{Branch $y=0$ (and hence $x \neq 0$)}
\label{sec:y=0}

In this case the nonzero brackets in addition to those in
\eqref{eq:conformal} are
\begin{equation}
  \begin{split}
    [S_+, \Vz{a}] &= \alpha_3 x \Vp{a}\\
    [S_+ , \Vm{a}] &= \alpha_2 x \Vz{a}\\
    [\Vz{a}, \Vm{b}] &= \alpha_1 x \delta_{ab} S_-.
  \end{split}
\end{equation}
Via automorphism of the type \eqref{eq:rescale-auto}, we can bring any
nonzero $\alpha_i x$ to $1$.  Therefore we have eight isomorphism
classes of graded conformal algebras: the static graded conformal algebra and seven
nontrivial deformations, depending on the whether or not each of
$\alpha_{1,2,3}$ vanishes.  The isomorphism classes of such Lie
algebras are tabulated in Table~\ref{tab:isom-class-conf-1}.

\begin{table}[h!]
  \centering
  \caption{Isomorphism classes of graded conformal algebras ($y=0$)}
  \label{tab:isom-class-conf-1}
  \setlength{\extrarowheight}{2pt}  
  \rowcolors{2}{blue!10}{white}
  \begin{tabular}{*{3}{>{$}c<{$}}|*{3}{>{$}l<{$}}}\toprule
    \alpha_1 & \alpha_2 & \alpha_3 & \multicolumn{3}{c}{Nonzero Lie brackets in addition to \eqref{eq:conformal}} \\\midrule
    0 & 0 & 0 & & & \\
    1 & 0 & 0 & [V_0,V_-] = S_- & & \\
    0 & 1 & 0 & & [S_+, V_-] = V_0 & \\
    0 & 0 & 1 & & & [S_+,V_0] = V_+ \\
    1 & 1 & 0 & [V_0,V_-] = S_-& [S_+, V_-] = V_0 & \\
    1 & 0 & 1 & [V_0,V_-] = S_- & & [S_+,V_0] = V_+\\
    0 & 1 & 1 & & [S_+, V_-] = V_0 & [S_+,V_0] = V_+ \\
    1 & 1 & 1 & [V_0,V_-] = S_- & [S_+, V_-] = V_0 & [S_+,V_0] = V_+ \\\bottomrule
  \end{tabular}
\end{table}

\subsubsection{Branch $xy\neq 0$}
\label{sec:xy-neq-0}

We can now use automorphisms of the type \eqref{eq:rescale-auto} in
order to rescale the generators and bring the structure constants to a
normal form.  Under the automorphism \eqref{eq:rescale-auto}, we find
\begin{equation}
  \begin{aligned}[m]
    \alpha_1 x & \mapsto \frac{\alpha_-}{\gamma\beta_-} \alpha_1 x\\
    \alpha_2 x & \mapsto \frac{\gamma}{\alpha_+\beta_-} \alpha_2 x\\
    \alpha_3 x & \mapsto \frac{\beta_+}{\gamma\alpha_+} \alpha_3 x
  \end{aligned}
  \qquad\text{and}\qquad
  \begin{aligned}[m]
    \alpha_1 y & \mapsto \frac{\alpha_+}{\gamma\beta_+} \alpha_1 y\\
    \alpha_2 y & \mapsto \frac{\gamma}{\alpha_-\beta_+} \alpha_2 y\\
    \alpha_3 y & \mapsto \frac{\beta_-}{\gamma\alpha_-} \alpha_3 y.
  \end{aligned}
\end{equation}

Table~\ref{tab:normal-form} shows to what normal form we can bring
$\alpha_i x$ and $\alpha_i y$ depending on whether or not each
$\alpha_i$ vanishes.  The notation is such that $\varepsilon$ is the
sign of $\alpha_1\alpha_3 x y$, which is an invariant: indeed, under
the above rescalings $\alpha_1 \alpha_3 x y \mapsto \gamma^{-2}
\alpha_1\alpha_3 x y$, so the sign cannot change..

\begin{table}[h!]
  \centering
  \caption{Normal forms for structure constants}
  \label{tab:normal-form}
  \begin{tabular}{*{3}{>{$}c<{$}}|*{6}{>{$}c<{$}}}\toprule
    \alpha_1 & \alpha_2 & \alpha_3 & \alpha_1 x & \alpha_1 y & \alpha_2 x & \alpha_2 y & \alpha_3 x & \alpha_3 y\\\midrule
    0 & 0 & 0 & 0 & 0 & 0 & 0 & 0 & 0 \\
    1 & 0 & 0 & 1 & 1 & 0 & 0 & 0 & 0 \\
    0 & 1 & 0 & 0 & 0 & 1 & 1 & 0 & 0 \\
    0 & 0 & 1 & 0 & 0 & 0 & 0 & 1 & 1 \\
    1 & 1 & 0 & 1 & 1 & 1 & 1 & 0 & 0 \\
    1 & 0 & 1 & \varepsilon & \varepsilon & 0 & 0  & 1 & 1 \\
    0 & 1 & 1 & 0 & 0 & 1 & 1 & 1 & 1 \\
    1 & 1 & 1 & \varepsilon & \varepsilon  & 1 & 1 & 1 & 1 \\\bottomrule
  \end{tabular}
\end{table}

The resulting isomorphism classes of Lie algebras are tabulated in
Table~\ref{tab:isom-class-conf-2}, where we recognise the Lie algebras
\eqref{eq:simple-conformal}.

\begin{table}[h!]
  \centering
  \caption{Isomorphism classes of graded conformal algebras ($xy\neq0$)}
  \label{tab:isom-class-conf-2}
  \setlength{\extrarowheight}{2pt}  
  \rowcolors{2}{blue!10}{white}
  \resizebox{\textwidth}{!}{
  \begin{tabular}{*{3}{>{$}c<{$}}|*{6}{>{$}l<{$}}}\toprule
    \alpha_1 & \alpha_2 & \alpha_3 & \multicolumn{6}{c}{Nonzero Lie brackets in addition to \eqref{eq:conformal}}\\\midrule
    0 & 0 & 0 & & & & & & \\
    1 & 0 & 0 & & & & & [V_0,V_\pm] = S_\pm & \\
    0 & 1 & 0 & & [S_\pm, V_\mp] = V_0 & &  & & \\
    0 & 0 & 1 & & & [S_\pm, V_0] = V_\pm & & & \\
    1 & 1 & 0 & & [S_\pm, V_\mp] = V_0 & & & [V_0,V_\pm] = S_\pm & [V_+,V_-]= (J + D) \\
    1 & 0 & 1 & & & [S_\pm, V_0] = V_\pm & [V_0,V_0]=-\varepsilon J & [V_0,V_\pm] = \varepsilon S_\pm & \\
    0 & 1 & 1 & [S_+,S_-]=D & [S_\pm, V_\mp] = V_0 & [S_\pm, V_0] = V_\pm & & & \\
    1 & 1 & 1 & [S_+,S_-]=D & [S_\pm, V_\mp] = V_0 & [S_\pm, V_0] = V_\pm & [V_0,V_0]=-\varepsilon J & [V_0,V_\pm] = \varepsilon S_\pm & [V_+,V_-]=\varepsilon (J + D) \\\bottomrule
  \end{tabular}}
\end{table}

\subsection{Deformations for $d=4$}
\label{sec:deformations-d=4}

If $d=4$ all that happens is that $\so(4)$ is not simple and hence we
can decompose the rotation generators $J_{ab}$ into its (anti)
self-dual parts $J^\pm_{ab}$.  (The sign has nothing to do with the
$D$-weight, despite the notation.)  Following the above calculation
now in $d=4$ we see that the infinitesimal deformations do not change,
so that $\mu_1$ is formally as in $d\geq 5$ and hence so is
$\frac12 \NR{\mu_1,\mu_1}$.   The cochain $\mu_2$ changes in
principle, since now
\begin{equation}
  \mu_2 = u_1 \sigma^+\sigma^- D + u_2 \upsilon^+\upsilon^- D + u^+_3 \upsilon^+\upsilon^-
  J^+ + u^-_3 \upsilon^+\upsilon^- J^- + u^+_4 \tfrac12 \upsilon^0 \upsilon^0 J^+ + u^-_4
  \tfrac12 \upsilon^0 \upsilon^0 J^-.
\end{equation}
However, in order to cancel $\frac12 \NR{\mu_1,\mu_1}$ we must take
$u^+_3 = u^-_3=u_3$ and $u^+_4 = u^-_4=u_4$, where $u_3,u_4$ are as in
the $d\geq 5$ calculation, and thus $\mu_2$ is formally equal
to the one for $d\geq 5$.  The rest of the calculation is formally
identical to the $d\geq 5$ case, with identical results.

\subsection{Summary}
\label{sec:summary}

We can now summarise our results and list the isomorphism classes of
graded conformal algebras with $d\geq 4$.  These are displayed in
Table~\ref{tab:summary-d-geq-4}.  The first column is our label in
this paper and should not be taken too seriously.  The Lie algebra
$\hyperlink{ca15}{\GCA{15}^{(\varepsilon)}}$ is isomorphic to $\so(d{+}1,2)$ if $\varepsilon=-1$
and to $\so(d{+}2,1)$ if $\varepsilon=1$.

\begin{table}[h!]
  \centering
  \caption{Isomorphism classes of graded conformal algebras ($d\geq 4$)}
  \label{tab:summary-d-geq-4}
  \setlength{\extrarowheight}{2pt}  
  \rowcolors{2}{blue!10}{white}
  \begin{tabular}{l|*{6}{>{$}l<{$}}}\toprule
    Label & \multicolumn{6}{c}{Nonzero Lie brackets in addition to \eqref{eq:conformal}}\\\midrule
    \hypertarget{ca1}{$\GCA{1}$} & & & & & & \\
    \hypertarget{ca2}{$\GCA{2}$} & & & & & [V_0,V_-] = S_- & \\
    \hypertarget{ca3}{$\GCA{3}$} & & [S_+, V_-] = V_0 & & & & \\
    \hypertarget{ca4}{$\GCA{4}$} & & & [S_+,V_0] = V_+ & & & \\
    \hypertarget{ca5}{$\GCA{5}$} & & [S_+, V_-] = V_0 & & & [V_0,V_-] = S_- & \\
    \hypertarget{ca6}{$\GCA{6}$} & & & [S_+,V_0] = V_+& & [V_0,V_-] = S_- & \\
    \hypertarget{ca7}{$\GCA{7}$} & & [S_+, V_-] = V_0 & [S_+,V_0] = V_+ & & & \\
    \hypertarget{ca8}{$\GCA{8}$} & & [S_+, V_-] = V_0 & [S_+,V_0] = V_+ & & [V_0,V_-] = S_- & \\
    \hypertarget{ca9}{$\GCA{9}$} & & & & & [V_0,V_\pm] = S_\pm & \\
    \hypertarget{ca10}{$\GCA{10}$} & & [S_\pm, V_\mp] = V_0 & & & & \\
    \hypertarget{ca11}{$\GCA{11}$} & & & [S_\pm, V_0] = V_\pm & & & \\
    \hypertarget{ca12}{$\GCA{12}$} & & [S_\pm, V_\mp] = V_0 & & & [V_0,V_\pm] = S_\pm & [V_+,V_-]= J + D  \\
    \hypertarget{ca13}{$\GCA{13}^{(\varepsilon=\pm1)}$} & & & [S_\pm, V_0] = V_\pm & [V_0,V_0]=-\varepsilon J & [V_0,V_\pm] = \varepsilon S_\pm & \\
    \hypertarget{ca14}{$\GCA{14}$} & [S_+,S_-]=D & [S_\pm, V_\mp] = V_0 & [S_\pm, V_0] = V_\pm & & & \\
    \hypertarget{ca15}{$\GCA{15}^{(\varepsilon=\pm1)}$}& [S_+,S_-]=D & [S_\pm, V_\mp] = V_0 & [S_\pm, V_0] = V_\pm & [V_0,V_0]=-\varepsilon J & [V_0,V_\pm] = \varepsilon S_\pm & [V_+,V_-]=\varepsilon (J + D) \\\bottomrule
  \end{tabular}
\end{table}

\subsection{Invariant inner products}
\label{sec:invar-inner-prod}

Let us now consider whether any of the Lie algebras in
Table~\ref{tab:summary-d-geq-4} admit an invariant inner product; that
is, a non-degenerate symmetric bilinear form $\left<-,-\right>$ which
is ``associative'' in the sense that
\begin{equation}\label{eq:assoc}
  \left<[X,Y],Z\right> = \left<X,[Y,Z]\right>
\end{equation}
for all $X,Y,Z$ in the Lie algebra. As we will now see, with the
exception of the simple Lie algebras $\hyperlink{ca15}{\GCA{15}^{(\varepsilon)}}$, none
of the other Lie algebras in Table~\ref{tab:summary-d-geq-4} admit an
invariant inner product.

Indeed, invariance under $D$ and $J$ already says that for a graded
conformal algebra the only possible nonzero components of such an
inner product are
\begin{equation}\label{ip-comps}
  \left<S_+,S_-\right>, \qquad \left<V_+,V_-\right>, \qquad
  \left<V_0,V_0\right>, \qquad \left<D,D\right> \qquad\text{and}\qquad
  \left<J,J\right>.
\end{equation}
Concentrating on the first component and using \eqref{eq:assoc}, we
find
\begin{equation}
  \left<S_+,S_-\right> =  \left<[D,S_+],S_-\right> = \left<D,[S_+,S_-]\right>,
\end{equation}
so that unless a $D$ appears in $[S_+,S_-]$, the inner product
$\left<S_+,S_-\right>$ is zero and hence it is degenerate.  Similarly,
\begin{equation}
    \left<V_+,V_-\right> =  \left<[D,V_+],V_-\right> = \left<D,[V_+,V_-]\right>,
\end{equation}
so that again the inner product is degenerate unless $D$ appears in
$[V_+,V_-]$.  Inspecting Table~\ref{tab:summary-d-geq-4} we see that
only $\hyperlink{ca15}{\GCA{15}^{(\varepsilon)}}$ satisfies these conditions.  Since
these Lie algebras are simple, the Killing form is non-degenerate and
hence they are the only metric Lie algebras in
Table~\ref{tab:summary-d-geq-4}.

\subsection{Contractions}
\label{sec:contractions}

It is clear from the expression \eqref{eq:gen-def-dgeq5} that every
graded conformal algebra in $d\geq 4$ can be obtained as a contraction
of $\hyperlink{ca15}{\GCA{15}^{(\varepsilon)}}$.  Indeed, let $t \in (0,1]$ and consider the
following invertible linear transformation on the underlying vector
space of the Lie algebra:
\begin{equation}
  \varphi_t J = J, \qquad \varphi_t D = D, \qquad \varphi_t S_\pm =
  t^{a_\pm} S_\pm, \qquad \varphi_t V_\pm = t^{b_\pm} V_\pm
  \qquad\text{and}\qquad \varphi_t V_0 = t^c V_0~,
\end{equation}
for some real numbers $a_\pm, b_\pm, c$.  We define new Lie brackets
(which are isomorphic for all $t \in (0,1]$)
\begin{equation}
  [X,Y]_t = \varphi^{-1}_t [ \varphi_t X, \varphi_t Y].
\end{equation}
If the limit $t \to 0$ of $[-,-]_t$ exists, then it defines a Lie
algebra which is typically not isomorphic to the original Lie algebra
at $t=1$. The Lie algebra defined by $[-,-]_0$ is a contraction of the
Lie algebra defined by $[-,-]_1$. If we start at $t=1$ with
$\hyperlink{ca15}{\GCA{15}^{(\varepsilon)}}$, then by choosing the
weights $a_\pm, b_\pm, c$ judiciously we can arrive at all the other
graded conformal algebras $\hyperlink{ca1}{\GCA{1}}$ to
$\hyperlink{ca14}{\GCA{14}}$. Table~\ref{tab:contractions} gives
(non-unique) values for these weights for each of the non-simple
graded conformal algebras. Of course, contracting with zero weights
does not change the isomorphism class of the Lie algebra.

\begin{table}[h!]
  \centering
  \caption{Weights for contractions of simple conformal algebras}
  \label{tab:contractions}
  \begin{tabular}{l|*{5}{>{$}c<{$}}}\toprule
    \multicolumn{1}{c|}{Contraction} & a_+ & a_- & b_+ & b_- & c \\\midrule
    $\hyperlink{ca15}{\GCA{15}} \to \hyperlink{ca1}{\GCA{1}}$ & 1 & 1 & 1 & 1 & 1 \\
    $\hyperlink{ca15}{\GCA{15}} \to \hyperlink{ca2}{\GCA{2}}$ & 1 & 2 & 1 & 1 & 1 \\
    $\hyperlink{ca15}{\GCA{15}} \to \hyperlink{ca3}{\GCA{3}}$ & 3 & 2 & 2 & 0 & 3 \\
    $\hyperlink{ca15}{\GCA{15}} \to \hyperlink{ca4}{\GCA{4}}$ & 1 & 1 & 2 & 1 & 1 \\
    $\hyperlink{ca15}{\GCA{15}} \to \hyperlink{ca5}{\GCA{5}}$ & 1 & 1 & 1 & 0 & 1 \\
    $\hyperlink{ca15}{\GCA{15}} \to \hyperlink{ca6}{\GCA{6}}$ & 1 & 2 & 2 & 1 & 1 \\
    $\hyperlink{ca15}{\GCA{15}} \to \hyperlink{ca7}{\GCA{7}}$ & 1 & 1 & 3 & 1 & 2 \\\bottomrule
  \end{tabular}
  \hspace{3cm}
  \begin{tabular}{l|*{5}{>{$}c<{$}}}\toprule
    \multicolumn{1}{c|}{Contraction} & a_+ & a_- & b_+ & b_- & c \\\midrule
    $\hyperlink{ca15}{\GCA{15}} \to \hyperlink{ca8}{\GCA{8}}$ & 1 & 1 & 2 & 0 & 1 \\
    $\hyperlink{ca15}{\GCA{15}} \to \hyperlink{ca9}{\GCA{9}}$ & 2 & 1 & 1 & 0 & 1 \\
    $\hyperlink{ca15}{\GCA{15}} \to \hyperlink{ca10}{\GCA{10}}$ & 1 & 0 & 1 & 0 & 1 \\
    $\hyperlink{ca15}{\GCA{15}} \to \hyperlink{ca11}{\GCA{11}}$ & 1 & 0 & 2 & 1 & 1 \\
    $\hyperlink{ca15}{\GCA{15}} \to \hyperlink{ca12}{\GCA{12}}$ & 1 & 1 & 0 & 0 & 1 \\
    $\hyperlink{ca15}{\GCA{15}} \to \hyperlink{ca13}{\GCA{13}}$ & 1 & 0 & 1 & 0 & 0 \\
    $\hyperlink{ca15}{\GCA{15}} \to \hyperlink{ca14}{\GCA{14}}$ & 0 & 0 & 1 & 1 & 1 \\\bottomrule
  \end{tabular}
\end{table}

Contractions define a partial order in the set of isomorphism classes
of conformal Lie algebras and, as any partial order, they have an
associated Hasse diagram.  This diagram appears in
Figure~\ref{fig:contractions}, which also takes into account the
graded conformal algebras for $d=3$ and $d=2$.  The red vertices
depict the algebras unique to $d=3$.  Deleting these vertices and any
edges from it, one recovers the Hasse diagram for graded conformal
algebras in $d\geq 4$ and, as we shall see below, also for $d=2$.

\section{Deformations for $d=3$}
\label{sec:deformations-d=3}

We now extend the above classification to $d=3$.

\subsection{The deformation complex}
\label{sec:deformation-complex-deq3}

The existence of the Levi-Civita symbol $\epsilon_{abc}$ now embiggens
the deformation complex. There are now additional cochains in $C^1$
and $C^2$:
\begin{equation}
  \begin{split}
    C^1 &= C^1_{d\geq 4} \oplus \spn{\tfrac12 \upsilon^0 J}\\
    C^2 &= C^2_{d\geq 4} \oplus \spn{\upsilon^+\upsilon^-V_0, \upsilon^0\upsilon^+ V_+,
      \upsilon^0\upsilon^- V_-, \tfrac12 \upsilon^0 \upsilon^0 V_0, \tfrac12 \sigma^+ \upsilon^- J,
      \tfrac12 \sigma^- \upsilon^+ J},
  \end{split}
\end{equation}
where $C^{1,2}_{d \geq 4}$ are given by equation~\eqref{eq:cochains}
and where we continue using a shorthand notation where, for example,
\begin{equation}
  \upsilon^0 J := \epsilon_{abc} \upsilon^0_a \otimes J_{bc}, \qquad
  \upsilon^+\upsilon^-V_0 := \epsilon_{abc} \upsilon^+_a \wedge \upsilon^-_b \otimes
  \Vz{c},
\end{equation}
et cetera.

\subsection{Infinitesimal deformations}
\label{sec:infin-deform-1}

The differential on generators is unchanged, but now
$\d : C^1 \to C^2$ is no longer the zero map, since
\begin{equation}
  \d(\tfrac12 \upsilon^0 J) = - \upsilon^0 \upsilon^+ V_+ -  \upsilon^0 \upsilon^- V_- -  \upsilon^0 \upsilon^0 V_0,
\end{equation}
whose right-hand side spans the space $B^2$ of $2$-coboundaries.  The
$2$-cocycles are now
\begin{equation}
  Z^2 = \spn{\upsilon^0\upsilon^+S_+, \upsilon^0 \upsilon^- S_-, \sigma^+ \upsilon^- V_0,
    \sigma^-\upsilon^+ V_0,\sigma^+\upsilon^0 V_+,\sigma^-\upsilon^0 V_-, \upsilon^+\upsilon^-V_0, \upsilon^0\upsilon^+ V_+,
      \upsilon^0\upsilon^- V_-,  \tfrac12 \upsilon^0 \upsilon^0 V_0}.
\end{equation}
We may use the freedom to modify cocycles by coboundaries in order to
``normalise'' the cocycles and in this way choose a unique cocycle
representative for each cohomology class.  A convenient normalisation
condition is to demand that the coefficient of
$\tfrac12 \upsilon^0 \upsilon^0 V_0$ should be zero.  This can always be
achieved by adding a suitable multiple of $\d(\tfrac12 \upsilon^0 J)$.  It
is clear that every cohomology class has a unique normalised cocycle
and hence, in summary,
\begin{equation}
  H^2 \cong \spn{\upsilon^0\upsilon^+S_+, \upsilon^0 \upsilon^- S_-, \sigma^+ \upsilon^- V_0,
    \sigma^-\upsilon^+ V_0,\sigma^+\upsilon^0 V_+,\sigma^-\upsilon^0 V_-,
    \upsilon^+\upsilon^-V_0, \upsilon^0\upsilon^+ V_+, \upsilon^0\upsilon^- V_-}.
\end{equation}
The most general infinitesimal deformation is then given by
\begin{multline}
  \mu_1 = t_1^+ \upsilon^0\upsilon^+S_+ + t_1^- \upsilon^0 \upsilon^- S_- + t_2^+ \sigma^+
  \upsilon^- V_0 + t_2^- \sigma^-\upsilon^+ V_0 + t_3^+ \sigma^+\upsilon^0 V_+ +
  t_3^- \sigma^-\upsilon^0 V_-\\
  + t_4 \upsilon^+\upsilon^-V_0 + t_5^+ \upsilon^0\upsilon^+ V_+ + t_5^- \upsilon^0\upsilon^- V_-,
\end{multline}
for some real parameters $t_1^\pm, t_2^\pm, t_3^\pm, t_4, t_5^\pm$.

\subsection{Obstructions}
\label{sec:obstructions-1}

The first obstruction to integrating $\mu_1$ is given by
$\frac12\NR{\mu_1,\mu_1}$.  We can reuse some of the calculations in
Section~\ref{sec:obstructions} and embed Table~\ref{tab:NR-dot-1} into
Table~\ref{tab:NR-dot-d-3}.

\begin{table}[h!]
  \centering 
  \caption{Some components of $\bullet: C^2 \times C^2 \to C^3$}
  \label{tab:NR-dot-d-3}
  \setlength{\extrarowheight}{2pt}
  \resizebox{\textwidth}{!}{
    \begin{tabular}{>{$}l<{$}*{6}{|>{$}c<{$}}|*{3}{|>{$}c<{$}}}
    \multicolumn{1}{c|}{$\bullet$} & \upsilon^0\upsilon^+S_+ & \upsilon^0 \upsilon^- S_- & \sigma^+ \upsilon^- V_0 & \sigma^-\upsilon^+ V_0 & \sigma^+\upsilon^0 V_+ & \sigma^-\upsilon^0 V_- & \upsilon^+\upsilon^-V_0 & \upsilon^0\upsilon^+V_+ & \upsilon^0\upsilon^-V_-\\\midrule
    \upsilon^0\upsilon^+S_+ & & & \upsilon^0\upsilon^+\upsilon^-V_0 & & \upsilon^0 \upsilon^+ \upsilon^0 V_+ & & & & \\\hline
    \upsilon^0 \upsilon^- S_- & & & & \upsilon^0\upsilon^-\upsilon^+V_0 & & \upsilon^0 \upsilon^- \upsilon^0 V_- & & &  \\\hline
    \sigma^+ \upsilon^- V_0& - \sigma^+ \upsilon^+ \upsilon^- S_+ &&&&& \sigma^+ \sigma^- \upsilon^- V_- & & \sigma^+ \upsilon^+\upsilon^-V_+ & \sigma^+\upsilon^-\upsilon^-V_-\\\hline
    \sigma^-\upsilon^+ V_0 & & \sigma^- \upsilon^+ \upsilon^- S_- &&&- \sigma^+ \sigma^- \upsilon^+ V_+ & & & \sigma^- \upsilon^+\upsilon^+V_+ & \sigma^-\upsilon^+\upsilon^-V_- \\\hline
    \sigma^+\upsilon^0 V_+ & & & & \sigma^+ \sigma^- \upsilon^0 V_0 && & \sigma^+ \upsilon^0 \upsilon^- V_0 & \sigma^+ \upsilon^0 \upsilon^0 V_+ & \\\hline
    \sigma^-\upsilon^0 V_- & & & - \sigma^+ \sigma^- \upsilon^0 V_0 &&& & \sigma^- \upsilon^0 \upsilon^+ V_0 & & \sigma^- \upsilon^0 \upsilon^0 V_-\\\hline\hline
    \upsilon^+\upsilon^-V_0 &\upsilon^+\upsilon^+\upsilon^-S_+ & \upsilon^+ \upsilon^- \upsilon^- S_- &&& -\sigma^+\upsilon^+\upsilon^-V_+ & -\sigma^-\upsilon^+\upsilon^-V_- &&\upsilon^+\upsilon^-\upsilon^+V_+ & -\upsilon^+\upsilon^-\upsilon^-V_-\\\hline
    \upsilon^0\upsilon^+V_+ &- \upsilon^0\upsilon^0\upsilon^+S_+ &&& -\sigma^-\upsilon^0\upsilon^+V_0 &&&\shortstack{$\upsilon^+\upsilon^-\upsilon^0V_0$\\$-\upsilon^0\upsilon^-\upsilon^+V_0$} &\upsilon^0\upsilon^+\upsilon^0V_+ &\\\hline
    \upsilon^0\upsilon^-V_- && -\upsilon^0\upsilon^0\upsilon^-S_- & -\sigma^+\upsilon^0\upsilon^-V_0 & &&&\shortstack{$\upsilon^+\upsilon^-\upsilon^0V_0$\\$-\upsilon^0\upsilon^+\upsilon^-V_0$} && \upsilon^0\upsilon^-\upsilon^0V_- \\\bottomrule
  \end{tabular}} 
\end{table}

We find that $\tfrac12 \NR{\mu_1,\mu_1} = \mu_1 \bullet \mu_1$ is
given by
\begin{multline}
  \tfrac12 \NR{\mu_1,\mu_1} = 
  (t_1^+ t_2^+ -t_4  t_5^-) \upsilon^0\upsilon^+\upsilon^-V_0 + (t_1^+ t_3^+
  + (t_5^+)^2) \upsilon^0\upsilon^+\upsilon^0V_+ + (t_1^- t_2^- - t_4 t_5^+)
  \upsilon^0\upsilon^-\upsilon^+V_0 \\ + (t_1^- t_3^- + (t_5^-)^2)
  \upsilon^0\upsilon^-\upsilon^0V_- - t_1^+ t_2^+ \sigma^+ \upsilon^+ \upsilon^- S_+ + t_1^-
  t_2^-  \sigma^- \upsilon^+ \upsilon^- S_- - t_2^- t_3^+ \sigma^+ \sigma^-
  \upsilon^+ V_+ \\ + t_2^+ t_3^- \sigma^+ \sigma^- \upsilon^- V_- + (t_2^- t_3^+ -
  t_2^+ t_3^-) \sigma^+  \sigma^- \upsilon^0 V_0 + t_4 (t_5^+ + t_5^-)
  \upsilon^+\upsilon^-\upsilon^0V_0 + t_4 t_5^+ \upsilon^+\upsilon^-\upsilon^+V_+ \\ - t_4 t_5^-
  \upsilon^+\upsilon^-\upsilon^-V_- + t_1^+t_4 \upsilon^+\upsilon^+\upsilon^-S_++ t_1^- t_4
  \upsilon^+\upsilon^-\upsilon^-S_- - t_1^+ t_5^+ \upsilon^0\upsilon^0\upsilon^+S_+ - t_1^- t_5^-
  \upsilon^0\upsilon^0\upsilon^-S_-\\ + (t_3^+ t_4 - t_2^+ t_5^-)
  \sigma^+\upsilon^0\upsilon^-V_0 + t_3^+ t_5^+ \sigma^+\upsilon^0\upsilon^0V_+ + t_2^+
  t_5^- \sigma^+\upsilon^-\upsilon^-V_- + (t_2^+ t_5^+-t_3^+t_4)
  \sigma^+\upsilon^+\upsilon^-V_+\\
  + (t_3^- t_4 - t_2^- t_5^+)
  \sigma^-\upsilon^0\upsilon^+V_0 + t_3^- t_5^- \sigma^-\upsilon^0\upsilon^0V_- + t_2^-
  t_5^+ \sigma^-\upsilon^+\upsilon^+V_+ + (t_2^- t_5^--t_3^-t_4)
  \sigma^-\upsilon^+\upsilon^-V_-,
\end{multline}
which has to equal $\d\mu_2$ with
\begin{equation}
  \mu_2 = u_1 \sigma^+\sigma^- D + u_2 \upsilon^+\upsilon^- D + u_3 \upsilon^+\upsilon^- J
  + \tfrac12 u_4 \upsilon^0 \upsilon^0 J + \tfrac12 u_5^+ \sigma^+\upsilon^- J+
  \tfrac12 u_5^- \sigma^-\upsilon^+ J.
\end{equation}
We calculate
\begin{multline}
  \d\mu_2 = u_3 \upsilon^0\upsilon^+\upsilon^-V_0  - u_4 \upsilon^0\upsilon^+\upsilon^0V_+ + u_3
  \upsilon^0\upsilon^-\upsilon^+V_0 - u_4 \upsilon^0\upsilon^-\upsilon^0V_- - u_2 \sigma^+ \upsilon^+
  \upsilon^- S_+ + u_2 \sigma^- \upsilon^+ \upsilon^- S_-\\ - u_1 \sigma^+ \sigma^-
  \upsilon^+ V_+ + u_1 \sigma^+ \sigma^- \upsilon^- V_- + (u_2-u_3)
  (-\upsilon^+\upsilon^-\upsilon^+V_+ + \upsilon^+\upsilon^-\upsilon^-V_- )\\
  + u_5^+(\sigma^+\upsilon^+\upsilon^-V_+ + \sigma^+\upsilon^-\upsilon^-V_- +
  \sigma^+\upsilon^0\upsilon^-V_0) + u_5^- (\sigma^-\upsilon^+\upsilon^+V_+ + \sigma^-\upsilon^+\upsilon^-V_- +
  \sigma^-\upsilon^0\upsilon^+V_0).
\end{multline}
The equation $\d\mu_2 = \tfrac12 \NR{\mu_1,\mu_1}$ is equivalent to
the following conditions:
\begin{equation}\label{eq:integ-2nd-order}
  \begin{aligned}[m]
    u_1 &= t_2^+ t_3^- = t_2^- t_3^+\\
    u_2 = u_3 &= t_1^+ t_2^+ = t_1^- t_2^-\\
    u_4 &= - t_1^+ t_3^+ - (t_5^+)^2 = - t_1^- t_3^- - (t_5^-)^2\\
    u_5^\pm &= t_2^\pm t_5^\mp = t_2^\pm t_5^\pm - t_3^\pm t_4 = t_3^\pm t_4 - t_2^\pm t_5^\mp\\
  \end{aligned}
  \qquad\text{and}\qquad
  \begin{aligned}[m]
    t_1^\pm t_4 &= 0\\
    t_1^\pm t_5^\pm &= 0\\
    t_3^\pm t_5^\pm &= 0\\
    t_4 t_5^\pm &= 0,\\
  \end{aligned}
\end{equation}
after a little simplification.

\begin{table}[h!]
  \centering 
  \caption{More components of $\bullet: C^2 \times C^2 \to C^3$}
  \label{tab:NR-dot-too-d-3}
  \setlength{\extrarowheight}{2pt}
  \begin{tabular}{>{$}l<{$}*{6}{|>{$}c<{$}}}
    \multicolumn{1}{c|}{$\bullet$} & \sigma^+\sigma^- D & \upsilon^+\upsilon^- D & \upsilon^+\upsilon^- J & \tfrac12 \upsilon^0 \upsilon^0 J & \tfrac12 \sigma^+\upsilon^-J & \tfrac12 \sigma^-\upsilon^+J\\\midrule
    \upsilon^0\upsilon^+S_+ & \sigma^- \upsilon^0 \upsilon^+ D & & & & \tfrac12 \upsilon^0\upsilon^+\upsilon^-J &  \\\hline
    \upsilon^0 \upsilon^- S_- & - \sigma^+ \upsilon^0 \upsilon^- D & & & & &  \tfrac12 \upsilon^0\upsilon^-\upsilon^+J  \\\hline
    \sigma^+ \upsilon^- V_0 & & & & \sigma^+ \upsilon^0 \upsilon^- J & & \\\hline
    \sigma^-\upsilon^+ V_0 & & & & \sigma^- \upsilon^0 \upsilon^+ J & & \\\hline
    \sigma^+\upsilon^0 V_+ & & \sigma^+ \upsilon^0 \upsilon^- D &  \sigma^+ \upsilon^0 \upsilon^- J & & &  \tfrac12 \sigma^+\sigma^-\upsilon^0J\\\hline
    \sigma^-\upsilon^0 V_- & & - \sigma^- \upsilon^0 \upsilon^+ D & \sigma^- \upsilon^0 \upsilon^+ J & &- \tfrac12 \sigma^+\sigma^-\upsilon^0J & \\\hline\hline
    \upsilon^+\upsilon^-V_0 & & & & \shortstack{$\tfrac12 \upsilon^0\upsilon^+\upsilon^- J$\\$+ \frac12 \upsilon^0\upsilon^-\upsilon^+ J$} & & \\\hline
    \upsilon^0\upsilon^+V_+ & & \upsilon^0\upsilon^+\upsilon^-D & \shortstack{$-\tfrac12 \upsilon^0\upsilon^-\upsilon^+ J$\\$- \frac12 \upsilon^+\upsilon^-\upsilon^0 J$}& & & -\sigma^-\upsilon^0\upsilon^+J\\\hline
    \upsilon^0\upsilon^-V_- & & -\upsilon^0\upsilon^+\upsilon^-D & \shortstack{$\frac12 \upsilon^+\upsilon^-\upsilon^0 J$\\$-\tfrac12 \upsilon^0\upsilon^+\upsilon^- J$} & & - \sigma^+\upsilon^0\upsilon^-J& \\\bottomrule
  \end{tabular}
\end{table}

To determine the obstruction to integrability to third order we need
to calculate $\NR{\mu_1,\mu_2}$, for which we extend
Table~\ref{tab:NR-dot-too} to the larger
Table~\ref{tab:NR-dot-too-d-3} which includes the new cochains.
In arriving at some of the new entries we have used the identity
\begin{equation}
  J_{ab} = \tfrac12 \epsilon_{abc} \epsilon_{cde} J_{de}.
\end{equation}
As before, $\NR{\mu_1,\mu_2} = \mu_1 \bullet \mu_2$, since $\mu_2 \bullet \mu_1 = 0$.  We calculate
\begin{multline}
  \NR{\mu_1,\mu_2} = (t_1^+ u_1 - t_3^- u_2) \sigma^-\upsilon^0\upsilon^+D +  (t_3^+ u_2 - t_1^- u_1) \sigma^+\upsilon^0\upsilon^-D + (t_5^+-t_5^-) u_2  \upsilon^0\upsilon^+\upsilon^- D \\
  + (t_2^+ u_4 + t_3^+ u_3 - t_5^- u_5^+) \sigma^+ \upsilon^0 \upsilon^- J  + (t_2^- u_4 + t_3^- u_3 - t_5^+ u_5^-) \sigma^- \upsilon^0 \upsilon^+ J\\
  + \tfrac12 (t_4 u_4 - t_5^- u_3 + t_1^+ u_5^+)  \upsilon^0\upsilon^+\upsilon^-J + \tfrac12 (t_4 u_4 - t_5^+ u_3 + t_1^- u_5^-) \upsilon^0\upsilon^-\upsilon^+J \\
  + \tfrac12 (t_5^--t_5^+) u_3 \upsilon^+\upsilon^-\upsilon^0J + \tfrac12 (t_3^+ u_5^- - t_3^- u_5^+) \sigma^+\sigma^-\upsilon^0 J,
\end{multline}
which, after using conditions~\eqref{eq:integ-2nd-order}, reduces to
\begin{equation}
  \NR{\mu_1,\mu_2} = - t_2^+ \left((t_5^+)^2+(t_5^-)^2\right) \sigma^+ \upsilon^0 \upsilon^- J - t_2^- \left((t_5^+)^2+(t_5^-)^2\right) \sigma^- \upsilon^0 \upsilon^+ J.
\end{equation}
This cannot be cancelled with $\d\mu_3$ unless it actually vanishes,
which implies the extra conditions
\begin{equation}
  t_2^\pm \left((t_5^+)^2 + (t_5^-)^2\right) = 0.
\end{equation}
This being the case, the obstruction is overcome with $\mu_3 =0$ and
since $\NR{\mu_2,\mu_2}$ is identically zero, also all higher $\mu_i$
vanish.  In summary, the most general integrable deformation is given
by
\begin{multline}
  \mu = t_1^+ \upsilon^0\upsilon^+S_+ + t_1^- \upsilon^0 \upsilon^- S_- + t_2^+ \sigma^+
  \upsilon^- V_0 + t_2^- \sigma^-\upsilon^+ V_0 + t_3^+ \sigma^+\upsilon^0 V_+ + t_3^- \sigma^-\upsilon^0 V_-
  + t_4 \upsilon^+\upsilon^-V_0\\ + t_5^+ \upsilon^0\upsilon^+ V_+ + t_5^- \upsilon^0\upsilon^- V_- +
  t_2^+ t_3^- \sigma^+\sigma^- D + t_1^+ t_2^+ (\upsilon^+\upsilon^- D +
  \upsilon^+\upsilon^- J)\\ - \tfrac12 \left(t_1^+t_3^+ + (t_5^+)^2\right) \upsilon^0
  \upsilon^0 J + \tfrac12 t_2^+ t_5^- \sigma^+ \upsilon^- J + \tfrac12 t_2^- t_5^+ \sigma^- \upsilon^+ J,
\end{multline}
subject to the conditions
\begin{equation}
  \begin{aligned}[m]
    t_2^+ t_3^- &= t_2^- t_3^+\\
    t_1^+ t_2^+ &= t_1^- t_2^-\\
    t_1^+ t_3^+ - t_1^- t_3^- &= (t_5^- - t_5^+) (t_5^- + t_5^+)\\
    t_2^\pm \left((t_5^+)^2 + (t_5^-)^2\right) &= 0\\
    t_3^\pm t_4 &= 2 t_2^\pm t_5^\mp\\
  \end{aligned}
  \qquad\qquad
  \begin{aligned}[m]
    t_2^\pm t_5^\pm &= 3 t_2^\pm t_5^\mp\\
    t_1^\pm t_4 &= 0\\
    t_1^\pm t_5^\pm &= 0\\
    t_3^\pm t_5^\pm &= 0\\
    t_4 t_5^\pm &= 0.\\
  \end{aligned}
\end{equation}
It follows from these conditions that
\begin{equation}
  t_2^\pm t_5^\pm = 0, \qquad t_2^\pm t_5^\mp = 0
  \qquad\text{and}\qquad t_3^\pm t_4 = 0,
\end{equation}
and hence that $u_5^\pm = 0$.  We may therefore summarise again the
discussion by saying that the most general integrable deformation is 
\begin{multline}\label{eq:deformation-d-3}
  \mu = t_1^+ \upsilon^0\upsilon^+S_+ + t_1^- \upsilon^0 \upsilon^- S_- + t_2^+ \sigma^+
  \upsilon^- V_0 + t_2^- \sigma^-\upsilon^+ V_0 + t_3^+ \sigma^+\upsilon^0 V_+ + t_3^- \sigma^-\upsilon^0 V_-
  + t_4 \upsilon^+\upsilon^-V_0\\ + t_5^+ \upsilon^0\upsilon^+ V_+ + t_5^- \upsilon^0\upsilon^- V_- +
  t_2^+ t_3^- \sigma^+\sigma^- D + t_1^+ t_2^+ (\upsilon^+\upsilon^- D +
  \upsilon^+\upsilon^- J) - \tfrac12 \left(t_1^+t_3^+ + (t_5^+)^2\right) \upsilon^0
  \upsilon^0 J,
\end{multline}
subject to the conditions
\begin{equation}
  \label{eq:integrability-d-3}
  \begin{aligned}[m]
    t_2^+ t_3^- &= t_2^- t_3^+\\
    t_1^+ t_2^+ &= t_1^- t_2^-\\
    t_1^+ t_3^+ - t_1^- t_3^- &= (t_5^- - t_5^+) (t_5^- + t_5^+)\\
  \end{aligned}
  \qquad\qquad
  \begin{aligned}[m]
    t_1^\pm t_5^\pm &= 0\\
    t_2^\pm t_5^\pm = t_2^\mp t_5^\pm &= 0\\
    t_3^\pm t_5^\pm &= 0\\
  \end{aligned}
  \qquad\qquad
  \begin{aligned}[m]
    t_4 t_1^\pm &= 0\\
    t_4 t_3^\pm &= 0\\
    t_4 t_5^\pm &= 0.\\
  \end{aligned}
\end{equation}

The case $t_4 = t_5^\pm = 0$ was already treated in
Section~\ref{sec:obstructions} and proceeding in the same manner we
arrive at the $d=3$ version of Table~\ref{tab:summary-d-geq-4}.  Hence
those Lie algebras exist for all $d\geq 3$.  It remains to classify
graded conformal algebras which are unique to $d=3$ and to this end  we will
assume from now on that at least one of $t_4$, $t_5^+$ and $t_5^-$ is
different from zero.

The last equation ($t_4 t_5^\pm = 0$) in \eqref{eq:integrability-d-3}
implies that if $t_4 \neq 0$ then $t_5^\pm =0$ and, viceversa, if at
least one of $t_5^\pm$ is different from zero, then $t_4 = 0$.  This
breaks up the problem into several branches:
\begin{enumerate}
\item $t_4\neq 0$;
\item $t_5^\pm \neq 0$;
\item $t_5^+ \neq 0$ and $t_5^- = 0$; and
\item $t_5^- \neq 0$ and $t_5^+ = 0$.
\end{enumerate}
Since the last two branches are related by the automorphism
\eqref{eq:z2-auto}, it is enough to consider one of them.

\subsubsection{The branch $t_4 \neq 0$}
\label{sec:branch-t_4-neq-0}

In this case $t_1^\pm = t_3^\pm = t_5^\pm = 0$ and the deformation is
\begin{equation}
  \mu = t_2^+ \sigma^+ \upsilon^- V_0 + t_2^- \sigma^-\upsilon^+ V_0 + t_4 \upsilon^+\upsilon^-V_0,
\end{equation}
with Lie brackets
\begin{equation}
  [S_+, \Vm{a}] = t_2^+ \Vz{a}, \qquad [S_-, \Vp{a}] = t_2^- \Vz{a}
  \qquad\text{and}\qquad [\Vp{a},\Vm{b}] = t_4 \epsilon_{abc} \Vz{c},
\end{equation}
for any $t_2^\pm$ and $t_4\neq 0$.  We can redefine $V_0$ and, without
loss of generality, assume that $t_4 = 1$.  If nonzero, the other two
parameters $t_2^\pm$ can also be set to $1$ by redefining $S_\pm$,
respectively.  The cases $(t_2^+,t_2^-)=(1,0)$ and
$(t_2^+,t_2^-)=(0,1)$ are related by the
automorphism~\eqref{eq:z2-auto}, so they give isomorphic Lie algebras.
In summary, we have three isomorphism classes of graded conformal algebras in
this branch, which are listed in Table~\ref{tab:isom-class-conf-3}.

\begin{table}[h!]
  \centering
  \caption{Isomorphism classes of graded conformal algebras ($d=3, t_4\neq 0$)}
  \label{tab:isom-class-conf-3}
  \setlength{\extrarowheight}{2pt}  
  \rowcolors{2}{blue!10}{white}
  \begin{tabular}{*{2}{>{$}c<{$}}|*{3}{>{$}l<{$}}}\toprule
    t_2^+ & t_2^- & \multicolumn{3}{c}{Nonzero Lie brackets in addition to \eqref{eq:conformal}} \\\midrule
    0 & 0 & & & [V_+,V_-] = V_0\\
    1 & 0 & [S_+,V_-] = V_0 & & [V_+,V_-] = V_0\\
    1 & 1 & [S_+,V_-] = V_0 & [S_-,V_+] = V_0 & [V_+,V_-] = V_0\\\bottomrule
  \end{tabular}
\end{table}

\subsubsection{The branch $t_4 = 0$ and $t_5^+ t_5^-\neq0$}
\label{sec:branch-t_4-t5pt5pm-0}

In this case, $t_1^\pm = t_2^\pm = t_3^\pm = t_4 = 0$ and also
$(t_5^+)^2 = (t_5^-)^2 \neq 0$.  The deformation becomes
\begin{equation}
  \mu = t_5^+ \upsilon^0\upsilon^+ V_+ + t_5^- \upsilon^0\upsilon^- V_- - \tfrac12 (t_5^+)^2 \upsilon^0 \upsilon^0 J.
\end{equation}
If $t_5^+ = t_5^-$, then we can rescale $V_0$ to arrive at
\begin{equation}
  \mu = \upsilon^0\upsilon^+ V_+ + \upsilon^0\upsilon^- V_- - \tfrac12 \upsilon^0 \upsilon^0 J,
\end{equation}
whereas if $t_5^- = - t_5^+$, then under the same rescaling,
\begin{equation}
  \mu = \upsilon^0\upsilon^+ V_+ - \upsilon^0\upsilon^- V_- - \tfrac12 \upsilon^0 \upsilon^0 J.
\end{equation}
In summary, we have two isomorphism classes of Lie algebras
\begin{equation}
  [\Vz{a}, \Vp{b}] = \epsilon_{abc} \Vp{c}, \qquad
  [\Vz{a}, \Vm{b}] = \varepsilon \epsilon_{abc} \Vm{c}
  \qquad\text{and}\qquad
  [\Vz{a}, \Vz{b}] = - J_{ab},
\end{equation}
where $\varepsilon= \pm 1$.

\subsubsection{The branch $t_4 = 0$ and $t_5^-=0$}
\label{sec:branch-t_4-t5m-0}

In this case $t_5^+ \neq 0$ and hence $t_1^+ = t_2^\pm = t_3^+ = 0$.
The remaining integrability condition is
\begin{equation}
  t_1^- t_3^- = (t_5^+)^2.
\end{equation}
Therefore the deformation becomes
\begin{equation}
  \mu = t_1^- \upsilon^0 \upsilon^- S_- + t_3^- \sigma^-\upsilon^0 V_- + t_5^+
  \upsilon^0\upsilon^+ V_+ - \tfrac12 t_1^- t_3^- \upsilon^0 \upsilon^0 J,
\end{equation}
where $t_1^- t_3^- = (t_5^+)^2 \neq 0$.  The Lie brackets are
\begin{equation}
  \begin{split}
    [\Vz{a},\Vm{b}] &= t_1^- \delta_{ab} S_-\\
    [S_-, \Vz{a}] &= t_3^- \Vm{a}\\
    [\Vz{a}, \Vp{b}] &= t_5^+ \epsilon_{abc} \Vp{c}\\
    [\Vz{a},\Vz{b}] &= - t_1^- t_3^- J_{ab}.
  \end{split}
\end{equation}
We can rescale generators $V_0 \mapsto \frac1{t_5^+} V_0$, $S_-
\mapsto \frac{t_1^-}{t_5^+} S_-$ and arrive at
\begin{equation}
  \begin{split}
    [\Vz{a},\Vm{b}] &= \delta_{ab} S_-\\
    [S_-, \Vz{a}] &= \Vm{a}\\
    [\Vz{a}, \Vp{b}] &= \epsilon_{abc} \Vp{c}\\
    [\Vz{a},\Vz{b}] &= - J_{ab}.
  \end{split}
\end{equation}

\subsection{Isomorphism classes of deformations}
\label{sec:isom-class-lie}

We summarise the isomorphism classes of graded conformal algebras
which are unique to $d=3$: they all involve the $\epsilon_{abc}$
symbol in the Lie brackets.  They are listed in
Table~\ref{tab:summary-d-eq-3}, where the first column is the label
used in this paper and which continues from
Table~\ref{tab:summary-d-geq-4}.

\begin{table}[h!]
  \centering
  \caption{Isomorphism classes of graded conformal algebras unique to $d=3$}
  \label{tab:summary-d-eq-3}
  \setlength{\extrarowheight}{2pt}  
  \rowcolors{2}{blue!10}{white}
  \begin{tabular}{l|*{5}{>{$}l<{$}}}\toprule
    Label & \multicolumn{5}{c}{Nonzero Lie brackets in addition to \eqref{eq:conformal}}\\\midrule
    \hypertarget{ca16}{$\GCA{16}$} & & & & [V_+,V_-] = V_0 & \\
    \hypertarget{ca17}{$\GCA{17}$} & [S_+, V_-] = V_0 & & & [V_+,V_-] = V_0 & \\
    \hypertarget{ca18}{$\GCA{18}$} & [S_\pm, V_\mp] = V_0 & & & [V_+,V_-] = V_0 & \\
    \hypertarget{ca19}{$\GCA{19}$} & [S_-,V_0] = V_- & [V_0,V_+] = V_+ & [V_0, V_-] = S_- & & [V_0,V_0] = -J\\
    \hypertarget{ca20}{$\GCA{20}^{(\varepsilon=\pm1)}$} & & [V_0,V_+] = V_+ & [V_0, V_-] = \varepsilon V_- & & [V_0,V_0] = -J \\\bottomrule
  \end{tabular}
\end{table}

It is not hard to see that none of these Lie algebras admit an
invariant inner product.  Although now for $d=3$ invariance under $J$ and
$D$ allows for a component $\left<V_0,J\right>$ in addition to those
in equation~\eqref{ip-comps}, the argument in
Section~\ref{sec:invar-inner-prod} still holds and non-degeneracy
requires $D$ appearing in $[S_+,S_-]$ and $[V_+,V_-]$, which is not
the case for any of the Lie algebras in
Table~\ref{tab:summary-d-eq-3}.

\subsection{Contractions}
\label{sec:contractions-1}

\begin{figure}[h!]
  \centering
  \begin{tikzpicture}[>=latex, shorten >=2.5pt, shorten <=2.5pt, x=2cm,y=2cm]
    %
    %
    %
    %
    \coordinate [label=above:{\tiny $\hyperlink{ca15}{\GCA{15}}$}] (xv) at (1,4);
    \coordinate [label=above left:{\tiny $\hyperlink{ca14}{\GCA{14}}$}] (xiv) at (0,3);
    \coordinate [label=above left:{\tiny $\hyperlink{ca13}{\GCA{13}}$}] (xiii) at (-1,3);
    \coordinate [label=above right:{\tiny $\hyperlink{ca12}{\GCA{12}}$}] (xii) at (2,3);
    \coordinate [label=above left:{\tiny $\hyperlink{ca11}{\GCA{11}}$}] (xi) at (-3,2);
    \coordinate [label=below right:{\tiny $\hyperlink{ca10}{\GCA{10}}$}] (x) at (3,2);
    \coordinate [label=below left:{\tiny $\hyperlink{ca9}{\GCA{9}}$}] (ix) at (0,2);
    \coordinate [label=above left:{\tiny $\hyperlink{ca8}{\GCA{8}}$}] (viii) at (1,3);
    \coordinate [label=above right:{\tiny $\hyperlink{ca7}{\GCA{7}}$}] (vii) at (1,2);
    \coordinate [label=above left:{\tiny $\hyperlink{ca6}{\GCA{6}}$}] (vi) at (-1,2);
    \coordinate [label=right:{\tiny $\hyperlink{ca5}{\GCA{5}}$}] (v) at (2,2);
    \coordinate [label=below left:{\tiny $\hyperlink{ca4}{\GCA{4}}$}] (iv) at (-1,1);
    \coordinate [label=below right:{\tiny $\hyperlink{ca3}{\GCA{3}}$}] (iii) at (2,1);
    \coordinate [label=below right:{\tiny $\hyperlink{ca2}{\GCA{2}}$}] (ii) at (0,1);
    \coordinate [label=below:{\tiny $\hyperlink{ca1}{\GCA{1}}$}] (i) at (0,0);
    \coordinate [label=right:{\tiny $\hyperlink{ca16}{\GCA{16}}$}] (xvi) at (4,1);
    \coordinate [label=right:{\tiny $\hyperlink{ca17}{\GCA{17}}$}] (xvii) at (4,2);
    \coordinate [label=above:{\tiny $\hyperlink{ca18}{\GCA{18}}$}] (xviii) at (4,3);
    \coordinate [label=left:{\tiny $\hyperlink{ca19}{\GCA{19}}$}] (xix) at (-2,2);
    \coordinate [label=left:{\tiny $\hyperlink{ca20}{\GCA{20}}$}] (xx) at (-2,1);
    %
    %
    \draw [->,line width=0.5pt,color=blue] (xv) -- (xiii); 
    \draw [->,line width=0.5pt,color=blue] (xv) -- (xiv); 
    \draw [->,line width=0.5pt,color=blue] (xv) -- (xii); 
    \draw [->,line width=0.5pt,color=blue] (xv) -- (viii); 
    \draw [->,line width=0.5pt,color=blue] (xviii) -- (xvii); 
    \draw [->,line width=0.5pt,color=blue] (xviii) -- (x); 
    \draw [->,line width=0.5pt,color=blue] (xii) -- (x); 
    \draw [->,line width=0.5pt,color=blue] (xii) -- (ix); 
    \draw [->,line width=0.5pt,color=blue] (viii) -- (vii); 
    \draw [->,line width=0.5pt,color=blue] (viii) -- (vi); 
    \draw [->,line width=0.5pt,color=blue] (viii) -- (v); 
    \draw [->,line width=0.5pt,color=blue] (xiv) -- (xi); 
    \draw [->,line width=0.5pt,color=blue] (xiv) -- (x); 
    \draw [->,line width=0.5pt,color=blue] (xiii) -- (xi); 
    \draw [->,line width=0.5pt,color=blue] (xiii) -- (ix); 
    \draw [->,line width=0.5pt,color=blue] (xvii) -- (xvi); 
    \draw [->,line width=0.5pt,color=blue] (xvii) -- (iii); 
    \draw [->,line width=0.5pt,color=blue] (x) -- (iii); 
    \draw [->,line width=0.5pt,color=blue] (v) -- (iii); 
    \draw [->,line width=0.5pt,color=blue] (v) -- (ii); 
    \draw [->,line width=0.5pt,color=blue] (vii) -- (iii); 
    \draw [->,line width=0.5pt,color=blue] (vii) -- (iv); 
    \draw [->,line width=0.5pt,color=blue] (ix) -- (ii); 
    \draw [->,line width=0.5pt,color=blue] (vi) -- (ii); 
    \draw [->,line width=0.5pt,color=blue] (vi) -- (iv); 
    \draw [->,line width=0.5pt,color=blue] (xix) -- (ii); 
    \draw [->,line width=0.5pt,color=blue] (xix) -- (iv); 
    \draw [->,line width=0.5pt,color=blue] (xi) -- (iv); 
    \draw [->,line width=0.5pt,color=blue] (xx) -- (i); 
    \draw [->,line width=0.5pt,color=blue] (iv) -- (i); 
    \draw [->,line width=0.5pt,color=blue] (ii) -- (i); 
    \draw [->,line width=0.5pt,color=blue] (iii) -- (i); 
    \draw [->,line width=0.5pt,color=blue] (xvi) -- (i);
    %
    %
    \foreach \point in {i,ii,iii,iv,v,vi,vii,viii,ix,x,xi,xii,xiii,xiv,xv}
    \filldraw [color=blue!70!black,fill=blue!50!white] (\point) circle (2pt);
    \foreach \point in {xvi,xvii,xviii,xix,xx}
    \filldraw [color=red!70!black,fill=red!50!white] (\point) circle (2pt);
  \end{tikzpicture}  
  \caption{Contractions between graded conformal algebras: red dots are unique to $d=3$.}
  \label{fig:contractions}
\end{figure}
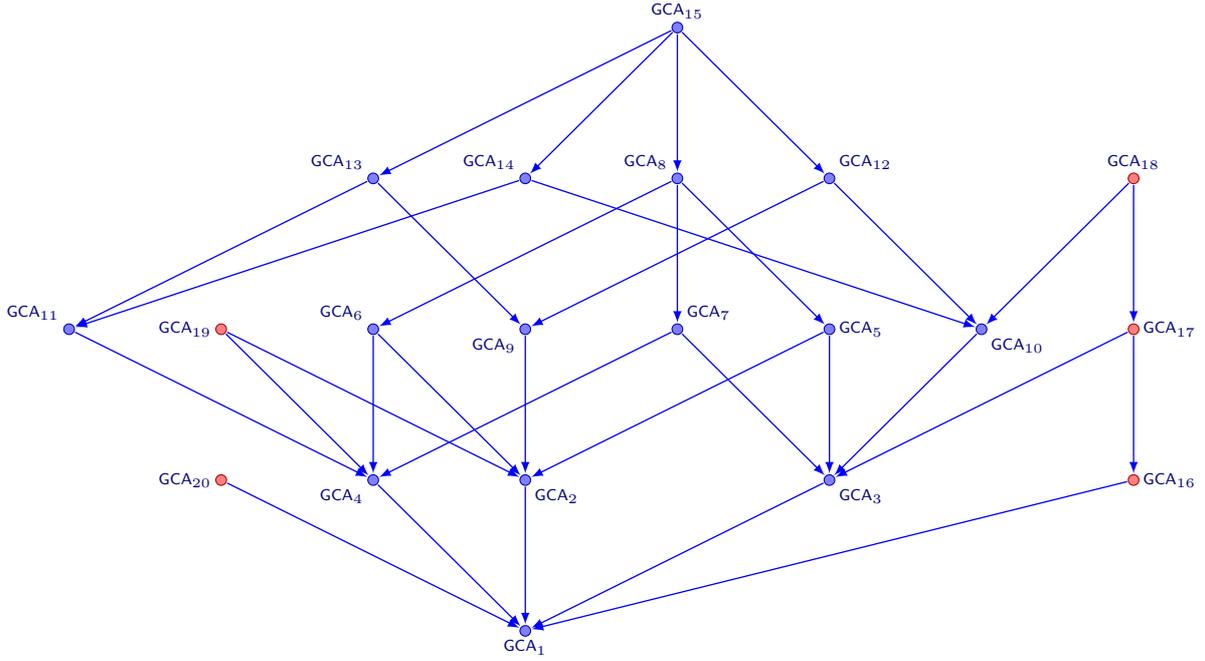

No Lie algebra in Table~\ref{tab:summary-d-geq-4} can contract to a
Lie algebra in Table~\ref{tab:summary-d-eq-3} because all Lie algebras
in this latter table contain the Levi-Civita symbol $\epsilon_{abc}$
in at least one bracket, whereas no Lie algebra in the former table
does. Any new contraction must therefore be from some Lie algebras in
Table~\ref{tab:summary-d-eq-3} to a Lie algebra in either table. All
Lie algebras in Table~\ref{tab:summary-d-eq-3} contract to the static
graded conformal algebra $\hyperlink{ca1}{\GCA{1}}$. One can work out
the contractions relating the different algebras and display the
result as a directed graph, as in
Figure~\ref{fig:contractions}. Contractions can be composed, so the
actual graph of contractions is the transitive closure of the graph in
the figure.  In Table~\ref{tab:contractions-too} we give the values of
the weights $a_\pm, b_\pm, c$ of $S_\pm$, $V_\pm$ and $V_0$ which are
responsible for the different contractions, in the language of the
discussion in Section~\ref{sec:contractions}.  In the right-most part
of the table, the contractions below the horizontal line involve
graded conformal algebras unique to $d=3$.

\begin{table}[h!]
  \centering
  \caption{Weights for contractions between graded conformal algebras}
  \label{tab:contractions-too}
  \resizebox{\textwidth}{!}{
  \begin{tabular}{l|*{5}{>{$}c<{$}}}\toprule
    \multicolumn{1}{c|}{Contraction} & a_+ & a_- & b_+ & b_- & c \\\midrule
    $\hyperlink{ca15}{\GCA{15}} \to \hyperlink{ca14}{\GCA{14}}$ & 0 & 0 & 1 & 1 & 1 \\
    $\hyperlink{ca15}{\GCA{15}} \to \hyperlink{ca13}{\GCA{13}}$ & 1 & 0 & 1 & 0 & 0 \\
    $\hyperlink{ca15}{\GCA{15}} \to \hyperlink{ca12}{\GCA{12}}$ & 1 & 1 & 0 & 0 & 1 \\
    $\hyperlink{ca15}{\GCA{15}} \to \hyperlink{ca8}{\GCA{8}}$ & 1 & 1 & 2 & 0 & 1 \\
    $\hyperlink{ca14}{\GCA{14}} \to \hyperlink{ca11}{\GCA{11}}$ & 1 & 1 & 1 & 1 & 0 \\
    $\hyperlink{ca14}{\GCA{14}} \to \hyperlink{ca10}{\GCA{10}}$ & 1 & 1 & 0 & 0 & 1 \\
    $\hyperlink{ca13}{\GCA{13}} \to \hyperlink{ca11}{\GCA{11}}$ & 0 & 0 & 1 & 1 & 1 \\
    $\hyperlink{ca13}{\GCA{13}} \to \hyperlink{ca9}{\GCA{9}}$ & 1 & 1 & 0 & 0 & 1 \\
    $\hyperlink{ca12}{\GCA{12}} \to \hyperlink{ca10}{\GCA{10}}$ & 0 & 0 & 1 & 1 & 1 \\
    $\hyperlink{ca12}{\GCA{12}} \to \hyperlink{ca9}{\GCA{9}}$ & 1 & 1 & 1 & 1 & 0 \\
    $\hyperlink{ca11}{\GCA{11}} \to \hyperlink{ca4}{\GCA{4}}$ & 0 & 1 & 0 & 0 & 0 \\\bottomrule
  \end{tabular}
  \qquad\qquad
  \begin{tabular}{l|*{5}{>{$}c<{$}}}\toprule
    \multicolumn{1}{c|}{Contraction} & a_+ & a_- & b_+ & b_- & c \\\midrule
    $\hyperlink{ca10}{\GCA{10}} \to \hyperlink{ca3}{\GCA{3}}$ & 0 & 1 & 0 & 0 & 0 \\
    $\hyperlink{ca9}{\GCA{9}} \to \hyperlink{ca2}{\GCA{2}}$ & 0 & 0 & 1 & 0 & 0 \\
    $\hyperlink{ca8}{\GCA{8}} \to \hyperlink{ca7}{\GCA{7}}$ & 1 & 0 & 2 & 0 & 1 \\
    $\hyperlink{ca8}{\GCA{8}} \to \hyperlink{ca6}{\GCA{6}}$ & 0 & 1 & 0 & 1 & 0 \\
    $\hyperlink{ca8}{\GCA{8}} \to \hyperlink{ca5}{\GCA{5}}$ & 0 & 2 & 0 & 1 & 1 \\
    $\hyperlink{ca7}{\GCA{7}} \to \hyperlink{ca4}{\GCA{4}}$ & 0 & 0 & 0 & 1 & 0 \\
    $\hyperlink{ca7}{\GCA{7}} \to \hyperlink{ca3}{\GCA{3}}$ & 1 & 0 & 0 & 0 & 1 \\
    $\hyperlink{ca6}{\GCA{6}} \to \hyperlink{ca4}{\GCA{4}}$ & 0 & 0 & 1 & 0 & 1 \\
    $\hyperlink{ca6}{\GCA{6}} \to \hyperlink{ca2}{\GCA{2}}$ & 0 & 1 & 0 & 0 & 1 \\
    $\hyperlink{ca5}{\GCA{5}} \to \hyperlink{ca3}{\GCA{3}}$ & 1 & 0 & 0 & 0 & 1 \\
    $\hyperlink{ca5}{\GCA{5}} \to \hyperlink{ca2}{\GCA{2}}$ & 0 & 1 & 0 & 1 & 0 \\\bottomrule
  \end{tabular}
  \qquad\qquad
  \begin{tabular}{l|*{5}{>{$}c<{$}}}\toprule
    \multicolumn{1}{c|}{Contraction} & a_+ & a_- & b_+ & b_- & c \\\midrule
    $\hyperlink{ca4}{\GCA{4}} \to \hyperlink{ca1}{\GCA{1}}$ & 1 & 0 & 0 & 0 & 0 \\
    $\hyperlink{ca3}{\GCA{3}} \to \hyperlink{ca1}{\GCA{1}}$ & 1 & 0 & 0 & 0 & 0 \\
    $\hyperlink{ca2}{\GCA{2}} \to \hyperlink{ca1}{\GCA{1}}$ & 0 & 0 & 0 & 0 & 1 \\\midrule
    $\hyperlink{ca20}{\GCA{20}} \to \hyperlink{ca1}{\GCA{1}}$ & 0 & 0 & 0 & 0 & 1 \\
    $\hyperlink{ca19}{\GCA{19}} \to \hyperlink{ca4}{\GCA{4}}$ & 0 & 0 & 0 & 1 & 1 \\
    $\hyperlink{ca19}{\GCA{19}} \to \hyperlink{ca2}{\GCA{2}}$ & 0 & 1 & 0 & 0 & 1 \\
    $\hyperlink{ca18}{\GCA{18}} \to \hyperlink{ca17}{\GCA{17}}$ & 0 & 1 & 0 & 0 & 0 \\
    $\hyperlink{ca18}{\GCA{18}} \to \hyperlink{ca10}{\GCA{10}}$ & 0 & 0 & 1 & 1 & 1 \\
    $\hyperlink{ca17}{\GCA{17}} \to \hyperlink{ca16}{\GCA{16}}$ & 1 & 0 & 0 & 0 & 0 \\
    $\hyperlink{ca17}{\GCA{17}} \to \hyperlink{ca3}{\GCA{3}}$ & 0 & 0 & 1 & 0 & 0 \\
    $\hyperlink{ca16}{\GCA{16}} \to \hyperlink{ca1}{\GCA{1}}$ & 0 & 0 & 1 & 0 & 0 \\\bottomrule
  \end{tabular}}
\end{table}

The simple conformal algebras $\so(d{+}1,2)$ and $\so(d{+}2,1)$ in
dimension $d{+}1$ are isomorphic to the Lie algebra of isometries of the
de~Sitter spacetimes in dimension $d{+}2$.   In other words, they are
kinematical Lie algebras with ($d{+}1$)-dimensional space isotropy.  The
kinematical contractions of these algebras are contained among the
kinematical Lie algebras in \cite{Figueroa-OFarrill:2017tcy} (since
$d+1 \geq 4$ for $d\geq 3$).  It is therefore a natural question to
ask whether any of the other Lie algebras in
Table~\ref{tab:summary-d-geq-4} can be interpreted as a kinematical
Lie algebra in $d+1$, although perhaps with $\so(d,1)$ isotropy as
opposed to $\so(d{+}1)$.  Let $\mu = (a,\natural)$, where $a=1,\dots,d$,
and let $J_{\mu\nu}$, $B_\mu$, $P_\mu$ and $H$ generate a kinematical
Lie algebra.  This means that, in particular, the following brackets
exist:
\begin{equation}
  \begin{split}
    [J_{\mu\nu},J_{\rho\sigma}] &= \eta_{\nu\rho} J_{\mu\sigma} - \eta_{\mu\rho} J_{\nu\sigma} - \eta_{\nu\sigma} J_{\mu\rho} + \eta_{\mu\sigma} J_{\nu\rho}\\
    [J_{\mu\nu},B_\rho] &= \eta_{\nu\rho} B_\mu - \eta_{\mu\rho} B_\nu\\
    [J_{\mu\nu},P_\rho] &= \eta_{\nu\rho} P_\mu - \eta_{\mu\rho} P_\nu,
\end{split}
\end{equation}
where $\eta_{ab} = \delta_{ab}$ and $\eta_{\natural\natural} = \pm 1$.
Breaking the symmetry down to $\so(d)$, we find that
\begin{enumerate}
\item $J_{ab}$ generate an $\so(d)$ subalgebra,
\item $J_{a\natural}$, $B_a$ and $P_a$ are vectors,
\item $B_\natural$, $P_\natural$ and $H$ are scalars,
\item and we have the following additional Lie brackets:
  \begin{equation}
    \begin{split}
      [J_{a\natural}, J_{b\natural}] &= - \eta_{\natural\natural} J_{ab}\\
      [J_{a\natural}, B_b] &= - \delta_{ab} B_\natural\\
      [J_{a\natural}, B_\natural] &= \eta_{\natural\natural} B_a\\
      [J_{a\natural}, P_b] &= - \delta_{ab} P_\natural\\
      [J_{a\natural}, P_\natural] &= \eta_{\natural\natural} P_a.
    \end{split}
  \end{equation}
\end{enumerate}
In summary, there is a vector $V$ such that (ignoring signs,...)
$[V,V] = J$ and additional vectors $V',V''$ and scalars $S',S''$ such
that $[V,V'] = S'$, $[V,V'']=S''$, $[V,S'] = V'$ and $[V,S''] = V''$.
Inspecting Tables~\ref{tab:summary-d-geq-4} and
\ref{tab:summary-d-eq-3}, we see that apart from the de Sitter
algebras, the only other Lie algebras satisfying this condition are
$\hyperlink{ca13}{\GCA{13}^{(\varepsilon)}}$.  To identify this Lie algebra, we let
$J_{a\natural} := \Vz{a}$, $B_a := \Vp{a}$, $P_a := \Vm{a}$, $B_\natural
:= - \varepsilon S^+$, $P_\natural := -\varepsilon S^-$ and $H := D$
and we see that this is kinematical with $\eta_{\natural\natural} =
\varepsilon$.  The only nonzero brackets in addition to the
kinematical ones are
\begin{equation}
  [H,B] = B \qquad\text{and}\qquad [H,P] = -P,
\end{equation}
which means that it is isomorphic to a Newton-Hooke algebra (but in $d+2$
dimensions) if $\varepsilon=1$ and to a pseudo-Newton-Hooke algebra if
$\varepsilon=-1$.

\section{Deformations for $d=2$}
\label{sec:deformations-d=2}

As explained in the context of kinematical Lie algebras in
\cite{Andrzejewski:2018gmz}, when $d=2$ it is convenient to work with
the complexified Lie algebra.  The reason is two-fold: first of all,
the vector representation of $\so(2)$ has a larger endomorphism ring
than that of $\so(d)$ for any $d>2$.  This is because $\so(2)$ is
abelian and extends the endomorphism ring from $\RR$ to $\CC$.  It is
convenient to complexify so that $\CC$ acts by complex
multiplication.  The second reason, which is the same reason in
disguise, is that over the complex numbers we can diagonalise the
action of $J$.

\subsection{The complex Lie algebra}
\label{sec:complex-d=2}

Let $\g_\CC$ be the complex Lie algebra spanned by $J,
D, \V_0 := \Vz{1}+i\Vz{2}, \Vbar_0 := \Vz{1}-i\Vz{2}, \V_\pm := \Vpm{1} + i
\Vpm{2}, \Vbar\pm := \Vpm{1} - i \Vpm{2}, S_\pm$, subject to the Lie
brackets
\begin{equation}
  \begin{aligned}[m]
    [J,\V_\pm] &= -i \V_\pm\\
    [J,\V_0] &= -i \V_0\\
    [J,\Vbar_\pm] &= i \Vbar_\pm\\
    [J,\Vbar_0] &= i \Vbar_0
  \end{aligned}
  \qquad\text{and}\qquad
  \begin{aligned}[m]
    [D,\V_\pm] &= \pm \V_\pm\\
    [D,\Vbar_\pm] &= \pm \Vbar_\pm\\
    [D,S_\pm] &= \pm S_\pm.\\
  \end{aligned}
\end{equation}
The complex Lie algebra $\g_\CC$ is the complexification of $\g$
which is the real Lie algebra fixed under the antilinear involutive
homomorphism $\star$ defined by
\begin{equation}
  \star J = J, \quad \star D = D, \quad \star \V_0 = \Vbar_0,\quad
  \star \V_\pm = \Vbar_\pm \quad\text{and}\quad \star S_\pm = S_\pm.
\end{equation}

\subsection{The deformation complex}
\label{sec:deform-compl-d=2}

The deformation complex for $\g_\CC$ has cochains
\begin{equation}
  C^p_\CC := \Hom(\Lambda^p \W_\CC, \g_\CC)^{\h_\CC},
\end{equation}
where $\W_\CC$ is the complex span of
$\V_0,\Vbar_0,\V_\pm,\Vbar_\pm,S_\pm$ and $\h_\CC$ is the complex
abelian Lie algebra spanned by $J,D$.  The differential $\d: C^p_\CC
\to C^{p+1}_\CC$ is defined on generators by
\begin{equation}
  \begin{split}
      \d J &= i \bnu^+ \V_+ - i \bnubar^+ \Vbar_+ + i \bnu^- \V_- - i
      \bnubar^- \Vbar_- + i \bnu^0 \V_0 - i \bnubar^0 \Vbar_0\\
      \d D &= - \bnu^+ \V_+ - \bnubar^+ \Vbar_+ + \bnu^- \V_- +
      \bnubar^- \Vbar_- - \sigma^+ S_+ + \sigma^- S_-\\
  \end{split}
\end{equation}
and zero on other generators, where $\bnu^0, \bnubar^0, \bnu^\pm,
\bnubar^\pm, \sigma^\pm$ are the canonical dual basis for $\W^*_\CC$.

We shall now enumerate the first few spaces of cochains.  To ensure
that we do no leave any cochains out, we can use character theory to
calculate the dimension of these spaces.  The $\h_\CC$ character of
$\W_\CC$ is given by
\begin{equation}
  \chi_{\W_\CC}(q,\tau) = \underbrace{(\tau + \tau^{-1})}_{S_\pm} + \underbrace{(\tau +
  \tau^{-1})(q + q^{-1})}_{\V_\pm,\Vbar_\pm} + \underbrace{(q + q^{-1})}_{\V_0,\Vbar_0},
\end{equation}
and that of $\g_\CC$ is given by
\begin{equation}
  \chi_{\g_\CC}(q,\tau) = 2 + (\tau + \tau^{-1}) + (\tau +
  \tau^{-1})(q + q^{-1}) + (q + q^{-1}),
\end{equation}
where the $2$ is due to $J$ and $D$.  The dimension of $C^p_\CC$ is
the constant term in
\begin{equation}
  \chi_{\Lambda^p\W^*_\CC \otimes \g_\CC}(q,\tau) =
  \chi_{\Lambda^p\W^*_\CC}(q,\tau)  \chi_{\g_\CC}(q,\tau).
\end{equation}
To calculate this, we observe that
$\chi_{\Lambda^p\W^*_\CC}(q,\tau) =
\chi_{\Lambda^p\W_\CC}(q^{-1},\tau^{-1}) =
\chi_{\Lambda^p\W_\CC}(q,\tau)$, where the last equality follows by
inspection.  The generating function for the characters of the
exterior powers is given by
\begin{equation}
  \sum_{p=0}^\infty t^p \chi_{\Lambda^p \W_\CC}(q,\tau) = \exp
  \left(-\sum_{\ell=1}^\infty \frac{(-t)^\ell}{\ell}\chi_{\W_\CC}(q^\ell,\tau^\ell)\right).
\end{equation}
Expanding and collecting terms, we find the following for the first
few spaces of cochains:
\begin{equation}
  \dim C^0_\CC = 2, \qquad \dim C^1_\CC = 8, \qquad \dim C^2_\CC = 20
  \qquad\text{and}\qquad \dim C^3_\CC = 36,
\end{equation}
which we proceed to enumerate:
\begin{equation}
  \begin{split}
    C^0_\CC &= \cspn{J,D}\\
    C^1_\CC &= \cspn{\bnu^+ \V_+, \bnubar^+\Vbar_+, \bnu^- \V_-,
      \bnubar^-\Vbar_-, \bnu^0 \V_0, \bnubar^0\Vbar_0, \sigma^+ S_+,
      \sigma^- S_-}\\
    C^2_\CC &= \cspn{\cc_1,\ccbar_1,\cc_2, \ccbar_2,
    \dots , \cc_8, \ccbar_8, \cc_9, \cc_{10}, \cc_{11},
      \cc_{12}}\\
    C^3_\CC &= \cspn{\bb_1,\bbbar_1, \bb_2, \bbbar_2, \dots, \bb_{16}, \bbbar_{16},
      \bb_{17}, \bb_{18}, \bb_{19}, \bb_{20}},
  \end{split}
\end{equation}
where (omitting $\otimes$ and $\wedge$),
\begin{equation}\label{eq:d=2-cochains-2}
  \begin{aligned}[m]
    \cc_1 &= \bnu^0 \bnubar^+ S_+\\
    \cc_2 &= \bnu^0 \bnubar^- S_-\\
    \cc_3 &= \sigma^+ \bnu^- \V_0
  \end{aligned}
  \qquad\qquad
  \begin{aligned}[m]
    \cc_4 &= \sigma^- \bnu^+ \V_0\\
    \cc_5 &= \sigma^+ \bnu^0 \V_+\\
    \cc_6 &= \sigma^- \bnu^0 \V_-
  \end{aligned}
  \qquad\qquad
  \begin{aligned}[m]
    \cc_7 &= \bnu^+ \bnubar^- J\\
    \cc_8 &= \bnu^+ \bnubar^- D\\
    \cc_9 &= \sigma^+ \sigma^- J
  \end{aligned}
  \qquad\qquad
  \begin{aligned}[m]
    \cc_{10} &= \sigma^+ \sigma^- D\\
    \cc_{11} &= i \bnu^0 \bnubar^0 J\\
    \cc_{12} &=  i \bnu^0 \bnubar^0 D,
  \end{aligned}
\end{equation}
and
\begin{equation}
  \begin{aligned}[m]
    \bb_1 &= \bnu^+ \bnu^- \bnubar^+ \V_+\\
    \bb_2 &= \bnu^+ \bnu^- \bnubar^- \V_-\\
    \bb_3 &= \bnu^+ \bnu^- \bnubar^0 \V_0\\
    \bb_4 &= \bnu^0 \bnu^+ \bnubar^- \V_0\\
    \bb_5 &= \bnu^0 \bnu^+ \bnubar^0 \V_+
  \end{aligned}
  \qquad\qquad
  \begin{aligned}[m]
    \bb_6 &= \bnu^0 \bnu^- \bnubar^+ \V_0\\
    \bb_7 &= \bnu^0 \bnu^- \bnubar^0 \V_-\\
    \bb_8 &= \sigma^+ \sigma^- \bnu^+ \V_+\\
    \bb_9 &= \sigma^+ \sigma^- \bnu^- \V_-\\
    \bb_{10} &= \sigma^+ \sigma^- \bnu^0 \V_0
  \end{aligned}
  \qquad\qquad
  \begin{aligned}[m]
    \bb_{11} &= \sigma^+ \bnu^+ \bnubar^- S_+\\
    \bb_{12} &= \sigma^- \bnu^- \bnubar^+ S_-\\
    \bb_{13} &= \sigma^+ \bnu^0 \bnubar^- J\\
    \bb_{14} &= \sigma^+ \bnu^0 \bnubar^- D\\
    \bb_{15} &= \sigma^- \bnu^0 \bnubar^+ J
  \end{aligned}
  \qquad\qquad
  \begin{aligned}[m]
    \bb_{16} &= \sigma^- \bnu^0 \bnubar^+ D\\
    \bb_{17} &= i \sigma^+ \bnu^0 \bnubar^0 S_+\\
    \bb_{18} &= i \sigma^- \bnu^0 \bnubar^0 S_-\\
    \bb_{19} &= i \sigma^+ \bnu^- \bnubar^- S_-\\
    \bb_{20} &= i \sigma^- \bnu^+ \bnubar^+ S_+.
  \end{aligned}
\end{equation}
Here $\ccbar_i = \cc_i$ for $i=9,10,11,12$ and $\bbbar_i = \bb_i$ for
$i=17,18,19,20$.  For the other cochains, complex conjugation is as
expected: e.g., $\ccbar_1 = \bnubar^0 \bnu^+ S_+ = - \bnu^+ \bnubar^0
S_+$, et cetera.

\subsection{Infinitesimal deformations}
\label{sec:infin-deform-d=2}

The subspace of real cochains is a subcomplex and we are interested in
its cohomology in degree 2.  We observe that $\d : C^1_\CC \to C^2_\CC$ is
the zero map, so that the cohomology $H^2_\CC$ coincides with the
space $Z^2_\CC$ of cocycles.  The map $\d : C^2_\CC \to C^3_\CC$ is
such that $\cc_1,\ccbar_1, \dots, \cc_6, \ccbar_6$ are cocycles, and
\begin{equation}\label{eq:b3-d=2}
  \begin{aligned}[m]
    \d \cc_7 & = i ( \bbbar_1 - \bb_2 + \bb_4 + \bbbar_6)\\
    \d \cc_8 & = \bbbar_1 - \bb_2 - \bb_{11} - \bbbar_{12}\\
    \d \cc_9 & = i (\bb_8 -\bbbar_8) + i (\bb_9 -\bbbar_9) + i (\bb_{10} -  \bbbar_{10})
  \end{aligned}
  \qquad\qquad
  \begin{aligned}[m]
    \d \cc_{10} & = -(\bb_8 +\bbbar_8) + (\bb_9 + \bbbar_9)\\
    \d \cc_{11} & = (\bb_5 + \bbbar_5) + (\bb_7 + \bbbar_7)\\
    \d \cc_{12} & = i(\bb_5-\bbbar_5) - i (\bb_7 - \bbbar_7) -
    \bb_{17} + \bb_{18},
  \end{aligned}
\end{equation}
which spans the space $B^3_\CC$ of $3$-coboundaries.  For future use,
we notice that there are some cochains in $C^3_\CC$ which do not
appear as components of any coboundary in $B^3_\CC$: $\bb_3, \bb_{13},
\bb_{14}, \bb_{15}, \bb_{16}$ (and their complex conjugates) and also
$\bb_{19}, \bb_{20}$.

The space of infinitesimal deformations is
\begin{equation}
  H^2_\CC \cong Z^2_\CC = \cspn{\cc_1,\ccbar_1, \dots, \cc_6, \ccbar_6}
\end{equation}
and hence the most general (real) infinitesimal deformation is given by
\begin{equation}
  \varphi^{(1)} = \sum_{i=1}^6 t_i \cc_i + \sum_{i=1}^6 \tbar_i
  \ccbar_i =: \psi + \psibar
\end{equation}
for some $t_i \in \CC$.

\subsection{Obstructions}
\label{sec:obstructions-2}

The first obstruction to integrability of the infinitesimal
deformation $\varphi^{(1)}$ is given by the cohomology class of
$\varphi^{(1)} \bullet \varphi^{(1)}$ in $H^3_\CC$, where
$\bullet$ is the complex-linear extension of the Lie-admissible
product introduced in Section~\ref{sec:obstructions}.  We tabulate
this product on the space of cocycles in Table~\ref{tab:NR-dot-d=2}.

\begin{table}[h!]
  \centering 
  \caption{Some components of the Nijenhuis--Richardson product $\bullet: C_\CC^2 \times C_\CC^2 \to C_\CC^3$}
  \label{tab:NR-dot-d=2}
  \setlength{\extrarowheight}{2pt}  
  \begin{tabular}{>{$}l<{$}*{12}{|>{$}c<{$}}}
    \multicolumn{1}{c|}{$\bullet$} & \cc_1 & \ccbar_1 & \cc_2 & \ccbar_2 & \cc_3 & \ccbar_3 & \cc_4 & \ccbar_4 & \cc_5 & \ccbar_5 & \cc_6 & \ccbar_6 \\\midrule
    \cc_1 & & & & & -\bb_6 & \bbbar_3 & & & & -\bbbar_5 & &  \\\hline
 \ccbar_1 & & & & & \bb_3 & -\bbbar_6 & & & -\bb_5  & & & \\\hline
    \cc_2 & & & & & & & -\bb_4 & -\bbbar_3 & & & & -\bbbar_7\\\hline
 \ccbar_2 & & & & & & & -\bb_3 & -\bbbar_4 & & & -\bb_7 & \\\hline
    \cc_3 & -\bbbar_{11} & & -i \bb_{19} & & & & & & & & \bb_9 & \\\hline
 \ccbar_3 & & -\bb_{11}  & & i \bb_{19} & & & & & & & & \bbbar_9 \\\hline
    \cc_4 & -i \bb_{20} & & \bb_{12} & & & & & & -\bb_8 & & & \\\hline
 \ccbar_4 & & i \bb_{20} & & \bbbar_{12} & & & & & & -\bbbar_8 & & \\\hline
    \cc_5 & & i \bb_{17} & & & & & \bb_{10} & & & & & \\\hline
 \ccbar_5 & -i \bb_{17} & & & & & & & \bbbar_{10} & & & & \\\hline
    \cc_6 & & & & i \bb_{18} & -\bb_{10} & & & & & & & \\\hline
 \ccbar_6 & & & -i \bb_{18} & & & -\bbbar_{10} & & & & & & \\\bottomrule
  \end{tabular}
\end{table}

Calculating the product $\varphi^{(1)} \bullet \varphi^{(1)}$
we can use that
\begin{equation}
    \varphi^{(1)} \bullet \varphi^{(1)}= (\psi + \psibar)
    \bullet (\psi + \psi\bar) = \psi\bullet \psi + \psi \bullet\psibar
    + \text{c.c.} 
\end{equation}
to arrive at
\begin{multline}
   \varphi^{(1)} \bullet \varphi^{(1)} = (t_3 \tbar_1 -
   \tbar_2 t_4) \bb_3 - t_2 t_4 \bb_4 - \tbar_1 t_5 \bb_5 - t_1 t_3
   \bb_6 - \tbar_2 t_6 \bb_7 - t_4 t_5 \bb_8 + t_3 t_6 \bb_9\\
   + (t_4 t_5 - t_3 t_6) \bb_{10} - \tbar_1 \tbar_3 \bb_{11} + t_2 t_4 \bb_{12} + i \tbar_1 t_5 \bb_{17} + i
   \tbar_2 t_6 \bb_{18} - i t_2 t_3 \bb_{19} - i t_1 t_4 \bb_{20} + \text{c.c.}
\end{multline}
We need to cancel this with a coboundary $\d\varphi^{(2)}$, where
\begin{equation}
  \varphi^{(2)} = u_1 \cc_7 + \ubar_1 \ccbar_7 + u_2 \cc_8 + \ubar_2
  \ccbar_8 + u_3 \cc_9 + u_4 \cc_{10} + u_5 \cc_{11} + u_6 \cc_{12},
\end{equation}
with $u_1,u_2 \in \CC$ and $u_3,\dots,u_6 \in \RR$.  Calculating
$\d\varphi^{(2)}$ we find
\begin{multline}
  \d\varphi^{(2)} = (\ubar_2 - i \ubar_1) \bb_1 - ( u_2 + i u_1) \bb_2
  + i u_1 \bb_4 + (u_5 + i u_6) \bb_5 - i \ubar_1 \bb_6 + (u_5 - i
  u_6) \bb_7 - (u_4 - i u_3) \bb_8 \\
  + (u_4 + i u_3) \bb_9 + i u_3
  \bb_{10} - u_2 \bb_{11} + \ubar_2 \bb_{12} - \tfrac12 u_6 \bb_{17} +
  \tfrac12 u_6 \bb_{18} + \text{c.c.}
\end{multline}

The first obstruction equation $\d\varphi^{(2)} = \varphi^{(1)} \bullet \varphi^{(1)}$
implies
\begin{equation}
  u_1 = i t_2 t_4, \quad u_2 = t_2 t_4, \quad u_3 = 0, \quad u_4 = t_3
  t_6, \quad u_5 = - \tbar_1 t_5 \quad\text{and}\quad u_6 = 0,
\end{equation}
and results in the following relations between the $t_i$:
\begin{equation}\label{eq:first-obst}
  t_3 t_6 = t_4 t_5 \in \RR, \quad \tbar_1 t_5 = \tbar_2 t_6 \in \RR,
  \quad \tbar_1 t_3 = \tbar_2 t_4, \quad \tbar_1 \tbar_3 = t_2 t_4, 
  \quad t_1 t_4 \in \RR \quad\text{and}\quad t_2 t_3 \in \RR.
\end{equation}

The second obstruction equation is $\d\varphi^{(3)} =
\varphi^{(1)} \bullet \varphi^{(2)} + \varphi^{(2)} \bullet
\varphi^{(1)}$, for which we need further components of the
Nijenhuis--Richardson Lie-admissible product.  Notice that
$\varphi^{(2)} \bullet \varphi^{(1)}$ since there are no generators
dual to $J$ or $D$, so that we only need $\varphi^{(1)} \bullet
\varphi^{(2)}$.  The relevant products are tabulated in
Table~\ref{tab:NR-dot-too-d=2}.

\begin{table}[h!]
  \centering 
  \caption{Further components of the Nijenhuis--Richardson product $\bullet: C_\CC^2 \times C_\CC^2 \to C_\CC^3$}
  \label{tab:NR-dot-too-d=2}
  \setlength{\extrarowheight}{2pt}  
  \begin{tabular}{>{$}l<{$}*{8}{|>{$}c<{$}}}
    \multicolumn{1}{c|}{$\bullet$} & \cc_7 & \ccbar_7 & \cc_8 & \ccbar_8 & \cc_9 & \cc_{10} & \cc_{11} & \cc_{12} \\\midrule
    \cc_1 & & & & & \bb_{15} & \bb_{16} & & \\\hline
 \ccbar_1 & & & & & \bbbar_{15} & \bbbar_{16} & & \\\hline
    \cc_2 & & & & & -\bb_{13} & -\bb_{14} & & \\\hline
 \ccbar_2 & & & & & -\bbbar_{13} & -\bbbar_{14} & & \\\hline
    \cc_3 & & & & & & & -i \bbbar_{13}  & -i \bbbar_{14} \\\hline
 \ccbar_3 & & & & & & & i \bb_{13} & i \bb_{14} \\\hline
    \cc_4 & & & & & & & -i \bbbar_{15}  & -i \bbbar_{16} \\\hline
 \ccbar_4 & & & & & & & i \bb_{15} & i \bb_{16} \\\hline
    \cc_5 & \bb_{13} & & \bb_{14} & & & & & \\\hline
 \ccbar_5 & & \bbbar_{13} & & \bbbar_{14} & & & & \\\hline
    \cc_6 & & -\bb_{15} & & -\bb_{16} & & & & \\\hline
 \ccbar_6 & -\bbbar_{15} & & -\bbbar_{16} & & & & & \\\bottomrule
  \end{tabular}
\end{table}

Using that $u_3 = u_6 = 0$, we find that
\begin{equation}
  \varphi^{(1)} \bullet \varphi^{(2)}= (u_1 t_5 + i \tbar_3 u_5)
  \bb_{13} + (t_5 u_2 - t_2 u_4) \bb_{14} + (i \tbar_4 u_5 - t_6
  \ubar_1) \bb_{15} + (t_1 u_4 - t_6 \ubar_2)\bb_{16}  + \text{c.c.}
\end{equation}
Using equation~\eqref{eq:first-obst} we see that
$\varphi^{(1)} \bullet \varphi^{(2)} = 0$ identically.  Therefore we
can take $\varphi^{(3)} = 0$.  Since
$\varphi^{(2)} \bullet \varphi^{(2)} = 0$, we also have
$\varphi^{(4)} = 0$ and indeed all higher $\varphi^{(i)} = 0$ as well.

In summary, the most general deformation is
\begin{equation}\label{eq:obst-d=2}
  \varphi = \sum_{i=1}^6 t_i \cc_i + \sum_{i=1}^6 \tbar_i \ccbar_i + i
  t_2 t_4 \cc_7 - i \tbar_2 \tbar_4 \ccbar_7 + t_2 t_4 \cc_8 + \tbar_2
  \tbar_4 \ccbar_8 + t_3 t_6 \cc_{10} - \tbar_1 t_5 \cc_{11}
\end{equation}
subject to the integrability conditions in equation
\eqref{eq:first-obst}.

\subsection{Isomorphism classes of deformations}
\label{sec:isom-class-deform}

The first three quadratic equations in \eqref{eq:first-obst} are
equivalent to the vanishing of the following determinants:
\begin{equation}
  \begin{vmatrix}
    t_4 & t_6 \\ t_3 & t_5
  \end{vmatrix}
  \qquad
  \begin{vmatrix}
    \tbar_1 & t_6 \\ \tbar_2 & t_5
  \end{vmatrix}
  \qquad
  \begin{vmatrix}
    \tbar_1 & t_4 \\ \tbar_2 & t_3
  \end{vmatrix},
\end{equation}
which is equivalent to the columns being collinear.  In other words, these
equations imply that there exists $(x,y) \in \CC^2\setminus\{(0,0)\}$
and $\alpha_1,\alpha_2,\alpha_3\in \CC$ such that
\begin{equation}\label{eq:sols-d=2}
  (\tbar_1,\tbar_2) = \alpha_1 (x,y),\qquad
  (t_4,t_3) = \alpha_2 (x,y) \qquad\text{and}\qquad
  (t_6, t_5) = \alpha_3 (x,y).
\end{equation}
The fourth quadratic equation in \eqref{eq:first-obst} says that
\begin{equation}\label{eq:quadric-d=2}
  t_1 t_3 = \tbar_2 \tbar_4 \iff (\alphabar_1 \alpha_2 - \alpha_1 \alphabar_2)
  \xbar y = 0.
\end{equation}
The reality conditions in  \eqref{eq:first-obst} say that the
following are real:
\begin{equation}
  \alpha_2\alpha_3 x y, \qquad \alpha_1\alpha_3 x y, \qquad \alphabar_1\alpha_2
  |x|^2 \qquad\text{and}\qquad \alphabar_1\alpha_2 |y|^2.
\end{equation}
Since $(x,y) \neq (0,0)$, $|x|^2 + |y|^2 \neq 0$, and therefore
$\alphabar_1\alpha_2 \in \RR$, which implies
equation~\eqref{eq:quadric-d=2}. 

In summary, the integrability conditions are equivalent to
equation~\eqref{eq:sols-d=2} and the reality conditions
\begin{equation}\label{eq:reality}
  \alphabar_1 \alpha_2 \in \RR, \qquad \alpha_2\alpha_3 x y \in \RR
  \qquad\text{and}\qquad \alpha_1\alpha_3 x y \in \RR.
\end{equation}
We will now analyse the different branches of solutions of these equations.

Since $(x,y) \neq (0,0)$, we have three possibilities:
\begin{enumerate}
\item $y=0$ (and hence $x \neq 0$),
\item $x=0$ (and hence $y \neq 0$), and
\item $xy \neq 0$.
\end{enumerate}
The first two are related by the involutive automorphism $\tau$ of $\g_\CC$,
defined by
\begin{equation}
  \tau : (J,D,\V_0,\V_\pm, S_\pm) \mapsto (J, -D, \V_0, \V_\mp, S_\mp),
\end{equation}
whose effect on the deformations is to exchange $x \leftrightarrow
y$.  Indeed, notice that on cochains $\tau$ exchanges
\begin{equation}
  \cc_1 \leftrightarrow \cc_2, \qquad \cc_3 \leftrightarrow \cc_4
  \qquad\text{and}\qquad \cc_5 \leftrightarrow \cc_6,
\end{equation}
which, at the level of the parameters in the infinitesimal
deformation, is equivalent to exchanging
\begin{equation}
  t_1 \leftrightarrow t_2, \qquad t_3 \leftrightarrow t_4
  \qquad\text{and}\qquad t_5 \leftrightarrow t_6,
\end{equation}
which, from equation~\eqref{eq:sols-d=2}, can be seen to be equivalent
to $x \leftrightarrow y$.

This leaves two branches, corresponding to (1) and (3).

\subsubsection{Branch $y=0$ and $x \neq 0$}
\label{sec:branch-y=0-x}

Here $t_2 = t_3 = t_5 = 0$ and $t_1 = \alphabar_1 \xbar$, $t_4 = \alpha_2
x$ and $t_6 = \alpha_3 x$, with $\alphabar_1\alpha_2 \in \RR$.  The
corresponding deformation is
\begin{equation}
  \varphi = \alphabar_1 \xbar \cc_1 + \alpha_2 x \cc_4  + \alpha_3 x \cc_6 + \text{c.c.}
\end{equation}

We still have the possibility of acting with automorphisms
\begin{equation}\label{eq:autos-d=2}
  \V_\pm \mapsto \lambda_\pm \V_\pm, \qquad \V_0 \mapsto \mu \V_0
  \qquad\text{and}\qquad S_\pm \mapsto \xi_\pm S_\pm,
\end{equation}
where $\lambda_\pm, \mu \in \CC^\times$ and $\xi_\pm \in \RR^\times$.
The effect on cochains is
\begin{equation}
  \cc_1 \mapsto \frac{\xi_+}{\mu\lambdabar_+} \cc_1, \qquad \cc_4 \mapsto
  \frac{\mu}{\xi_- \lambda_+} \cc_4 \qquad\text{and}\qquad \cc_6
  \mapsto \frac{\lambda_-}{\xi_- \mu} \cc_6,
\end{equation}
and hence on parameters
\begin{equation}
  t_1 \mapsto \frac{\mu\lambdabar_+}{\xi_+} t_1, \qquad t_4 \mapsto
  \frac{\xi_- \lambda_+}{\mu} t_4 \qquad\text{and}\qquad t_6
  \mapsto \frac{\xi_- \mu}{\lambda_-} t_6.
\end{equation}
We now claim that if $\alpha_i x$ is different from zero, we may bring
it to $1$.  Indeed, suppose that $\alpha_1 \neq 0$.  then let
$\mu = \frac{\xi_+}{\lambdabar_+ \alphabar_1 \xbar}$.  If
$\alpha_1 = 0$, then $\mu$ remains unconstrained.  Suppose that
$\alpha_2 \neq 0$, then if $\alpha_1 =0$ we choose
$\mu = \xi_- \lambda_+ \alpha_2 x$, whereas if $\alpha_1 \neq 0$, then
choose $\xi_- = \frac{\xi_+}{|\lambda_+|^2 |x|^2\alphabar_1
  \alpha_2}$, which is indeed real.  Finally, if $\alpha_3 \neq 0$,
then choose $\lambda_- = \xi_- \mu \alpha_3 x$.

In summary, we arrive at eight isomorphism classes of deformations in
this branch summarised in Table~\ref{tab:isom-class-d=2-branch-1}.

\begin{table}[h!]
  \centering
  \caption{Isomorphism classes of deformations ($d=2$, $y=0$)}
  \label{tab:isom-class-d=2-branch-1}
  \setlength{\extrarowheight}{2pt}  
  \rowcolors{2}{blue!10}{white}
  \begin{tabular}{*{3}{>{$}c<{$}}|>{$}l<{$}}\toprule
    \alpha_1 & \alpha_2 & \alpha_3 & \multicolumn{1}{c}{Deformation $\varphi$} \\\midrule
    0 & 0 & 0 & 0 \\
    1 & 0 & 0 & \bnu^0 \bnubar^+ S_+ - \bnu^+ \bnubar^0 S_+ \\
    0 & 1 & 0 &  \sigma^- \bnu^+ \V_0 +  \sigma^- \bnubar^+ \Vbar_0 \\
    0 & 0 & 1 & \sigma^- \bnu^0 \V_- + \sigma^- \bnubar^0 \Vbar_-\\
    1 & 1 & 0 & \bnu^0 \bnubar^+ S_+ +  \sigma^- \bnu^+ \V_0 - \bnu^+ \bnubar^0 S_+ + \sigma^- \bnubar^+ \Vbar_0\\
    1 & 0 & 1 & \bnu^0 \bnubar^+ S_+ + \sigma^- \bnu^0 \V_- - \bnu^+ \bnubar^0 S_+ +  \sigma^- \bnubar^0 \Vbar_-\\
    0 & 1 & 1 &  \sigma^- \bnu^+ \V_0 + \sigma^- \bnu^0 \V_- +  \sigma^- \bnubar^+ \Vbar_0 +  \sigma^- \bnubar^0 \Vbar_-\\
    1 & 1 & 1 & \bnu^0 \bnubar^+ S_+ +  \sigma^- \bnu^+ \V_0 + \sigma^- \bnu^0 \V_- - \bnu^+ \bnubar^0 S_+ +  \sigma^- \bnubar^+ \Vbar_0 +  \sigma^- \bnubar^0 \Vbar_-\\\bottomrule
  \end{tabular}
\end{table}

Working out the corresponding brackets, we find (up to the occasional
rescaling and after applying the automorphism $\tau$ in
\eqref{eq:z2-auto}) the same list of Lie algebras as in
Table~\ref{tab:isom-class-conf-1}.  In other words, for this branch at
least, the classification of $d=2$ agrees with that of $d>3$.

\subsubsection{Branch $xy \neq 0$}
\label{sec:branch-xy-neq}

In this branch we have
\begin{equation}
  \begin{aligned}[m]
    t_1 &= \alphabar_1\xbar\\
    t_3 &= \alpha_2 y\\
    t_5 &= \alpha_3 y
  \end{aligned}
  \qquad\text{and}\qquad
  \begin{aligned}[m]
    t_2 &= \alphabar_1\ybar\\
    t_4 &= \alpha_2 x\\
    t_6 &= \alpha_3 x
  \end{aligned}
  \qquad\text{subject to}\qquad
  \begin{aligned}[m]
    \alphabar_1 \alpha_2 & \in \RR\\
    \alpha_2 \alpha_3 x y  & \in \RR\\
    \alpha_1 \alpha_3 x y  & \in \RR,
  \end{aligned}
\end{equation}
and we have the possibility of applying the automorphisms in
\eqref{eq:autos-d=2}, whose effect on the $t_i$ is as follows:
\begin{equation}
  \begin{aligned}[m]
    t_1 & \mapsto \frac{\mu\lambdabar_+}{\xi_+} t_1\\
    t_3 & \mapsto \frac{\xi_+\lambda_-}{\mu} t_3\\
    t_5 & \mapsto \frac{\xi_+ \mu}{\lambda_+} t_5
  \end{aligned}
  \qquad\text{and}\qquad
  \begin{aligned}[m]
    t_2 & \mapsto \frac{\mu\lambdabar_-}{\xi_-} t_2\\
    t_4 & \mapsto \frac{\xi_-\lambda_+}{\mu} t_4\\
    t_6 & \mapsto \frac{\xi_- \mu}{\lambda_-} t_6.
  \end{aligned}
\end{equation}

It is then straightforward to go through the eight possibilities,
according to whether or not $\alpha_i = 0$ and bring these to normal
forms which are tabulated in Table~\ref{tab:isom-class-d=2-branch-2},
where an asterisk in the $\alpha_i$ column means that it is not zero.
As before, we have introduced $\varepsilon = \pm 1$, which
arises whenever $\alpha_1 \alpha_3 \neq 0$, and corresponds to the
sign of the nonzero (real) number $\alpha_1 \alpha_3 x y$.

\begin{table}[h!]
  \centering
  \caption{Isomorphism classes of deformations ($d=2$, $xy\neq0$)}
  \label{tab:isom-class-d=2-branch-2}
  \setlength{\extrarowheight}{2pt}  
  \rowcolors{2}{blue!10}{white}
  \begin{tabular}{*{3}{>{$}c<{$}}|>{$}l<{$}}\toprule
    \alpha_1 & \alpha_2 & \alpha_3 & \multicolumn{1}{c}{Deformation $\varphi$} \\\midrule
    0 & 0 & 0 & 0 \\
    \star & 0 & 0 & \cc_1 + \cc_2 + \text{c.c.}\\
    0 & \star & 0 & \cc_3 + \cc_4 + \text{c.c.}\\
    0 & 0 & \star & \cc_5 + \cc_6 + \text{c.c.}\\
    \star & \star & 0 & \cc_1 + \cc_2 + \cc_3 + \cc_4 + i \cc_7 + \cc_8 + \text{c.c.}\\
    \star & 0 & \star & (\cc_1 + \cc_2 + \varepsilon \cc_5 + \varepsilon \cc_6 + \text{c.c}) - \varepsilon \cc_{11}\\
    0 & \star & \star & (\cc_3 + \cc_4 + \cc_5 + \cc_6 + \text{c.c}) + \cc_{10}\\
    \star & \star & \star & (\varepsilon \cc_1 + \varepsilon \cc_2 + \cc_3 + \cc_4 + \cc_5 + \cc_6 + i \varepsilon \cc_7 + \varepsilon \cc_8 + \text{c.c.}) + \cc_{10} - \varepsilon \cc_{11} \\\bottomrule
  \end{tabular}
\end{table}

Substituting the definitions \eqref{eq:d=2-cochains-2} for the
cochains and working out the Lie brackets, we find (up to the
occasional rescaling) the same list of Lie algebras as in
Table~\ref{tab:isom-class-conf-2}.  In other words, the deformation
problem for $d=2$ has the same solution \emph{mutatis mutandis} as in
$d>3$, resulting in the same list of isomorphism classes of Lie
algebras in Table~\ref{tab:summary-d-geq-4}.

\section{Central extensions}
\label{sec:central-extensions}

It is a natural question to ask whether a given Lie algebra admits
central extensions, given the important rôle they play in
applications.  Central extensions of a Lie algebra\footnote{In this
  section, and for psychological reasons, $\g$ shall denote a general
  Lie algebra, not necessarily the static graded conformal Lie algebra
  as it did in the sections where we discussed deformations.}  $\g$
are classified by the second cohomology group $H^2(\g)$ with values in
the trivial one-dimensional representation.

\subsection{Central extensions for $d\geq 3$}
\label{sec:centr-extens-dgeq}

Let $d\geq 3$.  All graded conformal algebras have a semisimple
subalgebra $\h$ isomorphic to $\so(d)$: namely, the span of $J_{ab}$.
The factorisation theorem of Hochschild and Serre \cite{MR0054581}
implies that $H^2(\g) \cong H^2(\g,\h)$, the relative cohomology group
computed from the complex of $\h$-invariant cochains with no legs
along $\h$.  Since $\h$ lies in degree $0$ in every graded conformal
algebra $\g$, the complex $C^\bullet(\g,\h)$ breaks up into the
direct sum of subcomplexes of a fixed degree. Since the grading
element $D$ acts trivially on cohomology and acts reducibly in the
complex, we have further that $H^2(\g,\h) \cong H^{0,2}(\g,\h)$, which
can be computed from the degree-$0$ piece of the complex. This complex
is
\begin{equation}
  C^{0,p}(\g,\h)\cong \left(\Lambda^p(\g/\h)^*\right)^{\deg 0},
\end{equation}
with the differential induced from that of the Chevalley--Eilenberg
complex of $\g$.

For the graded conformal algebras under discussion and letting
$\delta$ denote the dual to $D$, we have that
\begin{equation}
  C^{0,1}(\g,\h) = \spn{\delta} \qquad\text{and}\qquad C^{0,2}(\g,\h)
  = \spn{\sigma^+\sigma^- := \sigma^+ \wedge \sigma^-, \upsilon^+ \upsilon^- := \upsilon_a^+
    \wedge \upsilon_a^-}.
\end{equation}
It is then a simple matter to go down Table~\ref{tab:summary-d-geq-4}
and calculate $\d\delta$, $\d(\sigma^+\sigma^-)$ and $\d(\upsilon^+\upsilon^-)$
to determine $H^2(\g)$ and hence the possible central extensions.

\subsubsection{$\GCA{1}$, $\GCA{3}$, $\GCA{10}$, $\GCA{16}$, $\GCA{17}$ and $\GCA{18}$}
\label{sec:ca-1-3-10-16-17-18}

These share the same differential:
\begin{equation}
  \d\delta = 0, \qquad \d\sigma^\pm = \mp \delta \sigma^\pm
  \qquad\text{and}\qquad \d\upsilon^\pm = \mp \delta \upsilon^\pm,
\end{equation}
which implies
\begin{equation}
  \d(\sigma^+\sigma^-) = 0 \qquad\text{and}\qquad \d(\upsilon^+\upsilon^-) = 0,
\end{equation}
and hence
\begin{equation}
  H^2 \cong \spn{\sigma^+\sigma^-, \upsilon^+\upsilon^-}.
\end{equation}
This results in the additional brackets
\begin{equation}
  [S_+,S_-] = Z_1 \qquad\text{and}\qquad [V_+,V_-] = Z_2,
\end{equation}
where we have introduced two central generators $Z_1$ and $Z_2$.

\subsubsection{$\GCA{2}$ and $\GCA{5}$}
\label{sec:ca-2-5}

These share the same differential:
\begin{equation}
  \d\delta = 0, \qquad \d\sigma^+ = - \delta\sigma^+, \qquad
  \d\sigma^- = \delta\sigma^- - \upsilon^0 \upsilon^-
  \qquad\text{and}\qquad \d\upsilon^\pm = \mp \delta \upsilon^\pm,
\end{equation}
so that
\begin{equation}
  \d(\sigma^+ \sigma^+) = \sigma^+\upsilon^0\upsilon^- \qquad\text{and}\qquad
  \d(\upsilon^+\upsilon^-) = 0,
\end{equation}
and hence
\begin{equation}
  H^2 \cong \spn{\upsilon^+\upsilon^-},
\end{equation}
with additional Lie brackets
\begin{equation}
  [V_+,V_-] = Z.
\end{equation}

\subsubsection{$\GCA{4}$ and $\GCA{7}$}
\label{sec:ca-4-7}

These share the same differential
\begin{equation}
  \d\delta = 0, \qquad \d\sigma^\pm = \mp \delta\sigma^\pm, \qquad
  \d\upsilon^- = \delta \upsilon^-
  \qquad\text{and}\qquad \d\upsilon^+ = -\delta \upsilon^+ - \sigma^+ \upsilon^0,
\end{equation}
so that
\begin{equation}
  \d(\sigma^+\sigma^-) = 0 \qquad\text{and}\qquad \d(\upsilon^+\upsilon^-) =
  -\sigma^+\upsilon^0\upsilon^-,
\end{equation}
so that
\begin{equation}
  H^2 \cong \spn{\sigma^+\sigma^-},
\end{equation}
with additional brackets
\begin{equation}
  [S_+,S_-] = Z.
\end{equation}

\subsubsection{$\GCA{6}$ and $\GCA{8}$}
\label{sec:ca-6-8}

These share the same differential
\begin{equation}
  \d\delta = 0, \qquad \d\sigma^+ = - \delta\sigma^+, \qquad
  \d\sigma^- = \delta\sigma^- - \upsilon^0 \upsilon^-, \qquad
  \d\upsilon^+ = -\delta \upsilon^+ - \sigma^=\upsilon^0
  \qquad\text{and}\qquad \d\upsilon^- = \delta \upsilon^-,
\end{equation}
so that
\begin{equation}
  \d(\sigma^+\sigma^-) = \sigma^+\upsilon^0\upsilon^- \qquad\text{and}\qquad
  \d(\upsilon^+\upsilon^-) = -\sigma^+\upsilon^0\upsilon^-,
\end{equation}
so that
\begin{equation}
  H^2 \cong \spn{\sigma^+\sigma^- + \upsilon^+\upsilon^-},
\end{equation}
with additional brackets
\begin{equation}
  [S_+,S_-] = Z \qquad\text{and}\qquad [V_+,V_-] = Z.
\end{equation}

\subsubsection{$\GCA{9}$}
\label{sec:ca-9}

Here the differential is given by
\begin{equation}
  \d \delta = 0, \qquad \d \sigma^\pm = \mp \delta \sigma^\pm -
  \upsilon^0\upsilon^\pm \qquad\text{and}\qquad \d\upsilon^\pm = \mp \delta\upsilon^\pm,
\end{equation}
so that
\begin{equation}
  \d (\sigma^+\sigma^-) = - \sigma^-\upsilon^0\upsilon^+ + \sigma^+\upsilon^0\upsilon^-
  \qquad\text{and}\qquad \d( \upsilon^+\upsilon^-) = 0,
\end{equation}
so that
\begin{equation}
  H^2 \cong \spn{\upsilon^+\upsilon^-},
\end{equation}
with additional brackets
\begin{equation}
  [V_+,V_-] = Z.
\end{equation}

\subsubsection{$\GCA{11}$}
\label{sec:ca-11}

Here the differential is given by
\begin{equation}
  \d \delta = 0, \qquad \d \sigma^\pm = \mp \delta \sigma^\pm 
\qquad\text{and}\qquad \d\upsilon^\pm = \mp \delta\upsilon^\pm - \sigma^\pm \upsilon^0,
\end{equation}
so that
\begin{equation}
  \d (\sigma^+\sigma^-) = 0 \qquad\text{and}\qquad \d( \upsilon^+\upsilon^-) = -
  \sigma^+\upsilon^0\upsilon^- + \sigma^-\upsilon^0\upsilon^+,
\end{equation}
so that
\begin{equation}
  H^2 \cong \spn{\sigma^+\sigma^-},
\end{equation}
with additional brackets
\begin{equation}
  [S_+,S_-] = Z.
\end{equation}

\subsubsection{$\GCA{12}$}
\label{sec:ca-12}

Here the differential is given by
\begin{equation}
  \d \delta = -\upsilon^+\upsilon^-, \qquad \d \sigma^\pm = \mp \delta
  \sigma^\pm - \upsilon^0\upsilon^\pm \qquad\text{and}\qquad \d\upsilon^\pm = \mp \delta\upsilon^\pm,
\end{equation}
so that
\begin{equation}
  \d (\sigma^+\sigma^-) =  - \sigma^-\upsilon^0\upsilon^+ + \sigma^+ \upsilon^0 \upsilon^-
  \qquad\text{and}\qquad \d( \upsilon^+\upsilon^-) = 0,
\end{equation}
so that the only $2$-cocycle is $\upsilon^+\upsilon^-$, which is also a
coboundary, so that $H^2 = 0$ and $\hyperlink{ca12}{\GCA{12}}$ admits
no (nontrivial) central extensions.

\subsubsection{$\GCA{13}^{(\varepsilon)}$}
\label{sec:ca-13}

The differential is given by
\begin{equation}
  \d\delta = 0, \qquad \d\sigma^\pm = \mp \delta\sigma^+ - \varepsilon
  \upsilon^0 \upsilon^\pm  \qquad\text{and}\qquad \d\upsilon^\pm = \mp \delta\upsilon^\pm -
  \sigma^\pm \upsilon^0,
\end{equation}
so that
\begin{equation}
  \d(\sigma^+\sigma^-) = \varepsilon \sigma^+\upsilon^0\upsilon^- -  \varepsilon \sigma^-\upsilon^0\upsilon^+ \qquad\text{and}\qquad
  \d(\upsilon^+\upsilon^-) = -\sigma^+\upsilon^0\upsilon^- + \sigma^-\upsilon^0\upsilon^+,
\end{equation}
and hence
\begin{equation}
  H^2 \cong \spn{\sigma^+\sigma^- + \varepsilon \upsilon^+\upsilon^-},
\end{equation}
with additional brackets
\begin{equation}
  [S_+,S_-] = Z \qquad\text{and}\qquad [V_+,V_-] = \varepsilon Z.
\end{equation}

This central extension is isomorphic to the ``pseudo-Bargmann'' Lie
algebra which has recently appeared in \cite{Hartong:2017bwq}, where
it is exhibited as a contraction of a trivial central extension
of the de Sitter algebras in $d+2$ dimensions.

\subsubsection{$\GCA{14}$}
\label{sec:ca-14}

Here the differential is given by
\begin{equation}
  \d \delta = -\sigma^+\sigma^-, \qquad \d \sigma^\pm = \mp \delta
  \sigma^\pm \qquad\text{and}\qquad \d\upsilon^\pm = \mp \delta\upsilon^\pm -
  \sigma^\pm \upsilon^0,
\end{equation}
so that
\begin{equation}
  \d (\sigma^+\sigma^-) =  0 \qquad\text{and}\qquad \d( \upsilon^+\upsilon^-) =
  - \sigma^+\upsilon^0\upsilon^- + \sigma^-\upsilon^0 \upsilon^+,
\end{equation}
so that the only $2$-cocycle is $\sigma^+\sigma^-$, which is also a
coboundary, so that $H^2 = 0$ and $\hyperlink{ca14}{\GCA{14}}$ admits no
(nontrivial) central extensions.

Lie algebra $\hyperlink{ca14}{\GCA{14}}$ is isomorphic to the galilean conformal algebra
of \cite{Bagchi:2009my}, which was known not to admit a central
extension.  The construction in \cite{Bagchi:2009my} is precisely the
contraction described above from the de Sitter algebras.

\subsubsection{$\GCA{15}^{(\varepsilon)}$}
\label{sec:ca-15}

These are semisimple Lie algebras and, by the second Whitehead Lemma, the second
cohomology $H^2 = 0$.

\subsubsection{$\GCA{19}$}
\label{sec:ca-19}

Here the differential is given by
\begin{equation}
  \begin{split}
    \d \delta &= 0\\
    \d \sigma^+ &= - \delta \sigma^+\\
    \d\sigma^- &= \delta\sigma^- - \upsilon^0\upsilon^-\\
    \d\upsilon_a^+  &= - \delta \upsilon_a^+ - \epsilon_{abc} \upsilon^0_b \upsilon^+_c\\
    \d\upsilon^- &= \delta\upsilon^- - \sigma^- \upsilon^0,
  \end{split}
\end{equation}
so that
\begin{equation}
  \d (\sigma^+\sigma^-) = \sigma^+\upsilon^0\upsilon^- \qquad\text{and}\qquad
  \d( \upsilon^+\upsilon^-) = \sigma^-\upsilon^0\upsilon^+ - \upsilon^0\upsilon^+\upsilon^-,
\end{equation}
so that $H^2 = 0$ and this conformal algebra admits no (nontrivial)
central extensions.  This result also follows with little or no
calculation by using Hochschild--Serre relative to the semisimple
subalgebra spanned by $J$ and $V_0$ which is isomorphic to
$\so(3)\oplus \so(3)$.  In this case, the relative complex has no
cochains in degree $0$.

\subsubsection{$\GCA{20}^{(\varepsilon)}$}
\label{sec:ca-20}

The differential is given by
\begin{equation}
  \begin{split}
    \d\delta &= 0\\
    \d\sigma^\pm &= \mp \delta\sigma^+\\
    \d\upsilon_a^+ &= - \delta\upsilon_a^+ - \epsilon_{abc} \upsilon^0_b \upsilon^+_c\\
    \d\upsilon_a^- &=  \delta\upsilon_a^- = \delta\upsilon_a^- - \varepsilon \epsilon_{abc} \upsilon^0_b \upsilon^-_c,
  \end{split}
\end{equation}
so that
\begin{equation}
  \d(\sigma^+\sigma^-) = 0 \qquad\text{and}\qquad
  \d(\upsilon^+\upsilon^-) = (\varepsilon-1) \upsilon^0\upsilon^+\upsilon^-
\end{equation}
and hence
\begin{equation}
  H^2 \cong
  \begin{cases}
    \spn{\sigma^+\sigma^-, \upsilon^+\upsilon^-} & \varepsilon = 1 \\
    \spn{\sigma^+\sigma^-} & \varepsilon = -1,
  \end{cases}
\end{equation}
with additional brackets
\begin{equation}
  \begin{cases}
    [S_+,S_-] = Z_1 \quad\text{and}\quad [V_+,V_-] = Z_2 &  \varepsilon=1\\
    [S_+,S_-] = Z & \varepsilon=-1.
  \end{cases}
\end{equation}

The results are summarised in Table~\ref{tab:summary-cent-ext-d-geq-3}.

\begin{table}[h!]
  \centering
  \caption{Central extensions of graded conformal algebras ($d\geq3$)}
  \label{tab:summary-cent-ext-d-geq-3}
  \setlength{\extrarowheight}{2pt}  
  \rowcolors{2}{blue!10}{white}
  \begin{tabular}{l|*{2}{>{$}l<{$}}|c}\toprule
    Label & \multicolumn{2}{c|}{Additional central brackets} & $\dim H^2$\\\midrule
    $\hyperlink{ca1}{\GCA{1}}$ & [V_+,V_-] = Z_1 & [S_+,S_-] = Z_2 & 2\\
    $\hyperlink{ca2}{\GCA{2}}$ & [V_+,V_-] = Z & & 1\\
    $\hyperlink{ca3}{\GCA{3}}$ & [V_+,V_-] = Z_1 & [S_+,S_-] = Z_2 & 2\\
    $\hyperlink{ca4}{\GCA{4}}$ & & [S_+,S_-] = Z & 1\\
    $\hyperlink{ca5}{\GCA{5}}$ & [V_+,V_-] = Z & & 1\\
    $\hyperlink{ca6}{\GCA{6}}$ & [V_+,V_-] = Z & [S_+,S_-] = Z & 1\\
    $\hyperlink{ca7}{\GCA{7}}$ & & [S_+,S_-] = Z & 1\\
    $\hyperlink{ca8}{\GCA{8}}$ & [V_+,V_-] = Z & [S_+,S_-] = Z & 1\\
    $\hyperlink{ca9}{\GCA{9}}$ & [V_+,V_-] = Z & & 1\\
    $\hyperlink{ca10}{\GCA{10}}$ & [V_+,V_-] = Z_1 & [S_+,S_-] = Z_2 & 2\\
    $\hyperlink{ca11}{\GCA{11}}$ & & [S_+,S_-] = Z & 1\\
    $\hyperlink{ca12}{\GCA{12}}$ & & & 0\\
    $\hyperlink{ca13}{\GCA{13}^{(\varepsilon=\pm1)}}$ & [V_+,V_-] = \varepsilon Z & [S_+,S_-] = Z & 1\\
    $\hyperlink{ca14}{\GCA{14}}$ & & & 0\\
    $\hyperlink{ca15}{\GCA{15}^{(\varepsilon=\pm1)}}$ & & & 0 \\\midrule
    $\hyperlink{ca16}{\GCA{16}}$ & [V_+,V_-] = Z_1 & [S_+,S_-] = Z_2 & 2\\
    $\hyperlink{ca17}{\GCA{17}}$ & [V_+,V_-] = Z_1 & [S_+,S_-] = Z_2 & 2\\
    $\hyperlink{ca18}{\GCA{18}}$ & [V_+,V_-] = Z_1 & [S_+,S_-] = Z_2 & 2\\
    $\hyperlink{ca19}{\GCA{19}}$ & & & 0\\
    $\hyperlink{ca20}{\GCA{20}^{(\varepsilon=+1)}}$ & [V_+,V_-]=Z_1 & [S_+,S_-]=Z_2 & 2 \\
    $\hyperlink{ca20}{\GCA{20}^{(\varepsilon=-1)}}$ & & [S_+,S_-]=Z & 1 \\\bottomrule
  \end{tabular}
\end{table}

\subsection{Invariant inner products}
\label{sec:metric-d-geq-3-centr-ext}

It often happens that a Lie algebra does not admit an invariant inner
product, yet a central extension of it does.  For this to happen, the
degeneracy of the invariant inner product in the original Lie algebra
should be curable by adding central elements.  For the kind of Lie
algebras under consideration, this happens if and only if the original
Lie algebra has an invariant symmetric bilinear form with a
one-dimensional kernel spanned by $D$.  It will turn out that there is
only one additional metric Lie algebra: a one-dimensional central
extension of $\hyperlink{ca16}{\GCA{16}}$, which only exists for $d=3$.
In this section we provide the details.

First of all we notice that (suppressing indices)
\begin{equation}
  \left<V_+,V_-\right> = \left<V_+,[V_-,J]\right> = \left<[V_+,V_-],J\right>,
\end{equation}
so that unless $\left<[V_+,V_-],J\right>\neq 0$, any invariant
bilinear form is degenerate.  Clearly no central term in $[V_+,V_-]$
contributes to $\left<[V_+,V_-],J\right>$, so the only way
$\left<[V_+,V_-],J\right>$ can be nonzero is if $[V_+,V_-] = J +
\cdots$ or, if $d=3$, $[V_+,V_-] = V_0 + \cdots$.  This only occurs
for $\hyperlink{ca12}{\GCA{12}}$, $\hyperlink{ca15}{\GCA{15}^{(\varepsilon)}}$, $\hyperlink{ca16}{\GCA{16}}$, $\hyperlink{ca17}{\GCA{17}}$ and
$\hyperlink{ca18}{\GCA{18}}$.  The Lie algebras
$\hyperlink{ca12}{\GCA{12}}$ and
$\hyperlink{ca15}{\GCA{15}^{(\varepsilon)}}$ do not admit central
extensions, so we concentrate on the rest.

No central extension of the Lie algebras $\hyperlink{ca17}{\GCA{17}}$ and $\hyperlink{ca18}{\GCA{18}}$ admit
an invariant inner product.  Indeed, omitting indices,
\begin{equation}
  \left<V_0,V_0\right> = \left<V_0,[V_0,J]\right> =
  \left<[V_0,V_0],J\right> = 0,
\end{equation}
and
\begin{equation}
  \left<J,V_0\right> = \left<J,[S_+,V_-]\right> =
  \left<[J,S_+],V_-\right> = 0.
\end{equation}
Therefore $\left<V_0,-\right> = 0$ and any invariant symmetric
bilinear form is degenerate.

Finally, let us consider central extensions of $\hyperlink{ca16}{\GCA{16}}$, whose Lie
brackets, in addition to \eqref{eq:conformal}, are (omitting indices)
\begin{equation}
  [V_+,V_-] = V_0 + Z_1 \qquad\text{and}\qquad [S_+,S_-] = Z_2.
\end{equation}
The two-dimensional central extension does not admit an invariant
inner product, but imposing a linear relation between $Z_1$ and $Z_2$
results in a metric Lie algebra.  We may describe it as a double
extension \cite{MedinaRevoy,FSalgebra} of the abelian Lie algebra $\a$
with generators $V_+,V_-,S_+,S_-$ with (trivially invariant) inner
product
\begin{equation}
  \left<\Vp{a},\Vm{b}\right> =  \alpha \delta_{ab}
  \qquad\text{and}\qquad \left<S_+,S_-\right> = \beta,
\end{equation}
for $\alpha,\beta \neq 0$.  The Lie algebra $\g \cong \co(3)$ spanned
by $R_a = -\tfrac12 \epsilon_{abc} J_{bc}$ and $D$ acts on $\a$
via skew-symmetric derivations:
\begin{equation}
  [R_a, \Vpm{b}] = \epsilon_{abc} \Vpm{c}, \qquad [D, \Vpm{a}]
  = \pm \Vpm{a} \qquad\text{and}\qquad [D,S_\pm] = \pm S_\pm.
\end{equation}
To construct the double extension, we add generators $\Vz{a}$ dual
to $R_a$ and $Z$ dual to $D$, together with the dual pairings
\begin{equation}
  \left<R_a, \Vz{b}\right> = \delta_{ab} \qquad\text{and}\qquad
  \left<D,Z\right> = 1,
\end{equation}
and extend the brackets of $\a$ by
\begin{equation}
  [\Vp{a},\Vm{b}] = \alpha \epsilon_{abc} \Vz{c} + \alpha
  \delta_{ab} Z \qquad\text{and}\qquad [S_+,S_-] = \beta Z.
\end{equation}
Choosing $\alpha = 1$, we see that this is a quotient of the universal
central extension of $\hyperlink{ca16}{\GCA{16}}$ where $Z_1=Z$ and $Z_2 = \beta Z$.  The
most general invariant inner product is given by
\begin{equation}
  \begin{aligned}[m]
    \left<\Vp{a},\Vm{b}\right> &=  \delta_{ab}\\
    \left<S_+,S_-\right> &= \beta\\
    \left<R_a, \Vz{b}\right> &= \delta_{ab}\\
  \end{aligned}
  \qquad\text{and}\qquad
  \begin{aligned}[m]
    \left<D,Z\right> &= 1\\
    \left<D,D\right> &= \lambda\\
    \left<R_a, R_b\right> &= \mu \delta_{ab},\\
  \end{aligned}
\end{equation}
for all $\lambda,\mu \in \RR$ and $\beta \in \RR^\times$.

\subsection{Central extensions for $d=2$}
\label{sec:centr-extens-d=2}

In $d=2$ the calculation differs because $\so(2)$ is not semisimple
and the factorisation theorem does not apply.  Nevertheless the
two-dimensional abelian subalgebra $\r = \spn{J,D}$ acts reducibly and
trivially on cohomology, and therefore we may work with the subcomplex
of $\r$-invariant cochains.  The 1- and 2-cochains are as follows:
\begin{equation}
  \begin{split}
    C^1 &= \spn{\rho, \delta}\\
    C^2 &= \spn{\rho\delta, \delta_{ab} \upsilon^+_a\upsilon^-_b,
      \epsilon_{ab} \upsilon^+_a\upsilon^-_b, \epsilon_{ab}
      \upsilon^0_a\upsilon^0_b, \sigma^+\sigma^-}.
  \end{split}
\end{equation}
The calculations are routine and we only list the result, which is
summarised in Table~\ref{tab:summary-cent-ext-d-eq-2}.

\begin{table}[h!]
  \centering
  \caption{Central extensions of graded conformal algebras ($d=2$)}
  \label{tab:summary-cent-ext-d-eq-2}
  \setlength{\extrarowheight}{2pt}  
  \rowcolors{2}{blue!10}{white}
  \begin{tabular}{l|*{4}{>{$}l<{$}}|c}\toprule
    Label & \multicolumn{4}{c|}{Additional central brackets} & $\dim H^2$\\\midrule
    $\hyperlink{ca1}{\GCA{1}}$ & [J,D]=Z_1 & [\Vp{a},\Vm{b}] = \delta_{ab} Z_2 + \epsilon_{ab} Z_3 & [\Vz{a},\Vz{b}] = \epsilon_{ab} Z_4 & [S_+,S_-] = Z_5 & 5\\
    $\hyperlink{ca2}{\GCA{2}}$ & [J,D]=Z_1 & [\Vp{a},\Vm{b}] = \delta_{ab} Z_2 + \epsilon_{ab} Z_3 & [\Vz{a},\Vz{b}] = \epsilon_{ab} Z_4 & & 4\\
    $\hyperlink{ca3}{\GCA{3}}$ & [J,D]=Z_1 & [\Vp{a},\Vm{b}] = \delta_{ab} Z_2 + \epsilon_{ab} Z_3 & & [S_+,S_-] = Z_4 & 4\\
    $\hyperlink{ca4}{\GCA{4}}$ & [J,D]=Z_1 & & [\Vz{a},\Vz{b}] = \epsilon_{ab} Z_2 & [S_+,S_-] = Z_3 & 3\\
    $\hyperlink{ca5}{\GCA{5}}$ & [J,D]=Z_1 & [\Vp{a},\Vm{b}] = \delta_{ab} Z_2 + \epsilon_{ab} Z_3 & & & 3\\
    $\hyperlink{ca6}{\GCA{6}}$ & [J,D]=Z_1 & [\Vp{a},\Vm{b}] = \delta_{ab} Z_2 & [\Vz{a},\Vz{b}] = \epsilon_{ab} Z_3 & [S_+,S_-] = - Z_2 & 3\\
    $\hyperlink{ca7}{\GCA{7}}$ & [J,D]=Z_1 & [\Vp{a},\Vm{b}] = \epsilon_{ab} Z_2 & [\Vz{a},\Vz{b}] = -\epsilon_{ab} Z_2 & [S_+,S_-] = Z_3 & 3\\
    $\hyperlink{ca8}{\GCA{8}}$ & [J,D]=Z_1 & [\Vp{a},\Vm{b}] = \delta_{ab} Z_2 + \epsilon_{ab} Z_3 & [\Vz{a},\Vz{b}] = - \epsilon_{ab} Z_3 & [S_+,S_-] = Z_2 & 3\\
    $\hyperlink{ca9}{\GCA{9}}$ & [J,D]=Z_1 & [\Vp{a},\Vm{b}] = \delta_{ab} Z_2 + \epsilon_{ab} Z_3 & [\Vz{a},\Vz{b}] = \epsilon_{ab} Z_4 & & 4\\
    $\hyperlink{ca10}{\GCA{10}}$ & [J,D]=Z_1 & [\Vp{a},\Vm{b}] = \delta_{ab} Z_2 + \epsilon_{ab} Z_3 & & [S_+,S_-] = Z_4 & 4\\
    $\hyperlink{ca11}{\GCA{11}}$ & [J,D]=Z_1 & & [\Vz{a},\Vz{b}] = \epsilon_{ab} Z_2 & [S_+,S_-] = Z_3 & 3\\
    $\hyperlink{ca12}{\GCA{12}}$ & & & & & 0\\
    $\hyperlink{ca13}{\GCA{13}^{(\varepsilon=\pm1)}}$ & & & & & 0\\
    $\hyperlink{ca14}{\GCA{14}}$ & & [\Vp{a},\Vm{b}] = \epsilon_{ab} Z_1 & [\Vz{a},\Vz{b}] = -\epsilon_{ab} Z_1 & [S_+,S_-] = Z_2 & 2 \\
    $\hyperlink{ca15}{\GCA{15}^{(\varepsilon=\pm1)}}$ & & & & & 0 \\\bottomrule
  \end{tabular}
\end{table}

\subsection{Invariant inner products}
\label{sec:invar-inner-prod-1}

A natural question is whether there are any centrally extended Lie
algebras for $d=2$ which admit an invariant inner product.  Again,
there are some Lie algebras which do not admit central extensions:
$\hyperlink{ca12}{\GCA{12}}$, $\hyperlink{ca13}{\GCA{13}^{(\varepsilon)}}$ and $\hyperlink{ca15}{\GCA{15}^{(\varepsilon)}}$,
and hence we will not consider them further.  In $d=2$ an invariant
inner product must have $\left<V_0,V_0\right>$ and
$\left<V_+,V_-\right>$ nonzero, but (omitting indices)
\begin{equation}
  \left<V_+,V_-\right> = \left<V_+, [V_-,J]\right> = \left<[V_+,V_-],
    J\right>
\end{equation}
and, similarly,
\begin{equation}
  \left<V_0,V_0\right> = \left<V_0, [V_0,J]\right> = \left<[V_0,V_0],
    J\right>,
\end{equation}
where now any of $J,D,Z_i$ can have nonzero inner product with $J$.

Every central extension of $\hyperlink{ca4}{\GCA{4}}$ and $\hyperlink{ca11}{\GCA{11}}$ has $[V_+,V_-]=0$
and every central extension of $\hyperlink{ca10}{\GCA{10}}$ has $[V_0,V_0]=0$, hence they
cannot be metric.  Neither can any central extension of $\hyperlink{ca7}{\GCA{7}}$ or
$\hyperlink{ca14}{\GCA{14}}$ because (omitting indices)
\begin{equation}
  \left<V_0,V_0\right> = \left<V_0, [S_+,V_-]\right> =
  \left<[V_-,V_0],S_+\right>= 0.
\end{equation}
Similarly, no central extension of $\hyperlink{ca3}{\GCA{3}}$ or $\hyperlink{ca5}{\GCA{5}}$ can be metric
because
\begin{equation}
  \left<V_0,V_0\right>=  \left<V_0, [S_+,V_-]\right> =
  \left<[V_0,S_+],V_-\right>= 0.
\end{equation}
No central extension of $\hyperlink{ca2}{\GCA{2}}$ can be metric:
\begin{equation}
  \left<S_+,S_-\right> = \left<S_+, [V_0,V_-]\right> =
  \left<[S_+,V_0],V_-\right> = 0,
\end{equation}
and neither can any central extension of $\hyperlink{ca6}{\GCA{6}}$:
\begin{equation}
  \left<S_+,S_-\right> = \left<S_+, [V_0,V_-]\right> =
  \left<[V_-,S_+],V_0\right> = 0,
\end{equation}
or of $\hyperlink{ca9}{\GCA{9}}$:
\begin{equation}
  \left<S_+,S_-\right> = \left<[V_0,V_+],S_-\right> =
  \left<V_0, [V_+,S_0]\right> = 0.
\end{equation}

This leaves only $\hyperlink{ca1}{\GCA{1}}$ and $\hyperlink{ca8}{\GCA{8}}$ to consider.  We will see that in
each case there is a quotient of the universal central extension which
is metric.  We will do this by exhibiting them as suitable double
extensions \cite{MedinaRevoy,FSalgebra}.

Let $\a$ denote the abelian Lie algebra spanned by $S_\pm, V_\pm, V_0$
and with inner product
\begin{equation}\label{eq:a-metric-ca1}
  \left<S_+,S_-\right> = \alpha, \qquad \left<\Vp{a},\Vm{b}\right> =
  \beta \delta_{ab} \qquad\text{and}\qquad \left<\Vz{a},\Vz{b}\right>
  = \gamma \delta_{ab},
\end{equation}
for some $\alpha,\beta,\gamma \in \RR^\times$.  Then $J$ and $D$ act
on $\a$ via \eqref{eq:conformal} preserving the inner product.
Introduce central elements $Z$ and $Z'$ dual to $J$ and $D$,
respectively, so that
\begin{equation}\label{eq:dual-pairing-ca1}
  \left<Z,J\right> = 1 \qquad\text{and}\qquad \left<Z',D\right> = 1,
\end{equation}
and centrally extend $\a$ as follows:
\begin{equation}
  [\Vz{a},\Vz{b}] = \gamma \epsilon_{ab} Z, \qquad [\Vp{a},\Vm{b}] =
  \beta \epsilon_{ab} Z + \beta \delta_{ab} Z' \qquad\text{and}\qquad
  [S_+,S_-] = \alpha Z'.
\end{equation}
If we normalise the inner product by setting $\beta = 1$, then we see
that this metric Lie algebra corresponds to the quotient of the
universal central extension of $\hyperlink{ca1}{\GCA{1}}$ in
Table~\ref{tab:summary-cent-ext-d-eq-2} where $Z_1 = 0$, $Z_2 = Z'$,
$Z_3 = Z$, $Z_4 = \gamma Z$ and $Z_5 = \alpha Z'$.  The most general
invariant inner product (up to scale) is given by
equations~\eqref{eq:a-metric-ca1} and \eqref{eq:dual-pairing-ca1} and
\emph{any} symmetric bilinear form on the span of $J$ and $D$.

Finally, let $\g$ denote the Lie algebra spanned by $S_\pm, V_\pm,
V_0$ subject to the brackets (omitting indices)
\begin{equation}
  [S_+,V_-] = V_0, \qquad [S_+,V_0] = V_+ \qquad\text{and}\qquad
  [V_0,V_-] = S_-.
\end{equation}
The Lie algebra $\g$ is metric relative to the following inner product
\begin{equation}\label{eq:g-metric-ca8}
  \left<S_+,S_-\right> = 1, \qquad \left<\Vz{a},\Vz{b}\right> = -
  \delta_{ab} \qquad\text{and}\qquad \left<\Vp{a},\Vp{b}\right> = \delta_{ab}.
\end{equation}
The generators $J$ and $D$ act on $\g$ via \eqref{eq:conformal}
preserving both the brackets and the inner product.  Therefore we can
introduce dual generators $Z$, $Z'$ to $J$ and $D$, respectively, with
the corresponding dual pairing 
\begin{equation}
  \label{eq:dual-pairing-ca8}
  \left<J,Z\right> = 1 \qquad\text{and}\qquad \left<D,Z'\right> = 1,
\end{equation}
and centrally extend $\g$ via
\begin{equation}
  [\Vz{a},\Vz{b}] = -\epsilon_{ab} Z, \qquad [\Vp{a},\Vm{b}] =
  \epsilon_{ab} Z + \delta_{ab} Z' \qquad\text{and}\qquad
  [S_+,S_-] = Z'.  
\end{equation}
The resulting Lie algebra is metric relative to the inner product
given by \eqref{eq:g-metric-ca8} and \eqref{eq:dual-pairing-ca8} and
\emph{any} symmetric bilinear form in the span of $J$ and $D$.
Comparing with Table~\ref{tab:summary-cent-ext-d-eq-2}, we see that
this is the quotient of the universal central extension of
$\hyperlink{ca8}{\GCA{8}}$ where $Z_1 = 0$, $Z_2 = Z'$ and $Z_3 = Z$.

\section{Generalised conformal algebras}
\label{sec:gener-conf-algebr}

A more general notion of conformal algebra results from dropping the
requirement that the Lie algebra be graded by the action of $D$. In
other words, a modified definition of conformal algebra might be the
following:

\begin{definition}\label{def:genca}
  By a \textbf{generalised conformal Lie algebra} with
  $d$-dimensional space isotropy we mean a real $\tfrac12
  (d+2)(d+3)$-dimensional Lie algebra with generators $J_{ab} = -
  J_{ba}$, with $1\leq a,b \leq d$, spanning a Lie subalgebra $\r
  \cong \so(d)$; that is,
  \begin{equation}\label{eq:genca-1}
    [J_{ab}, J_{cd}] = \delta_{bc} J_{ad} -  \delta_{ac} J_{bd} -
    \delta_{bd} J_{ac} +  \delta_{ad} J_{bc},
  \end{equation}
  and $3d+3$ generators: rotational vectors $V_a^i$, for $i=1,2,3$, and
  rotational scalars $S^A$, for $A=1,2,3$.  In other words, under
  $\so(d)$ they transform like
  \begin{equation}\label{eq:genca-2}
      [J_{ab}, V^i_c] = \delta_{bc} V^i_a - \delta_{ac} V^i_b \qquad\text{and}\qquad
      [J_{ab}, S^A] = 0.
  \end{equation}
  The rest of the brackets between $V^i_a$ and $S^I$ are only subject
  to the Jacobi identity: in particular, they must be
  $\r$-equivariant.
\end{definition}

In this definition we have dropped the condition that $D$ is a grading
element.  It is not clear why one should call such an algebra
conformal, since all it shares with the simple conformal algebras is
the existence of an $\so(d)$ subalgebra which acts in the same way on
the additional generators.  But these algebras form a large class of
Lie algebras with relations to the other candidates for the conformal
fellowship.

The classification problem for generalised conformal algebras is much more
involved than for the graded conformal algebras treated in the bulk of
this paper and we will not solve it here. Nevertheless let us make a
few comments. If $d=0$, a conformal Lie algebra has dimension $3$ and
hence this agrees with the Bianchi classification of real
three-dimensional Lie algebras \cite{Bianchi} (see \cite{MR1900159}
for an English translation). In contrast, the graded conformal
algebras for $d=0$ are Bianchi VI$_0$ and Bianchi VIII. If $d=1$ there
are no rotations and any real Lie algebra of dimension $6$ is
conformal. Six-dimensional real Lie algebras have been classified in a
series of papers starting with Morozov \cite{MR0130326}, who
classified nilpotent six-dimensional Lie algebras, then Mubarakzjanov
\cite{MR0155872}, who classified solvable six-dimensional solvable Lie
algebras with five-dimensional nil-radical and Turkowski
\cite{MR1054322}, who classified six-dimensional solvable Lie algebras
with a four-dimensional nil-radical. Having thus classified the
solvable six-dimensional Lie algebras, it remains to classify the
non-solvable ones, which can be done in principle via the Levi
decomposition, although I am not aware of any published list. The case
of the graded conformal algebras for $d=1$ has not been studied.

\subsection{The deformation complex for $d\geq 5$}
\label{sec:deformation-complex}

In this section we will take $d \geq 5$; that being the generic range
of dimensions.  As before, the deformation theory approach to the
classification of generalised conformal Lie algebras consists in a
perturbative approach to the solution of the corresponding
Maurer--Cartan equation for the difference between the Lie brackets of
the Lie algebras we want to classify and the ones of the static Lie
algebra, here denoted $\g$.  This approach consists in the first
instance in the calculation of the relative Lie algebra cohomology
group $H^2(\g,\r;\g)$, with $\r$ the rotational subalgebra, which
classifies the infinitesimal deformations.  The obstructions to
integrating infinitesimal deformations live in $H^3(\g,\r;\g)$;
although it is seldom necessary or convenient to calculate that
cohomology group.  What we will need to compute is the
Nijenhuis--Richardson bracket on the cochains
\begin{equation}
  [\![-,-]\!] : C^2(\g,\r;\g) \times C^2(\g,\r;\g) \to C^3(\g,\r;\g),
\end{equation}
which is symmetric in this degree.

In this section we will determine the first few spaces
$C^\bullet(\g,\r;\g)$ of cochains, the action of the differential and
the Nijenhuis--Richardson bracket.

We start by introducing a convenient notation to perform the necessary
calculations.  We will let $V_a^i$, for $i=1,2,3$, stand for $B_a$,
$P_a$ and $K_a$, respectively.  Similarly we will let $S^A$, for
$A=1,2,3$, stand for $D$, $H$ and $L$, respectively.  In this
notation, the static conformal Lie algebra $\g$ is the real span of
$J_{ab}$, $V_a^i$ and $S^A$ and subject to the following brackets:
\begin{equation}
  \begin{split}
    [J_{ab}, J_{cd}] &= \delta_{bc} J_{ad} -  \delta_{ac} J_{bd} - \delta_{bd} J_{ac} +  \delta_{ad} J_{bc}\\
    [J_{ab}, V_c^i] &= \delta_{bc} V^i_a - \delta_{ac} V^i_b\\
    [J_{ab}, S^A] &= 0,
  \end{split}
\end{equation}
with all other brackets vanishing.

We will let $\W$ denote the real vector space spanned by the
$V_a^i$ and the $S^A$.  The canonical dual basis for $\W^*$ is denoted
$\upsilon_{ai}$ and $\sigma_A$, respectively.  We may use the
rotational invariant $\delta_{ab}$ to raise and lower the vector
$\so(d)$ indices with impunity, but the automorphism group of the
deformation complex, which is isomorphic to $\GL(3,\RR) \times \GL(3,\RR)$
acting in the natural way on the three copies of the vector
representation of $\so(d)$ and on the three copies of the scalar
representation of $\so(d)$, does distinguish between these
representations and their duals, hence the need to be careful with
where we position the indices.

The cochain complex $C^\bullet := C^\bullet(\g,\r;\g)$ is isomorphic
to $\left(\Lambda^\bullet \W^* \otimes \g\right)^\r$ and we proceed to
enumerate the first few spaces.  Omitting the $\wedge$ and $\otimes$ products, we see that
\begin{enumerate}[start=0]
\item $C^0$ is spanned by the three cochains $S^A$, for $A=1,2,3$;
\item $C^1$ is spanned by the $18 = 9 + 9$ cochains $\sigma_A S^B$ and
  $\upsilon_{ai}  V_a^j$, where we are summing over $a$ in order to
  arrive at a rotational scalar;
\item $C^2$ is spanned by the $51 = 9 + 27 + 6 + 9$ cochains $\sigma_A
  \sigma_B S^C$, $\sigma_A \upsilon_{ai} V_a^j$, $\upsilon_{ai}
  \upsilon_{bj} J_{ab}$ and $\upsilon_{ai} \upsilon_{aj} S^A$; and
\item $C^3$ is spanned by the $102 = 3 + 27 + 18 + 27 + 27$ cochains
  $\sigma_A \sigma_B \sigma_C S^D$, $\sigma_A \sigma_B \upsilon_{ai}
  V_a^j$, $\sigma_A \upsilon_{ai} \upsilon_{bj} J_{ab}$, $\sigma_A
  \upsilon_{ai} \upsilon_{aj} S^B$ and $\upsilon_{ai} \upsilon_{aj}
  \upsilon_{bk} V_b^\ell$.
\end{enumerate}

The differential $\d: C^\bullet \to C^{\bullet + 1}$ is uniquely
defined by its action on the generators:
\begin{equation}
  \d S^A = \d V_a^i = \d \sigma_A = \d \upsilon_{ai} = 0
  \qquad\text{and}\qquad \d J_{ab} = \upsilon_{ai} V_b^i - \upsilon_{bi} V_a^i.
\end{equation}

\subsection{Infinitesimal deformations}
\label{sec:infin-deform-2}

It follows from the above expression for the differential that
$\d: C^0 \to C^1$ and $\d: C^1 \to C^2$ are the zero maps and thus the
space of infinitesimal deformations is naturally isomorphic to the
kernel of $\d: C^2 \to C^3$.  Computing the differential on the above
basis for $C^2$ we find
\begin{equation}
  \label{eq:diff}
    \d(\sigma_A \sigma_B S^C) = 0 \qquad 
    \d(\sigma_A \upsilon_{ai} V_a^j) = 0 \qquad
    \d(\upsilon_{ai} \upsilon_{bj} J_{ab}) =  \upsilon_{ak} \left(
      \upsilon_{ai} \upsilon_{bj} + \upsilon_{aj} \upsilon_{bi}
    \right) V_b^k \qquad \d(\upsilon_{ai} \upsilon_{aj} S^A) = 0.
\end{equation}
Therefore $H^2(\g,\r;\g)$ is a real $45$-dimensional vector space
spanned by
\begin{equation}
  \sigma_A \sigma_B S^C, \qquad \sigma_A \upsilon_{ai} V_a^j \qquad\text{and}\qquad \upsilon_{ai} \upsilon_{aj} S^A.
\end{equation}
The general infinitesimal deformation is given by
\begin{equation}
  \label{eq:inf-def}
  \varphi_1 = t_C^{AB} \sigma_A \sigma_B S^C + t^A{}^i_j  \sigma_A \upsilon_{ai} V_a^j + t_A^{ij} \upsilon_{ai} \upsilon_{aj} S^A,
\end{equation}
for real parameters $t_C^{AB}= - t_C^{BA}$, $t^A{}^i_j$ and
$t_A^{ij}=-t_A^{ji}$.

\subsection{Obstructions to integrability}
\label{sec:obstr-integr}

In order to explore the integrability properties of the infinitesimal
deformations we will need to calculate the Nijenhuis--Richardson
bracket on $C^2$.  This is tabulated below in Table~\ref{tab:NR-dot},
where we have abbreviated the notation by omitting the $\wedge$ and
$\otimes$ products.

\begin{table}[h!]
  \setlength{\extrarowheight}{4pt}
  \centering
  \caption{Nijenhuis--Richardson $\bullet$}
  \label{tab:NR-dot}
  \begin{tabular}{>{$}c<{$}|*{4}{>{$}c<{$}}}
    \bullet & \sigma_D \sigma_E S^F & \sigma_D \upsilon_{ck} V^\ell_c & \upsilon_{ck} \upsilon_{d\ell} J_{cd} & \upsilon_{ck} \upsilon_{c\ell} S^D\\\hline
    \sigma_A \sigma_B S^C  & \sigma_A \sigma_B (\delta^C_D \sigma_E - \delta^C_E \sigma_D) S^F & \delta_D^C \sigma_A \sigma_B \upsilon_{ck} V^\ell_c & \zero & \zero \\
    \sigma_A \upsilon_{ai} V^j_a & \zero & \delta^j_k \sigma_A \sigma_D \upsilon_{ai} V^\ell_a & \sigma_A \upsilon_{ai} (\delta^j_k \upsilon_{c\ell} + \delta^j_\ell \upsilon_{ck}) J_{ac} & \sigma_A \upsilon_{ai} (\delta^j_k \upsilon_{a\ell} - \delta^j_\ell \upsilon_{ak}) S^D \\
    \upsilon_{ai} \upsilon_{bj} J_{ab}  & \zero & \zero & \zero & \zero \\
    \upsilon_{ai} \upsilon_{bj} S^A  & \upsilon_{ai} \upsilon_{aj} (\delta^A_D \sigma_E - \delta^A_E \sigma_D) S^F & \delta^A_D \upsilon_{ai} \upsilon_{aj} \upsilon_{ck} V^\ell_c & \zero & \zero \\
  \end{tabular}
\end{table}

From Table~\ref{tab:NR-dot} it is easy to write down the
Nijenhuis--Richardson bracket, which in this degree is the
anticommutator of the (non-associative) $\bullet$ product.  The
nonzero brackets are given by
\begin{equation}
  \label{eq:NR-bracket}
  \begin{split}
    [\![\sigma_A\sigma_B S^C, \sigma_D\sigma_E S^F]\!] &= \sigma_A\sigma_B (\delta^C_D\sigma_E - \delta^C_E\sigma_D) S^F + \sigma_D\sigma_E (\delta^F_A\sigma_B - \delta^F_B\sigma_A) S^C \\    
    [\![\sigma_A\sigma_B S^C, \sigma_D\upsilon_{ck}V^\ell_c]\!] &= \delta^C_D \sigma_A\sigma_B \upsilon_{ck} V^\ell_c\\
    [\![\sigma_A\sigma_B S^C, \upsilon_{ck} \upsilon_{c\ell} S^D]\!] &= \upsilon_{ck} \upsilon_{c\ell} (\delta^D_A \sigma_B - \delta^D_B \sigma_A) S^C\\
    [\![\sigma_A \upsilon_{ai} V^j_a, \sigma_D\upsilon_{ck}V^\ell_c]\!] &= \sigma_A\sigma_D (\delta^j_k \upsilon_{ai} V^\ell_a - \delta^\ell_i \upsilon_{ak} V^j_a)\\
    [\![\sigma_A \upsilon_{ai} V^j_a, \upsilon_{ck} \upsilon_{d\ell} J_{cd} ]\!] &= \sigma_A \upsilon_{ai} (\delta^j_k \upsilon_{c\ell} + \delta^j_\ell \upsilon_{ck}) J_{ac}\\
    [\![\sigma_A \upsilon_{ai} V^j_a, \upsilon_{ck} \upsilon_{c\ell} S^D]\!] &= \sigma_A \upsilon_{ai} (\delta^j_k \upsilon_{a\ell} - \delta^j_\ell \upsilon_{ak}) S^D + \delta^D_A \upsilon_{ck} \upsilon_{c\ell} \upsilon_{ai} V^j_a\\
  \end{split}
\end{equation}

Let $\varphi_1$ be the general infinitesimal deformation in
equation~\eqref{eq:inf-def}. The first obstruction to integrating
$\varphi_1$ is the cohomology class of
$\tfrac12 [\![\varphi_1,\varphi_1]\!]$ in $H^3(\g,\r;\g)$. From
equation~\eqref{eq:NR-bracket} we calculate \begin{multline}
  \label{eq:obstruction-1}
  \tfrac12 [\![\varphi_1,\varphi_1]\!] = 2 t^{AB}_C t^{CE}_F \sigma_A\sigma_B\sigma_E S^F +
  \left( t^{AB}_C t^C{}^k_\ell + \tfrac12 t^A{}^k_j t^B{}^j_\ell - \tfrac12 t^B{}^k_j t^A{}^j_\ell \right) \sigma_A \sigma_B \upsilon_{ak} V_a^\ell\\
  + 2 \left(t^{AB}_C t^{k\ell}_A + \tfrac12 t^B{}^k_i t_C^{i\ell} - \tfrac12 t^B{}^\ell_i t_C^{ik}\right) \sigma_B \upsilon_{bk} \upsilon_{b\ell} S^C
  + t^A{}^i_j t_A^{k\ell} \upsilon_{ai} \upsilon_{b k} \upsilon_{b \ell} V_a^j
\end{multline}
From equation~\eqref{eq:diff}, the only way that this can be $\d\varphi_2$
for some $\varphi_2 \in C^2$, is if the following equations are satisfied:
\begin{equation}
  \label{eq:obstreqns-1}
  \begin{split}
    t^{[AB}_C t^{E]C}_F &= 0\\
    t^A{}^k_j t^B{}^j_\ell - t^B{}^k_j t^A{}^j_\ell &= - 2 t^{AB}_C t^C{}^k_\ell\\
    t^B{}^k_i t^{i\ell}_C - t^B{}^\ell_i t_C^{ik} &= - 2 t^{AB}_C t_A^{k\ell}\\
   2 t^A{}^i_j t_A^{k\ell} &= u^{\ell i} \delta^k_j - u^{ki}\delta^\ell_j  \qquad \exists u^{ij} = u^{ji},
  \end{split}
\end{equation}
where $\varphi_2 = \tfrac12 u^{ij} \upsilon_{ai} \upsilon_{bj} J_{ab}$.
Assuming these equations, the next obstruction is the class in
$H^3(\g,\r;\g)$ of
\begin{equation}
  [\![\varphi_1,\varphi_2]\!] = t^A{}^i_j u^{j\ell} \sigma_A \upsilon_{ai} \upsilon_{c\ell} J_{ac}.
\end{equation}
The only way that this can be a coboundary is if it vanishes
identically, which becomes the second-order obstruction
\begin{equation}
  \label{eq:obstreqns-2}
  u^{j\ell} t^A{}^i_j +   u^{ji} t^A{}^\ell_j = 0.
\end{equation}
If this is the case, then we can take $\varphi_3 = 0$ and since
$[\![\varphi_2,\varphi_2]\!] = 0$ identically, we can also take
$\varphi_{n>3} = 0$ and the deformation integrates.  In summary, the
integrability domain is the locus of the equations
\eqref{eq:obstreqns-1} and \eqref{eq:obstreqns-2}.

Contracting the last equation in \eqref{eq:obstreqns-1} with $\delta^j_k$, we can solve
\begin{equation}
  u^{\ell i} =t^A{}^i_k t_A^{k \ell},
\end{equation}
which implies for consistency the symmetry in $\ell \leftrightarrow i$ of the RHS:
\begin{equation}
  \label{eq:obstructions-3}
  t^A{}^i_k t_A^{k \ell} =   t^A{}^\ell_k t_A^{k i}.
\end{equation}

In summary, the most general deformation is given by
\begin{equation}
  \label{eq:def}
  \varphi = t_C^{AB} \sigma_A \sigma_B S^C + t^A{}^i_j\sigma_A
  \upsilon_{ai} V_a^j + t_A^{ij} \upsilon_{ai} \upsilon_{aj} S^A +
  \tfrac12 t^A{}^j_k t_A^{ki} \upsilon_{ai} \upsilon_{bj} J_{ab},
\end{equation}
for some parameters $t_C^{AB}$, $t^A{}^i_j$ and $t_A^{ij}$ and where
we have omitted $\otimes$ and $\wedge$.  This deformation is
integrable provided the following polynomial equations of second and
third degree are satisfied:
\begin{equation}
  \label{eq:obstructions}
  \begin{split}
    t^{[AB}_C t^{E]C}_F &= 0\\
    t^A{}^k_j t^B{}^j_\ell - t^B{}^k_j t^A{}^j_\ell &= - 2 t^{AB}_C t^C{}^k_\ell\\
    t^B{}^k_i t^{i\ell}_C - t^B{}^\ell_i t_C^{ik} &= - 2 t^{AB}_C t_A^{k\ell}\\
    t^A{}^i_j t_A^{j \ell} &=   t^A{}^\ell_j t_A^{j i}\\
    2 t^A{}^i_j t_A^{k\ell} &= t^A{}^i_m t_A^{m\ell} \delta^k_j - t^A{}^i_m t_A^{mk} \delta^\ell_j\\
     t^A{}^i_j t^B{}^j_k t^{k\ell}_B + t^A{}^\ell_j t^B{}^j_k t^{ki}_B &= 0.
  \end{split}
\end{equation}
The integrability locus is the common zero of these equations and the
moduli space of conformal algebras (for $d\geq 4$) is the quotient of
the integrability locus by the natural action of $\GL(3,\RR)
\times \GL(3,\RR)$ on the lower and upper case indices, respectively.

Some of the equations in \eqref{eq:obstructions} admit a natural
Lie-theoretical interpretation.  For example, the first equation says
that $t^{AB}_C$ are the structure constants of a real
three-dimensional Lie algebra $\b$. The second equation says that
$-\tfrac12 t^A{}^i_j$ define a three-dimensional real representation
$E$, say, of $\b$. The third equation says that $t_A^{ij}$ defines a
$\b$-equivariant map $\Lambda^2 E \to \b$. The fourth equation can be
rewritten as
\begin{equation}\label{eq:unimodular}
    0 = t^A{}^i_j t_A^{j \ell} - t^A{}^\ell_j t_A^{j i} = \delta^A_B
    \left( t^B{}^i_j t_A^{j \ell} - t^B{}^\ell_j t_A^{j i}\right)  =
    \delta^A_B \left( t^B{}^i_j t_A^{j \ell} + t^B{}^\ell_j t_A^{i
        j}\right) = \delta^A_B \left(2 t^{BC}_A t_C^{i\ell} \right) =
    2 t^{AC}_A t_C^{i\ell},
\end{equation}
where in the penultimate equality we used the equivariance of
$t_A^{ij}$ under $\b$. Notice that this equation is identically zero
if $\b$ is unimodular. We therefore see that in order to solve the
obstruction conditions requires, as a first step, classifying the
three-dimensional representations of all the Bianchi Lie algebras.

As with deformations of kinematical and graded conformal Lie algebras,
these are also the results for $d=4$.  In $d=4$ we find that $\so(4)$
is not simple, but only semisimple and hence we may split the
rotational generators $J_{ab}$ into their self-dual and anti-self-dual
components.  This does not alter the calculations of the infinitesimal
deformations and the additional freedom it gives in the form of
$\varphi_2$ is not used in overcoming the obstruction to integrability
to second-order.  Therefore the results above for $d\geq 5$ actually
apply to all $d\geq 4$.

Finally, we end this section with the observation that holographic conformal
algebras (defined in the introduction) are special cases of the
generalised conformal algebras in Definition~\ref{def:genca}.  Indeed, a
generalised conformal algebra is holographic if and only if it contains an
$\so(d{+}1)$-subalgebra, so that as a vector space, a holographic
conformal algebra $\g$ decomposes as
\begin{equation}
  \g = \underbrace{\so(d) \oplus \V}_{\cong \so(d{+}1)} \oplus \underbrace{(\V
    \oplus S)}_{\so(d{+}1)-\text{vector}} \oplus \underbrace{(\V \oplus
  S)}_{\so(d{+}1)-\text{vector}} \oplus \underbrace{S}_{\so(d{+}1)-\text{scalar}}.
\end{equation}
As shown in Section~\ref{sec:contractions-1}, the only graded
conformal algebras which are holographic are the de~Sitter algebras
($\hyperlink{ca15}{\GCA{15}^{(\varepsilon)}}$) and the Newton--Hooke
algebra ($\hyperlink{ca13}{\GCA{13}^{(\varepsilon=1)}}$).

\section{Generalised Lifshitz algebras}
\label{sec:gener-grad-conf}

We shall now consider a generalisation of the graded conformal
algebras which results from demanding the existence of a grading, but not
necessarily with the same weights as in Definition~\ref{def:gca}.
This generalises the Lifshitz algebra extended by boosts (see, e.g.,
\cite{Gibbons:2009me}).

\begin{definition}\label{def:gla}
  A \textbf{generalised Lifshitz algebra} (with $d$-dimensional
  space isotropy) is a real Lie algebra $\g$ of dimension $\tfrac12
  (d+1)(d+2) + 1$ satisfying the following properties:
  \begin{enumerate}
  \item $\g$ has a Lie subalgebra $\h \cong \co(d)$,
    and
  \item as a vector space,
    $\g = \h \oplus \V_\alpha \oplus \V_\beta \oplus S_\gamma$, where
    $\V_{\alpha,\beta}$ are copies of the $d$-dimensional vector
    representation of $\so(d)$ with weights $\alpha,\beta$ relative to
    $D$ and $S_\gamma$ is a copy of the scalar representation of
    $\so(d)$ and weight $\gamma$ relative to $D$.
  \end{enumerate}
\end{definition}

It follows from the definition, that a generalised Lifshitz algebra is
nothing but a kinematical Lie algebra $\k$ (with $d$-dimensional space
isotropy) extended by a grading element $D$ which commutes with the
rotations.  Therefore to classify generalised Lifshitz algebras we
need only classify possible gradings of kinematical Lie algebras where
$\so(d)$ lies in degree $0$.  Every such grading defines a derivation
on $\k$ which commutes with the rotations and which is diagonalisable
over $\RR$.  Every such derivation integrates to an automorphism of
$\k$ which acts like the identity on the rotational subalgebra and
lies in the identity component of the group of automorphisms.
Automorphisms of kinematical Lie algebras have been discussed (except
for a few examples where it was not then necessary) in
\cite{Figueroa-OFarrill:2018ilb} in the process of classifying
simply-connected homogeneous kinematical spacetimes.  It is then a
matter mostly of recontextualising those calculations in
\cite{Figueroa-OFarrill:2018ilb} to arrive at the results summarised
in Table~\ref{tab:glas}.

In the notation of Table~\ref{tab:glas} we do not write the $\so(d)$
indices explicitly.  We write $\J$, $\B$, $\P$ and $H$ for the
generators of the kinematical Lie algebra $\k$ and write the
kinematical Lie brackets as
\begin{equation}
  \label{eq:kin}
  [\J,\J] = \J \qquad [\J,\B] = \B \qquad [\J, \P] = \P
  \qquad\text{and}\qquad [\J, H] = 0.
\end{equation}
For $d\neq 2$, any other brackets can be reconstructed unambiguously
from the abbreviated expression since there is only one way to
reintroduce indices using only the $\so(d)$-invariant tensor
$\delta_{ab}$ and, when $d=3$ also $\epsilon_{abc}$ on the right hand
side of the brackets. For example,
\begin{equation}
  [H, \B] = \P \quad\text{stands for}\quad [H, B_a] = P_a \qquad\text{and}\qquad
  [\B,\P] = H + \J \quad\text{for}\quad [B_a, P_b] = \delta_{ab} H + J_{ab}.
\end{equation}
In $d=3$ we may also have brackets of the form
\begin{equation}
 [\P,\P] = \P \qquad\text{which we take to mean}\qquad [P_a,P_b] =
 \epsilon_{abc} P_c.
\end{equation}
If $d=2$, then $\epsilon_{ab}$ is rotationally invariant and can
appear in Lie brackets.  So we will write, e.g.,
\begin{equation}
  [H, \B] = \B + \Pt  \qquad\text{for}\qquad  [H, B_a] = B_a +
  \epsilon_{ab} P_b,
\end{equation}
and
\begin{equation}
  [\B,\B] = \Ht \qquad\text{for}\qquad [B_a, B_b] = \epsilon_{ab} H,
\end{equation}
et cetera.  Also, whenever $\J$ appears it denotes $J_{ab}$.  In
$d=1$, it is tacitly assumed that we set any $\J$ to zero.

In Table~\ref{tab:glas} we write the isomorphism class of the Lie
algebra (if known) using the notation in Table~\ref{tab:notation}.

\begin{table}[h!]
  \centering
  \caption{Notation for Lie algebras}
  \label{tab:notation}
  \begin{tabular}{>{$}c<{$}|l} \toprule
    \multicolumn{1}{c|}{Notation} & \multicolumn{1}{c}{Name}\\\midrule
    \s & static\\
    \n_+ & (elliptic) Newton\\
    \n_- & (hyperbolic) Newton\\
    \e & euclidean\\\bottomrule
  \end{tabular}
  \hspace{3cm}
  \begin{tabular}{>{$}c<{$}|l} \toprule
    \multicolumn{1}{c|}{Notation} & \multicolumn{1}{c}{Name}\\\midrule
    \p & Poincaré\\
    \so & orthogonal\\
    \g & galilean\\
    \c & Carroll\\ \bottomrule
  \end{tabular}
\end{table}

Table~\ref{tab:glas} is subdivided into three sections separated by
horizontal rules: kinematical Lie algebras which exist for generic
$d$, those which are unique to $d=3$ and those which are unique to
$d=2$.  In $d=1$ there are accidental isomorphisms: $\c \cong \g$,
$\so(2,1) \cong \so(1,2)$, $\e \cong \n_+$ and $\p \cong \n_-$.  In
addition, for $d=2$, $\n_+$ admits more gradings than for $d\neq 2$,
which is why we have listed it separately in $d=2$.  The notation for
the grading is such that $w_X$ denotes the degree (or ``weight'') of
the generator $X$ and $\alpha,\beta,\gamma \in \RR$.

\begin{table}[h!]
  \centering
  \caption{Generalised Lifshitz algebras with $d$-dimensional space isotropy}
  \label{tab:glas}
  \rowcolors{2}{blue!10}{white}
  \resizebox{\textwidth}{!}{
  \begin{tabular}{l|>{$}c<{$}|>{$}l<{$}|*{5}{>{$}l<{$}}|*{3}{>{$}c<{$}}} \toprule
    \multicolumn{1}{c|}{GLA\#} & d & \multicolumn{1}{c|}{$\cong$} &  \multicolumn{5}{c|}{Nonzero Lie brackets in addition to $[\J,\J] = \J$, $[\J, \B] = \B$ and $[\J,\P] = \P$} & w_{\B} & w_{\P} & w_H \\\midrule
    $\GLA1$ & \geq 0 & \s & & & & & & \alpha & \beta & \gamma \\
    $\GLA2$ & \geq 1 & \g & [H,\B] = -\P & & & & & \alpha  & \alpha + \gamma & \gamma \\
    $\GLA3^{(\chi\in(-1,1))}$ & \geq 1 &  & [H,\B] = \chi \B & [H,\P] = \P & & & & \alpha & \beta & 0 \\
    $\GLA4$ & \geq 1 &  & [H,\B] = \B & [H,\P] = \P & & & & \alpha & \beta & 0\\
    $\GLA5$ & \geq 1 & \n_- & [H,\B] = - \B & [H,\P] = \P & & & & \alpha  & \beta & 0 \\
    $\GLA6$ & \geq 1 & & [H,\B] = -\P & [H, \P] = \B + 2\P & & & & \alpha & \alpha & 0 \\
    $\GLA7^{(\chi>0)}$ & \geq 1 & & [H,\B] = \chi \B + \P & [H,\P] = \chi \P - \B &  & & & \alpha & \alpha & 0 \\
    $\GLA8$ & \geq 1 (\neq 2) & \n_+ & [H,\B] = \P & [H,\P] = - \B & & & & \alpha & \alpha & 0 \\
    $\GLA9$ & \geq 2 & \c & & & & [\B,\P] = H & & \alpha & \beta & \alpha + \beta \\
    $\GLA{10}^{(\varepsilon=\pm1)}$ & \geq 2 & \choice{\p}{\e} & [H,\B] = -\varepsilon \P & &  [\B,\B]= \varepsilon \J & [\B,\P] = H & & 0 & \beta & \beta \\
    $\GLA{11}$ & \geq 2 & \so(d{+}1,1) & [H,\B] = \B & [H,\P] = -\P & &  [\B,\P] = H + \J & & \alpha & -\alpha & 0\\
    $\GLA{12}^{(\varepsilon=\pm1)}$ & \geq 1 & \choice{\so(d,2)}{\so(d{+}2)} & [H,\B] = -\varepsilon \P & [H,\P] = \varepsilon \B &  [\B,\B]= \varepsilon \J & [\B,\P] = H &  [\P,\P] = \varepsilon \J & 0 & 0 & 0 \\ \midrule
    $\GLA{13}^{(\varepsilon=\pm1)}$ & 3 & & & & [\B,\B]= \B & & [\P,\P] = \varepsilon (\B-\J) & 0 & 0 & \gamma \\
    $\GLA{14}$ & 3 & & & & [\B,\B] = \B & & & 0 & \beta & \gamma \\
    $\GLA{15}$ & 3 & & & & [\B, \B] = \P & & & \alpha & 2\alpha & \gamma \\
    $\GLA{16}$ & 3 & & & [H ,\P] = \P & [\B,\B] = \B & & & 0 & \beta & 0 \\
    $\GLA{17}$ & 3 & & [H ,\B] = -\P & & [\B,\B] = \P & & & \alpha & 2 \alpha & \alpha \\
    $\GLA{18}$ & 3 & & [H ,\B] = \B & [H ,\P] = 2\P & [\B,\B] = \P & & & \alpha  & 2\alpha & 0 \\ \midrule
    $\GLA{19}^{(\xi,\chi)}$ & 2 & & [H,\B] = \B & [H,\P] = \xi \P - \chi \Pt & \multicolumn{3}{c|}{($\xi \in [-1,1),~\chi > 0$)} & \alpha & \beta & 0 \\
    $\GLA{20}$ & 2 & \n_+ & [H,\B] = \Bt & & & & & \alpha & \beta & 0 \\
    $\GLA{21}$ & 2 & & & & [\B, \B]= \Ht & & [\P, \P] = \Ht & \alpha & \alpha & 2\alpha \\
    $\GLA{22}$ & 2 & & [H,\B] = \Bt & & [\B,\B] = \Ht & & [\P,\P] = \J + \Ht & 0 & 0 & 0 \\
    $\GLA{23}$ & 2 & & & & [\B,\B] = \Ht & & & \alpha & \beta & 2\alpha \\
    $\GLA{24}$ & 2 & & [H,\B] = \P & & [\B,\B] = \Ht & & & \alpha & 3 \alpha & 2 \alpha \\
    $\GLA{25}^{(\varepsilon=\pm1)}$ & 2 & & [H,\B] = \varepsilon \Bt & & [\B,\B] = \Ht & & & 0 & \beta & 0 \\ \bottomrule
  \end{tabular}
  }
\end{table}

To each row in Table~\ref{tab:glas} there corresponds a generalised
Lifshitz algebra with additional generator $D$ satisfying:
\begin{equation}
  [D,\J] =0, \qquad [D,\B] = w_{\B} \B, \qquad [D,\P] = w_{\P} \P
  \qquad\text{and}\qquad [D,H] = w_H H,
\end{equation}
with the given $w_{\B}$, $w_{\P}$ and $w_H$.  We may rescale $D$
so that one of the nonzero $w_{\B}$, $w_{\P}$ and $w_H$ is equal to
$1$.

In a generalised Lifshitz algebra, the grading element is an (outer)
derivation of a kinematical Lie algebra.  To be able to interpret the
grading element as a dilatation, it seems reasonable to require that
the generalised Lifshitz algebra $\g$ can be ``extended'' (\emph{not}
in the algebraic sense of an extension) by additional generators:
namely, one $\so(d)$ scalar and/or one $\so(d)$ vector, in such a way
that the resulting algebra is still graded by the adjoint action of
$D$, but such that now $D$ appears in the right-hand side of a Lie
bracket.  This is an interesting problem which is beyond the scope of
this paper.

\section{Generalised Schrödinger algebras}
\label{sec:gener-schr-algebr}

As advocated in several places
\cite{Duval:1990hj,Balasubramanian:2008dm,Bagchi:2009my}, it is
worthwhile to think of the Schrödinger algebra as (the central
extension of) a conformal algebra.  The defining feature of Lie
algebras such as the Schrödinger algebra is that it should be the
central extension of a Lie algebra possessing an
$\so(d) \oplus \sl(2,\RR)$ subalgebra and, in addition, two (or
perhaps three) copies of the $\so(d)$ vector representation.  In other
words, comparing to the generalised conformal algebra in
Definition~\ref{def:genca}, one of the vector representations might be
missing (typically the one which can be interpreted as spatial special
conformal transformations) and the Bianchi Lie algebra spanned by the
three scalars is isomorphic to Bianchi VIII.  Of the graded conformal
algebras classified in this paper, the only ones where the scalars
span an $\sl(2,\RR)$ subalgebra are the galilean conformal algebra
($\hyperlink{ca14}{\GCA{14}}$) and the simple conformal algebras
($\hyperlink{ca15}{\GCA{15}^{(\varepsilon)}}$).

\begin{definition}\label{def:gsa}
  A \textbf{generalised Schrödinger algebra} (with $d$-dimensional
  space isotropy) is a real Lie algebra $\g$ of dimension $\frac12 d(d+3)+4$
  satisfying the following properties:
  \begin{enumerate}
  \item $\g$ has a Lie subalgebra $\h \cong \so(d) \oplus \sl(2,\RR)$,
    and
  \item as a vector space, $\g = \h \oplus (\V\otimes E) \oplus \RR Z$, where $\V$ is
    a copy of the $d$-dimensional vector representation of $\so(d)$,
    $E$ is a representation of $\sl(2,\RR)$ of dimension $2$ or $3$,
    and $Z$ is a central element.
  \end{enumerate}
\end{definition}

\subsection{The case $\dim E = 2$}
\label{sec:dim-e-eq-2}

We shall start with the case where $E$ is two-dimensional.  Observe
that $\sl(2,\RR)$ has precisely two inequivalent two-dimensional
representations: the fundamental representation and the trivial
representation.  This gives rise to two classes of generalised
Schrödinger algebras with $\dim E = 2$.

In more concrete terms, a generalised Schrödinger algebra (with
$\dim E =2$) admits a basis $(J_{ab},S^A,V^i_a,Z)$, where $J_{ab}$
span an $\so(d)$ subalgebra, $S^A$ are rotational scalars spanning an
$\sl(2,\RR)$ subalgebra, $V^i_a$, for $i=1,2$, are rotational vectors
and the rotational scalar $Z$ is central.  The (potentially) nonzero
brackets (for $d\geq 4$) are given by
\begin{equation}
  \label{eq:common-schr}
  \begin{split}
    [J_{ab}, J_{cd}] &= \delta_{bc} J_{ad} - \delta_{ac} J_{bd} - \delta_{bd} J_{ac} + \delta_{ad} J_{bc}\\
    [J_{ab}, V^i_c] &= \delta_{bc} V^i_a - \delta_{ac} V^i_b\\
    [S^A,S^B] &= f^{AB}{}_C S^C\\
    [S^A, V^i_a] &= t^{A\,i}{}_j V^j_a\\
  \end{split}
\end{equation}
where $f^{AB}{}_C$ are the structure constants of $\sl(2,\RR)$, and
where $S^A \mapsto t^{A\,i}{}_j$ is a two-dimensional representation of
$\sl(2,\RR)$, and in addition
\begin{equation}\label{eq:addit-schr}
  [V^i_a,V^j_b] = \delta_{ab} \epsilon_{ij} Z + u^{ij} J_{ab},\\
\end{equation}
where $u^{ij} = u^{ji}$.  Notice that we cannot have $S^A$
appearing in the $[V,V]$ brackets, for that would require the
existence of an $\sl(2,\RR)$-equivariant map $f: \Lambda^2E \to
\sl(2,\RR)$, but since $\Lambda^2E$ is always a trivial representation
(whether or not $E$ is trivial), its image under $f$ would have to be
central in $\sl(2,\RR)$ and $\sl(2,\RR)$, being simple, has trivial
centre.

In the first class of algebras, $E$ is the trivial representation and
hence $t^{A\,i}{}_j = 0$.  Subjecting the Lie bracket
\eqref{eq:addit-schr} to the $[V,V,V]$ Jacobi identity, we find that
\begin{equation}
  (u^{ij}\delta^k_\ell - u^{ki} \delta^j_\ell) \delta_{bc} \delta_{ae} + 
  (u^{jk}\delta^i_\ell - u^{ij} \delta^k_\ell) \delta_{ca} \delta_{be} + 
  (u^{ki}\delta^j_\ell - u^{jk} \delta^i_\ell) \delta_{ab} \delta_{ce} = 0,
\end{equation}
for all $i,j,k,\ell=1,2$ and $a,b,c,e = 1,\dots,d$, where $d\geq 4$.
Taking $a = b \neq c = e$, we are left with
\begin{equation}
  u^{ki}\delta^j_\ell = u^{jk} \delta^i_\ell,
\end{equation}
for all $i,j,k,\ell=1,2$.  Put $i\neq \ell = j$ and we see that
$u^{ki} = 0$ for all $k$ and all $i$.  Therefore,
\begin{equation}\label{eq:VV}
  [V^i_a,V^j_b] = \delta_{ab} \epsilon_{ij} Z.
\end{equation}

In the second class of algebras, $E$ is the fundamental representation
of $\sl(2,\RR)$.  Then the Jacobi identity implies, in particular,
that $u^{ij}$ is $\sl(2,\RR)$-invariant.  But $u \in \odot^2 E$, which
is a non-trivial irreducible representation of $\sl(2,\RR)$, and hence
the only invariant is $u=0$.  Therefore, the additional brackets are
those in \eqref{eq:addit-schr} with $u^{ij} = 0$ and $t^{A\,i}{}_j$
the fundamental representation of $\sl(2,\RR)$.  This is (the $d\geq
4$ avatar of) the Schrödinger algebra of
\cite{Hagen:1972pd,MR0400948}.

The situation for $d=3$ is slightly more involved for the first type
of algebras where $E$ is a trivial representation of $\sl(2,\RR)$, but
not for the second type of algebras where $E$ is the fundamental
representation. Treating the problem using the methods of deformation
theory, we look to classify the deformations of the Lie algebra $\g$
with all brackets zero except for those in \eqref{eq:common-schr}. The
deformation complex, by Hochschild--Serre, is the Chevalley--Eilenberg
complex of $\g$ with values in the adjoint representation, but
relative to the semisimple subalgebra
$\h \cong \so(3) \oplus \sl(2,\RR)$. If we write $\g = \h \oplus \W$,
where $\W$ is the span of the $V_a^i$, then the cochains in the
deformation complex are
$C^p := \left(\Lambda^p \W^* \otimes \g\right)^\h$.  The differential
in this complex is zero, so $H^p = C^p$ for all $p$.  If $E$ is the
fundamental representation of $\sl(2,\RR)$, it is an easy calculation
using the representation theory of $\so(3)$ and $\sl(2,\RR)$, that
$C^2 =0$.  Therefore for $E$ the fundamental representation, there are
no deformations of $\g$ and hence its central extension is again the
Schrödinger algebra.

For $E$ the trivial representation, however,
equation~\eqref{eq:addit-schr} is replaced with
\begin{equation}
  [V^i_a,V^j_b] = \delta_{ab} \epsilon_{ij} Z + u^{ij} J_{ab} +
  t^{ij}_k \epsilon_{abc} V^k_c,\\
\end{equation}
subject to the $[V,V,V]$ Jacobi identity.  There are three components
to the Jacobi identity: the one along $Z$, the one along $J$ and the
one along $V$ itself.  The component along $Z$ says
\begin{equation}\label{eq:jac-Z}
  t^{ij}_\ell \epsilon^{\ell k} + t^{jk}_\ell \epsilon^{\ell i} +
  t^{ki}_\ell \epsilon^{\ell j} = 0,
\end{equation}
which translates into
\begin{equation}\label{eq:sol-Z}
t^{11}_2 = t^{22}_1 = 0, \qquad t^{11}_1 = 2 t^{12}_2
\qquad\text{and}\qquad t^{22}_2 = 2 t^{12}_1.
\end{equation}
The component along $J$ gives
\begin{equation}
  \label{eq:jac-J}
  t^{ij}_\ell u^{\ell k} (\delta_{ac}\delta_{bd} - \delta_{ad}\delta_{bc}) +
  t^{jk}_\ell u^{\ell i} (\delta_{ba}\delta_{cd} - \delta_{bd}\delta_{ca}) + 
  t^{ki}_\ell u^{\ell j} (\delta_{cb}\delta_{ad} - \delta_{cd}\delta_{ab}) = 0,
\end{equation}
which taking \eqref{eq:sol-Z} into account gives two relations
\begin{equation}\label{eq:rel-J}
  2 t^{12}_1 u^{11} = t^{11}_1 u^{12} \qquad\text{and}\qquad 2
  t^{12}_1 u^{12} = t^{11}_1 u^{22}.
\end{equation}
Finally, the component of the $[V,V,V]$ Jacobi identity along $V$
gives
\begin{equation}
  \label{eq:jac-V}
  (u^{ij} \delta^k_\ell - t^{ij}_m t^{mk}_\ell)(\delta_{bc}  \delta_{ad} - \delta_{ac}\delta_{bd}) + 
  (u^{jk} \delta^i_\ell - t^{jk}_m t^{mi}_\ell)(\delta_{ca}  \delta_{bd} - \delta_{ba}\delta_{cd}) + 
  (u^{ki} \delta^j_\ell - t^{ki}_m t^{mj}_\ell)(\delta_{ab}
  \delta_{cd} - \delta_{cb}\delta_{ad}) = 0,
\end{equation}
which taking \eqref{eq:sol-Z} into account, simply allows us to solve
for the $u^{ij}$ in terms of the $t^{ij}_k$:
\begin{equation}
  u^{11} = \left(t^{12}_2\right)^2,\qquad   u^{12} = t^{12}_1 t^{12}_2
  \qquad\text{and}\qquad u^{22} = \left(t^{12}_1\right)^2.
\end{equation}
Inserting these back into equation~\eqref{eq:rel-J} we see that the
two equations are identically satisfied.

Letting $t^{12}_1 =: \lambda$ and $t^{12}_2 =: \mu$, the $[V,V]$
bracket becomes (in shorthand notation)
\begin{equation}
  [V^1,V^1] = \mu^2 J + 2 \mu V^1, \qquad [V^1,V^2] = Z + \lambda \mu
  J + \lambda V^1 + \mu V^2 \qquad\text{and}\qquad [V^2,V^2] =
  \lambda^2 J + 2 \lambda V^2.
\end{equation}
We have four cases depending on whether or not $\lambda$ and $\mu$ are
zero:
\begin{enumerate}
\item $\lambda = \mu = 0$: the only nonzero bracket is
  \begin{equation}
    [V^1_a, V^2_b] = \delta_{ab} Z;
  \end{equation}
\item $\lambda = 0$ and $\mu \neq 0$: the nonzero brackets are now (after
  rescaling $V^1$ by $\mu^{-1}$):
  \begin{equation}
    [V^1_a,V^1_b] = J_{ab} + 2 \epsilon_{abc} V^1_c
    \qquad\text{and}\qquad
    [V^1_a,V^2_b] = \delta_{ab} Z + \epsilon_{abc} V^2_c,
  \end{equation}
  but changing basis to $V^1_a \mapsto V^1_a + \frac12 \epsilon_{abc}
  J_{bc}$, we arrive at only one nonzero bracket: namely,
  \begin{equation}
    [V^1_a, V^2_b] = \delta_{ab} Z;
  \end{equation}
\item $\lambda \neq 0$ and $\mu = 0$ is isomorphic to the previous
  case via the change of basis $(J,V^1,V^2,Z) \mapsto (J,V^2,V^1,-Z)$;
  and
\item $\lambda \mu \neq 0$, where the nonzero brackets are now (after
  rescaling $V^1$ by $\mu^{-1}$ and $V^2$ by $\lambda^{-1}$):
  \begin{equation}
    \begin{split}
      [V^1_a,V^1_b] &= J_{ab} + 2 \epsilon_{abc} V^1_c\\
      [V^2_a,V^2_b] &= J_{ab} + 2 \epsilon_{abc} V^2_c\\
      [V^1_a,V^2_b] &= \delta_{ab} Z + J_{ab} + \epsilon_{abc} (V^1_c + V^2_c),
    \end{split}
  \end{equation}
  which, changing basis to $V^\pm :=\tfrac12 (V^1\pm V^2)$, simplifies
  to the following non-zero brackets:
  \begin{equation}
    [V^+_a,V^+_b] = J_{ab} + 2 \epsilon_{abc} V^+_c
    \qquad\text{and}\qquad
    [V^+_a,V^-_b] = -\tfrac12 \delta_{ab} Z + \epsilon_{abc} V^-_c,
  \end{equation}
  and, finally, changing basis $V^+_a \mapsto V^+_a + \frac12
  \epsilon_{abc} J_{bc}$ and redefining $Z$, we find that the only nonzero bracket
  remains
  \begin{equation}
    [V^+_a,V^-_b] = \delta_{ab} Z.
  \end{equation}
\end{enumerate}
In summary, we find again that just as for $d\geq 4$, there is only
one isomorphism class of Lie algebras, with additional bracket given
by equation~\eqref{eq:VV}.

\subsection{The case $\dim E = 3$}
\label{sec:dim-e-eq-3}

Consider now the case where $E$ is a three-dimensional representation
of $\sl(2,\RR)$.  Then we are essentially in the special case of
Definition~\ref{def:genca} where the Lie algebra spanned by the
scalars is isomorphic to $\sl(2,\RR)$ or, equivalently, Bianchi~VIII.
The representation $E$ is isomorphic to one of the following three
representations:
\begin{enumerate}
\item the trivial three-dimensional representation $\RR^3$;
\item the direct sum $F \oplus \RR$, where $\RR$ is the trivial
  one-dimensional representation and $F$ is the fundamental
  two-dimensional representation; or
\item the adjoint representation.
\end{enumerate}
Restricting to $d\geq 4$, we have that the most general deformation is
given by equation~\eqref{eq:def} and the obstructions to integrating
it are given by equation~\eqref{eq:obstructions}.  The first three
equations say that $t^{AB}_C$ are the structure constants of
$\sl(2,\RR)$ relative to some basis, $-\tfrac12 t^A{}^i_j$ define the
three-dimensional representation $E$ and $t_A^{ij}$ defines an
$\sl(2,\RR)$-equivariant linear map $\Lambda^2E \to \sl(2,\RR)$.  The
fourth obstruction equation vanishes because $\sl(2,\RR)$ is
unimodular, as shown by equation~\eqref{eq:unimodular}.

If $E$ is the trivial representation, then $t^A{}^i_j = 0$ and since
$\sl(2,\RR)$ has no centre, $t_A^{ij} = 0$ as well.  The remaining two
equations in \eqref{eq:obstructions} are automatically satisfied.
Hence the algebra is given by the common Lie brackets \eqref{eq:genca-1} and
\eqref{eq:genca-2} and in addition $[S^A, S^B] = t^{AB}_C S^C$
defining $\sl(2,\RR)$.  We may centrally extend this algebra by
generators $Z^{ij} = -Z^{ji}$ and brackets
\begin{equation}
  [V^i_a, V^j_b] = \delta_{ab} Z^{ij}.
\end{equation}

If $E = F \oplus \RR^2$, then $\Lambda^2 E \cong E$ as
$\sl(2,\RR)$-representations and since there is no
$\sl(2,\RR)$-equivariant map $E \to \sl(2,\RR)$, we see that
$t_A^{ij} = 0$.  The last two equations in \eqref{eq:obstructions} are
automatically satisfied.  Therefore the Lie algebra is spanned by
$J,V^0,V^+,V^-,S^0,S^+,S^-$, with Lie brackets given by
\eqref{eq:genca-1} and \eqref{eq:genca-2} and in addition
\begin{equation}
  [S^\pm,V^\mp] = V^\pm, \quad [S^0,V^\pm] = \pm V^\pm, \quad
  [S^0,S^\pm] = \pm 2 S^\pm \qquad\text{and}\qquad [S^+,S^-] = S^0.
\end{equation}
The Lie algebra admits a central extension with generator $Z$ and Lie
brackets
\begin{equation}
  [V^+_a, V^-_b] = \delta_{ab} Z.
\end{equation}
Notice that the subspace spanned by $V_0$ is an ideal and quotienting
by this ideal gives again the Schrödinger algebra.  This Lie algebra
is graded, but the weights are not those of a graded conformal algebra
as in Definition~\ref{def:gca}.  This suggests relaxing the definition
of a graded conformal algebra by allowing arbitrary conformal weights.
We discussed this in Section \ref{sec:gener-grad-conf} in the context of
generalised Lifshitz algebras.

Finally, if $E$ is the adjoint representation, then now $t_A^{ij}$ is
a constant multiple of the isomorphism $\Lambda^2 E \to E \cong
\sl(2,\RR)$.  For any value of this constant, one can check explicitly
that the last two equations in \eqref{eq:obstructions} are satisfied.
Therefore the resulting Lie algebra is isomorphic to one of the simple
conformal algebras $\hyperlink{ca15}{\GCA{15}^{(\varepsilon)}}$.
These algebras do not admit any nontrivial central extensions.

\section{Conclusions and open problems}
\label{sec:conclusions}

This paper is devoted to the slippery notion of a conformal algebra.
We have focussed primarily on the notion of a graded conformal
algebra with $d$-dimensional space isotropy (see
Definition~\ref{def:gca}) and we have classified them up to isomorphism
for all $d \geq 2$.  We have done this by classifying deformations of
the static graded conformal algebra defined in the introduction.  For
each such $d\neq 3$ there are 17 isomorphism classes, which are listed
in Table~\ref{tab:summary-d-geq-4}, which also applies when $d=2$.
For $d=3$ there are in addition another 6 isomorphism classes, which are listed in
Table~\ref{tab:summary-d-eq-3}.  Some of these Lie algebras are
related by contraction and this defines a partial order in the set of
isomorphism classes whose Hasse diagram is depicted in
Figure~\ref{fig:contractions}, which also applies to $d=2$.  With the
exception of the simple Lie algebras
$\hyperlink{ca15}{\GCA{15}^{(\varepsilon)}}$ (isomorphic to
$\so(d{+}2,1)$ or $\so(d{+}1,2)$, respectively), none of the other graded
conformal algebras admit an invariant inner product.  We then
classified the central extensions of these Lie algebras.  The
situation was different in $d\geq 3$ and $d=2$ and is summarised in
Tables~\ref{tab:summary-cent-ext-d-geq-3} and
\ref{tab:summary-cent-ext-d-eq-2}, respectively.  We then investigated
whether any central extended graded conformal algebras admit an
invariant inner product and we found that this was the case for
$\hyperlink{ca16}{\GCA{16}}$ (which exists only for $d=3$) and,
provided that $d=2$, also for $\hyperlink{ca1}{\GCA{1}}$ and
$\hyperlink{ca8}{\GCA{8}}$.

We then discussed some other notions of conformal algebras obtained by
relaxing and/or modifying some of the properties of the graded
conformal algebras.

In Section~\ref{sec:gener-conf-algebr} we discussed a class of Lie
algebras defined by dropping the condition that $D$ is a grading
element in Definition~\ref{def:gca}.  We call the resulting Lie
algebras \emph{generalised conformal algebras} (see
Definition~\ref{def:genca}), but it is questionable whether all such
Lie algebras are in any way conformal, since all they share with the
ur-example of conformal algebra (the algebra of conformal symmetries
of Minkowski spacetime) is that they have the generators transform in
the same way under the $\so(d)$ subalgebra of rotations.  Nevertheless
we present some preliminary results about their classification via
deformation theory for $d\geq 4$.  We will use some of these results
below in a restricted context.

In Section~\ref{sec:gener-grad-conf} we discuss what we call
\emph{generalised Lifshitz algebras} (see Definition~\ref{def:gla}),
which consist of a graded kinematical Lie algebra extended by the
grading element, but not requiring that the degrees are those of the
graded conformal algebras treated in the bulk of this paper.  The
results here are preliminary, but we present a classification of the
possible $\ZZ$-gradings of all kinematical Lie algebras, which is
contained in Table~\ref{tab:glas}.

In Section~\ref{sec:gener-schr-algebr} we discussed a class of
conformal Lie algebras related to the Schrödinger algebra.  The
Schrödinger algebra can be understood as a central extension of a
conformal algebra and this suggests the definition of a
\emph{generalised Schrödinger algebra} (see Definition~\ref{def:gsa}),
whose main property is the existence of an $\so(d) \oplus \sl(2,\RR)$
subalgebra under which the remaining (non-central) generators
transform according to $V \otimes E$, where $V$ is the vector
representation of $\so(d)$ and $E$ is a representation of $\sl(2,\RR)$
of dimension $2$ or $3$.   We classified the isomorphism classes of
generalised Schrödinger algebras with $\dim E = 2$ for $d\geq 3$ and
with $\dim E = 3$ for $d\geq 4$.  Apart from the Schrödinger algebra,
we find an (non-central) extension by a vector representation of
$\so(d)$ (trivial under $\sl(2,\RR)$) and centrally extended algebras
where $E$ is the trivial representation of $\sl(2,\RR)$.

There are a number of open problems remaining to complete the results
in the last three sections, among which we list the following in no
particular order:
\begin{itemize}
\item Classify the generalised conformal algebras of
  Section~\ref{sec:gener-conf-algebr}.  For $d\geq 4$ we need only
  solve the equations \eqref{eq:obstructions} and quotient by the
  action of the automorphisms of the static algebra.  For $d\leq 3$
  the deformation problem has to be looked anew.

\item Classify the generalised Schrödinger algebras of
  Section~\ref{sec:gener-schr-algebr} for $d \leq 3$ (for $\dim E =
  3$) and for $d=2$ (for $\dim E = 2$).

\item Study the addition of generators to the generalised Lifshitz
  algebras of Section~\ref{sec:gener-grad-conf}.  This is not
  unrelated to the classification of the generalised conformal
  algebras where one of the scalar generators is a grading element.

\item Classify filtered subdeformations of $\so(d{+}1,2)$ or
  $\so(d{+}2,1)$ containing an $\so(d)$ subalgebra.
\end{itemize}

A possible extension of this work is to determine the
($d{+}1$)-dimensional homogeneous manifolds of the corresponding groups,
as was done in \cite{Figueroa-OFarrill:2018ilb}.  This might give a
rich class of ``spacetimes'' in which to discuss conformal-like
theories.  Of course, of the graded conformal algebras treated in the
bulk of the paper, only $\hyperlink{ca15}{\GCA{15}^{(\varepsilon)}}$
can act on a ($d+1$)-dimensional (riemannian or lorentzian) manifold
via conformal Killing vectors: their dimension would mean that the
manifold is conformally flat and, hence, any such conformal algebra
would be isomorphic to $\hyperlink{ca15}{\GCA{15}^{(\varepsilon)}}$
for some value of $\varepsilon$.

\section*{Acknowledgments}
\label{sec:acknowledgments}

I am grateful to Jelle Hartong for interesting conversations on the
subject of this paper and to Marc Henneaux for a question after a
seminar I gave in Brussels which prompted me to think about this
subject in the first place.  Some of the early calculations were done
with Paul de Medeiros during a visit to the University of Stavanger in
July 2018 (supported by the Research Council of Norway, Toppforsk
grant no.~250367, held by Sigbjørn Hervik).  It is my pleasure to
thank Paul and Sigbjørn for the invitation and the hospitality.  This
research is partially supported by the grant ST/L000458/1 ``Particle
Theory at the Higgs Centre'' from the UK Science and Technology
Facilities Council.


\providecommand{\href}[2]{#2}\begingroup\raggedright\endgroup

\end{document}